\documentclass[12pt,a4paper]{article}

\usepackage{latexsym}
\usepackage{amsfonts}
\usepackage{amsmath}
\usepackage{amsthm}

\def\d{\partial}

\newcommand{\asy}{\longrightarrow}

\newcommand{\C}[3]{{C_{#1#2}}^{#3}}

\newcommand{\ddl}[2]{\frac{\d^L #2}{\d #1}}

\newcommand{\vddr}[2]{\frac{\delta^R #1}{\delta #2}}
\newcommand{\vddl}[2]{{\frac{\delta^L #1}{\delta #2}}}

\newcommand{\vdd}[2]{{\frac{\delta #1}{\delta #2}}}

\newcommand{\winkel}{\theta}
\newcommand{\Lmat}{L_\mathrm{matter}}

\def\half{{\frac{1}{2}}}

\newcommand{\mysection}[1]{\section{#1}\setcounter{equation}{0}}
\def\be{\begin{eqnarray}}
\def\ee{\end{eqnarray}}
\def\beann{\begin{eqnarray*}}
\def\eeann{\end{eqnarray*}}
\def\beq{\begin{equation}}
\def\eeq{\end{equation}}
\def\ba{\begin{array}}
\def\ea{\end{array}}
\def\ben{\begin{enumerate}}
\def\een{\end{enumerate}}
\def\bea{\begin{eqnarray}}
\def\eea{\end{eqnarray}}

\def\5{\bar }
\def\6{\partial }
\def\7{\hat }
\def\4{\tilde }

\def\cA{{\cal A}}

\def\cC{{\cal C}}

\def\cH{{\cal H}}

\def\cL{{\cal L}}

\def\s0#1#2{\mbox{\small{$\frac{#1}{#2}$}}}

\newtheorem{theorem}{Theorem}
\newtheorem{lemma}{Lemma}
\def\qed{\hbox{${\vcenter{\vbox{
\hrule height 0.4pt\hbox{\vrule width 0.4pt height 6pt
\kern5pt\vrule width 0.4pt}\hrule height 0.4pt}}}$}}

\begin{document}

\begin{titlepage}
\begin{flushright}
ULB-TH/01-19\\
MPI-MIS-94/2001\\
hep-th/0111246
\end{flushright}

\begin{centering}

\vspace{0.5cm}

{\bf{\Large Covariant theory of asymptotic symmetries,\\[6pt]
conservation laws and central charges}} \\

\vspace{1.5cm}

{\large Glenn Barnich$^{*}$}

\vspace{.5cm}

Physique Th\'eorique et Math\'ematique,\\ Universit\'e Libre de
Bruxelles,\\
Campus Plaine C.P. 231, B--1050 Bruxelles, Belgium

\vspace{1cm}

{\large Friedemann Brandt}

\vspace{.5cm}

Max-Planck-Institute for Mathematics in the Sciences,\\
Inselstra\ss e 22--26, D--04103 Leipzig, Germany

\end{centering}
\vspace{.5cm}

\begin{abstract}
Under suitable assumptions on the boundary conditions, 
it is shown that there is a bijective correspondence between
equivalence classes of asymptotic reducibility parameters and
asymptotically conserved $n-2$ forms in the context of 
Lagrangian gauge theories.
The asymptotic reducibility parameters can 
be interpreted as asymptotic Killing vector fields of
the background, with asymptotic behaviour determined by
a new dynamical condition. A universal formula for
asymptotically conserved $n-2$ forms in terms of the reducibility
parameters is derived. 
Sufficient conditions for finiteness of the charges built out of the 
asymptotically conserved $n-2$ forms and for the existence of
a Lie algebra $\mathfrak g$
among equivalence classes of asymptotic reducibility
parameters are given. 
The representation of 
$\mathfrak g$ in terms of the charges may be centrally extended. 
An explicit and covariant formula for the central charges is
constructed. They are shown to be 2-cocycles on the Lie  algebra
$\mathfrak g$. The general considerations and formulas are applied to
electrodynamics, Yang-Mills theory and Einstein gravity.  
\end{abstract}

\vfill

\footnotesize{$^*$Research Associate of the Belgium National Fund for
Scientific Research.}

\end{titlepage}

\tableofcontents

\mysection{Introduction and summary}

\subsection{The Noether-charge puzzle for gauge symmetries}

A physical motivation for studying asymptotic
symmetries and conservation laws is the problem of
a systematic construction
of meaningful charges related to gauge symmetries,
such as the electric charge in electrodynamics or energy
and momentum in general relativity.
This problem is closely related to the famous
Noether-charge puzzle for gauge symmetries.
It is encountered when one tries
to define the charge related to a gauge symmetry
``in the usual manner'',
by applying Noether's first theorem \cite{Noether:1918aa}
on the relation of symmetries and
conserved currents. The problem of such an
approach is that a Noether current 
associated to a gauge symmetry necessarily
vanishes on-shell (i.e., for 
{\em every} solution of the Euler-Lagrange equations
of motion), 
up to the divergence of an arbitrary
superpotential. This is a direct consequence of 
Noether's second theorem \cite{Noether:1918aa}
and was already pointed out by Noether herself.
Let us first review this problem and then indicate how it
can be resolved through asymptotic
symmetries and conservation laws.
 
We denote 
the fields of the theory by $\phi^i$, the Lagrangian by $L[\phi]$ and
a generating set of non trivial gauge symmetries by
$\delta_{f}\phi^i=R^i_\alpha(f^\alpha)=R^i_f$ for some operators 
$R^i_\alpha=\sum_{k=0}R^{i(\mu_1\dots\mu_k)}_\alpha\partial_{\mu_1}\dots
\partial_{\mu_k}$ and arbitrary local functions $f^\alpha$ 
(which should be understood as possibly field dependent gauge parameters).
By definition
of a gauge symmetry, the
$R^i_\alpha(f^\alpha)$ satisfy
\bea
R^i_\alpha(f^\alpha)
\vdd{L}{\phi^i}=\partial_\mu j^\mu_{f},\label{sec1basic}
\eea
for some set of local functions $j^\mu_{f}$, where
$\delta L/\delta\phi^i$ is the Euler-Lagrange derivative
of $L[\phi]$ with respect to $\phi^i$. The sets of
local functions $j^\mu_{f}$ are the
Noether
currents associated to the gauge symmetry $\delta_f$.
Noether's second theorem
states that there exist
associated identities
among the Euler-Lagrange equations of motions,%
\footnote{
A particularly simple derivation of (\ref{sec1ni})
is to take the Euler-Lagrange derivative
of (\ref{sec1basic}) with respect to $f^\alpha$ which is
possible because (\ref{sec1basic}) holds for all
functions $f^\alpha$; the result is (\ref{sec1ni}).
Other derivations  can be found for instance in 
\cite{Noether:1918aa,Olver:1993,Bak:1994gf,Julia:1998ys,Barnich:2000zw}.
}
\bea
R^{+i}_\alpha \vdd{L}{\phi^i}=0,\label{sec1ni}
\eea
where the operators
$R^{+i}_\alpha$ 
are obtained from the operators $R^i_\alpha$ 
defining the gauge transformations through
``integrations by parts'' and ``forgetting about the boundary
term''. Explicitly, they are defined for all collections of 
local functions 
$Q_i$ by $R^{+i}_\alpha(Q_i)=
\sum_{k=0}(-)^k\partial_{\mu_1}\dots\partial_{\mu_k}
[R^{i(\mu_1\dots\mu_k)}_\alpha Q_i]$.

Doing these integrations by parts without forgetting any
boundary terms, one obtains
\bea
\forall Q_i,f^\alpha:\quad
Q_iR^i_\alpha(f^\alpha)=f^\alpha
R^{+i}_\alpha(Q_i)+\partial_\mu S^{\mu i}_\alpha(Q_i,f^\alpha),
\label{1.3}
\eea
where the $S^{\mu i}_\alpha(\cdot,\cdot)$ are 
some bidifferential operators.
Choosing now $Q_i=\delta L/\delta\phi^i$ and using 
\eqref{sec1ni}, one obtains
\bea
\vdd{L}{\phi^i}R^i_\alpha(f^\alpha)=
\partial_\mu S^{\mu i}_\alpha(\vdd{L}{\phi^i},f^\alpha).\label{sec1cur}
\eea
Hence, $S^{\mu i}_\alpha(\delta L/\delta\phi^i,f^\alpha)$
is a particular Noether current satisfying (\ref{sec1basic}). This current
vanishes on-shell
because it is a linear combination (with field dependent
coefficients) of the Euler-Lagrange derivatives
$\delta L/\delta\phi^i$ and their derivatives. 
For any other current $j^\mu_{f}$ satisfying
(\ref{sec1basic}) one obtains from
(\ref{sec1basic}) and (\ref{sec1cur}):
\bea
\partial_\mu \Big(j^\mu_f-S^{\mu
  i}_{\alpha}(\vdd{L}{\phi^i},f^\alpha)\Big) =0.
\label{xxl}
\eea

Using the algebraic Poincar\'e lemma 
(see e.g. 
\cite{Vinogradov:1977,Takens:1979aa,Tulczyjew:1980aa,%
Anderson:1980aa,DeWilde:1981aa,Tsujishita:1982aa,Olver:1993,%
Brandt:1990gy,Wald:1990aa,Dubois-Violette:1991is,Dickey:1992aa}),
one concludes from (\ref{xxl}) that $j^\mu_{f}$
is given on-shell by the divergence of a
superpotential $k^{[\nu\mu]}_f$,
\bea
n>1:\quad
j^\mu_f=S^{\mu
  i}_{\alpha}(\vdd{L}{\phi^i},f^\alpha)-\partial_{\nu} k^{[\nu\mu]}_f\, ,
\label{trivcurrent}
\eea
where $n$ is the spacetime dimension%
\footnote{For simplicity and clarity
we asssume throughout 
the paper that the topologies of spacetime and 
of the field space are trivial, and that so are all
bundles possibly associated with the fields. 
If one drops this assumption, (\ref{trivcurrent})
still holds locally, but globally its right hand side
may contain in addition a topological conserved current,
and other equations of the paper may get
analogously modified.} (in one dimension
one has instead $j_f=S^{i}_{\alpha}(\delta L/\delta\phi^i,f^\alpha)
+C$ where $C$ is an arbitrary constant).
(\ref{trivcurrent}) is the general solution of (\ref{sec1basic}) for given
$L$ and $\delta_f$. Note that the 
superpotential is completely {\em arbitrary}
because it drops out of (\ref{sec1basic}) owing
to $\partial_\mu\partial_{\nu} k^{[\nu\mu]}_f=0$.
This implies in particular that the Noether charge
corresponding to $\delta_f$ is undefined
because it is given by the surface integral of an arbitrary $n-2$
form,
\bea
n>1:\quad
Q[\phi(x)]=\int_\Sigma j_f|_{\phi(x)}=
\int_{\partial \Sigma} k_f|_{\phi(x)}, 
\label{badcharge}
\eea
where $\phi(x)$ is a solution of the Euler-Lagrange equations of
motion, $\Sigma$ is an $n-1$ dimensional spacelike
surface of spacetime with
boundary $\partial \Sigma$, $j_f$ is the current $n-1$ form
and $k_f$ is the $n-2$ form associated to the superpotential
defined according to
\beann
& j_f=j^\mu_f (d^{n-1}x)_\mu\, ,\quad
k_f=k_f^{[\mu\nu]}(d^{n-2}x)_{\mu\nu}\, ,&
\\
&(d^{n-p}x)_{\mu_1\dots\mu_p}:=
\frac 1{p!(n-p)!}\, \epsilon_{\mu_1\dots\mu_n}
dx^{\mu_{p+1}}\dots dx^{\mu_n},\quad
\epsilon_{0\dots (n-1)}=1.&
\eeann

Equation (\ref{badcharge}) expresses the problem
described in the beginning, but at the same time it hints at
a resolution of this problem. Since (\ref{badcharge}) is the
flux of the superpotential through the boundary $\partial \Sigma$, 
it depends solely on the
properties of the superpotential near the boundary.
The situation is familiar from electrodynamics
where the electric charge reduces to the flux of the electric
field and the superpotential is
the field strength $F^{\mu\nu}$ itself.
This suggests to define charges of gauge symmetries
through corresponding superpotentials
rather than through currents, and
calls for an appropriate criterion which
allows one to single out these superpotentials, such
as $F^{\mu\nu}$ in electrodynamics.
Such a criterion is the 
requirement that the superpotentials
be asymptotically conserved $n-2$ forms, for
specific boundary conditions imposed on the fields.

This indicates that there may exist a general relation
between asymptotically conserved $n-2$ forms and
gauge symmetries in Lagrangian gauge theories.
The formulation of such a
relation is one of our central results. It states that, under
quite generic conditions,
every nontrivial asymptotically conserved $n-2$ form
is related to nontrivial 
asymptotic reducibility parameters, and vice versa.
Asymptotic reducibility parameters are the parameters $f^\alpha$ of
gauge transformations that vanish sufficiently fast when evaluated at 
a background field configuration characterizing (partly) the 
boundary conditions for the fields. 
Therefore asymptotic reducibility parameters may be interpreted 
as asymptotic Killing vectors of the background.
Asymptotic symmetries are generated by gauge transformations whose
parameters are asymptotic reducibility parameters. 

\subsection{Various approaches}

The link between asymptotically conserved charges and 
asymptotic reducibility parameters had been known for a long time
in general relativity (see e.g. 
\cite{Arnowitt:1962aa,Regge:1974zd,Abbott:1982ff} and references therein), 
where the gauge symmetry is diffeomorphism invariance and the 
asymptotic reducibility parameters are asymptotic Killing 
vectors of the background metric. 
Later it was realized that it also applies to other gauge theories,
such as Yang-Mills theory \cite{Abbott:1982jh}.

{}From a general point of view, a criterion for the construction of
asymptotic charges and their relation to asymptotic
symmetries was given in \cite{Regge:1974zd} in the context of the
Hamiltonian formalism. 
This criterion was subsequently used in
\cite{Brown:1986nw,Brown:1986ed}
to develop the canonical theory of the central charges that appear
in the representation of the Lie algebra of asymptotic symmetries in
terms of the Poisson brackets of the canonical generators.
The problem of 
defining and constructing asymptotically conserved currents and charges and of
establishing their correspondence with asymptotic symmetries 
in a manifestly covariant way has
received of lot of attention for quite some time. Recent approaches
are often referred to as 
the Lagrangian Noether method 
\cite{Bak:1994gf,Julia:1998ys,Silva:1998ii,%
Aros:1999id,Julia:2000er,Julia:2000sy,Francaviglia:2001ww}
or the covariant phase space
approach \cite{Iyer:1994ys,Wald:1999wa} 
(see also \cite{Anco:2001aa,Anco:2001bb}).  Let us also mention 
quasi-local techniques \cite{Brown:1993br} and conformal methods
(see e.g. \cite{Ashtekar1980,Ashtekar:1999jx} and references therein).

{}From a more technical point of view, the starting point for the
present paper is the recent investigation by 
Anderson and Torre \cite{Anderson:1996sc} (see also \cite{Torre:1997cd})
who have shown
that lower degree asymptotic conservation laws should be 
understood as suitable
asymptotic cohomology groups of the variational bicomplex pulled
back to the surface defined by the equations of motion.

Our approach is guided by the results derived in
\cite{Barnich:1995db} for exact global
reducibility identities and conserved $n-2$ forms.
For instance, in the case of pure Maxwell theory, 
the superpotential $F^{\mu\nu}$ is related in a precise way 
(through descent equations) to the global
reducibility of the electromagnetic gauge symmetry
which constrains the gauge parameter $f$ through $\partial_\mu f=0$,
and such a relation between global
reducibility identities and conserved $n-2$ form holds generally
for gauge theories \cite{Barnich:1995db}. For
interacting gauge theories, such as Yang-Mills theory
with a semi-simple gauge group or general relativity, there are no
non trivial reducibility identities and consequently no non trivial 
conserved $n-2$ forms. However,
nontrivial {\em asymptotic} reducibility identities and 
{\em asymptotically} conserved $n-2$ forms may well exist
in interacting gauge theories, if the theory near the boundary 
becomes asymptotically linear
when expanded around a suitable background. This is the
basic idea of
\cite{Anderson:1996sc,Torre:1997cd} that also underlies our paper.
It allows us to apply and extend the methods used
in \cite{Barnich:1995db}
to asymptotic quantities.

\subsection{Summary of results}

\subsubsection{Asymptotic reducibility parameters and asymptotic
  symmetries\label{s131}}

Suppose that
$\bar\phi(x)$ is a background solution of the Euler Lagrange equations
describing the theory near the
boundary. 
Let us decompose the fields according to $\phi=\bar\phi(x)+\varphi$
and suppose that there are some functions $\chi^i$ that specify the 
boundary conditions imposed on the field $\phi^i$
through $\varphi^i\longrightarrow O(\chi^i)$. Typically
one has $\chi^i=1/r^{m^i}$ for some number $m^i$
that may depend on the specific field.

We denote by $L^{\rm free}$ the Lagrangian of the theory
linearized around the background,  and by $\chi_i$ the asymptotic
behaviour of the corresponding field equations times the volume form,
when evaluated for generic fields $\varphi^i$ that satisfy the boundary
conditions,
\bea
\forall \varphi^i(x)\asy O(\chi^i):\ 
\frac{\delta L^{\rm free}}{\delta \varphi^i}\big|_{\varphi(x)}d^nx\asy
O(\chi_i). 
\eea
Let $\psi_i$ denote a field that behaves asymptotically as the
linearized field equations times the volume form. 
Asymptotic reducibility parameters are
functions $\tilde f^\alpha(x)$ which satisfy
\bea
\forall \psi_i\asy O(\chi_i):\ 
\psi_iR^i_\alpha|_{\bar\phi(x)}(\tilde f^\alpha)\longrightarrow
0.
\label{i1.8}
\eea
Furthermore let
$\chi_\alpha$ denote the asymptotic behaviour of the Noether
operators evaluated at the background when acting on a field $\psi_i$,
\bea
\forall \psi_i\asy O(\chi_i):\ R^{+i}_\alpha\big|_{\bar\phi(x)}(\psi_i)\asy
O(\chi_\alpha).
\eea
Then functions $\tilde f^\alpha$ that fall-off as 
$\tilde f^\alpha\asy o(1/\chi_\alpha)$
are automatically asymptotic reducibility parameters if we assume
that integrations by parts do not increase the asymptotic degree.  
Such asymptotic reducibility parameters are considered as trivial, 
and two sets
of asymptotic reducibility
parameters are called equivalent
($\sim$) if they differ by a trivial one.

By definition, asymptotic symmetries are 
gauge transformations $R^i_\alpha(\tilde f^\alpha)$ 
with asymptotic reducibility parameters.

Notice that (\ref{i1.8}) is {\em not} the condition that
asymptotic reducibility parameters yield gauge transformations which
preserve the boundary conditions
satisfied by the fields. Indeed, such a condition,
which one encounters in many articles on
asymptotic symmetries and conservation laws,
would read $R^i_\alpha|_{\bar\phi(x)}(\tilde f^\alpha)\asy O(\chi^i)$, 
whereas (\ref{i1.8}) requires 
$R^i_\alpha|_{\bar\phi(x)}(\tilde f^\alpha)\asy o(1/\chi_i)$
which involves the boundary
conditions of the asymptotic equations of motion
rather than those of the fields.
However, notice also that the two conditions have
important solutions in common, namely those which satisfy 
$R^i_\alpha|_{\bar\phi(x)}(\tilde f^\alpha)=0$ (with exact rather
than asymptotic equality). Adopting the
terminology of general relativity, these solutions
may be called ``Killing vectors of the background''.
Asymptotic reducibility parameters may thus be interpreted
as ``asymptotic Killing vectors of the background'' but one
should bear in mind that their
asymptotic behaviour is determined by the
asymptotic behaviour of the linearized equations 
of motion and not directly by the boundary conditions for the fields
themselves.

\subsubsection{Asymptotically conserved n--2 forms}

An $n-2$ form constructed of the
fields and their derivatives is called asymptotically conserved 
if its exterior derivative vanishes asymptotically for
the solutions of the field equations satisfying the boundary
conditions. When the theory is asymptotically linear,
asymptotically conserved $n-2$ forms
can be constructed as follows.
Let $\delta {L^{\rm free}}/{\delta \varphi^i}$ denote the 
``left hand sides'' of the
field equations linearized around the background and   
\bea
s_{\tilde f}^\mu[\varphi;\bar\phi(x)]=
S^{\mu i}_\alpha|_{\bar\phi(x)}(\vdd{L^{\rm
      free}}{\varphi^i},\tilde f^\alpha)\label{smalls}
\eea
the current defined by equation \eqref{1.3},
evaluated at the background
solution, for the linearized
field equations and for the asymptotic reducibility
parameters. The divergence of this current (multiplied by the volume
form)
vanishes asymptotically, 
$\partial_\mu s^\mu_{\tilde f}d^{n}x\longrightarrow 0$,  
when the $\tilde f^\alpha$ are asymptotic reducibility
parameters as one deduces from (\ref{sec1cur}).
This implies, under suitable assumptions on the
boundary conditions, 
that the $n-1$ form $s^\mu_{\tilde f}(d^{n-1}x)_\mu$ is
asymptotically the exterior derivative of an $n-2$ form that is a local
function in the fields $\varphi^i$,
\bea
s^\mu_{\tilde f}[\varphi;\bar\phi(x)](d^{n-1}x)_\mu\longrightarrow
-d_H\tilde k^{[\nu\mu]}_{\tilde
  f}[\varphi;\bar\phi(x)](d^{n-2}x)_{\nu\mu}
\label{SUPERPOT1}
\eea
with $d_H=dx^\mu\partial_\mu$ and $\partial_\mu$ the total
derivative, so that $d_H\tilde k_{\tilde
f}=\partial_\nu \tilde k^{[\mu\nu]}_{\tilde
f}(d^{n-1}x)_\mu$. 
The explicit expression of $\tilde k^{[\nu\mu]}_{\tilde f}$ is obtained
by applying the contracting homotopy of the algebraic Poincar\'e
lemma to $s^\mu_{\tilde f}$. 
In the case of equations of motion that are at most of second order
in derivatives,
this expression reduces to 
\bea
\tilde k^{[\mu\nu]}_{\tilde f}=
\frac{1}
{2}\varphi^i\frac{\partial^S s^{\nu}_{\tilde f}}{\partial
\varphi^i_{\mu}}
+\Big(\frac{2}{3}
\varphi^i_\lambda-\frac{1}{3}\varphi^i\partial_\lambda\Big)
\frac{\partial^S s^{\nu}_{\tilde f}}{\partial
\varphi^i_{\lambda\mu}}
-(\mu\leftrightarrow\nu)
\label{superpot2}
\eea
where lower indices of $\varphi^i$ represent derivatives, 
i.e., $\varphi^i_\mu$ and $\varphi^i_{\mu\nu}$ represent 
the first and second
order derivatives of $\varphi^i$, respectively, and 
the operation $\partial^S/\partial\varphi^i_{\mu_1\dots\mu_k}$
is defined according to
$\partial^S \varphi^i_{\mu_1\dots\mu_k}/\partial
\varphi^j_{\nu_1\dots\nu_k}
=\delta^i_j\delta^{\nu_1}_{(\mu_1}\dots\delta^{\nu_k}_{\mu_k)}$. Here,
the round parentheses denote symmetrization with weight one. For
example, 
${\6^S\varphi^i_{\mu\nu}}/{\6\varphi^j_{\rho\lambda}}=
1/2\ \delta^i_j(\delta_\mu^\rho\delta_\nu^\lambda+
\delta_\nu^\rho\delta_\mu^\lambda)$.

For the remainder of the introduction, we restrict ourselves to the
case where the asymptotic behaviour of the asymptotic reducibility
parameters is given by 
\bea
\tilde f^\alpha\asy O(1/\chi_\alpha).\label{iconvergencec} 
\eea
An asymptotic solution $\varphi_s(x)$ of the 
linearized equations of motion near the boundary is defined by 
\bea
\frac{\delta L^{\rm
    free}}{\delta\varphi^i}\big|_{\varphi_s(x)}d^nx\asy o(\chi_i).
\eea
In this case, the associated $n-2$ forms $\tilde k_{\tilde f}$ satisfy the 
asymptotic conservation law
\bea
d_H\tilde k_{\tilde
f}|_{\varphi_s(x)}
\longrightarrow 0\label{1.12}.
\eea
Defining a corresponding $n-2$ form of the full theory by 
\bea
k_{\tilde f}[\phi;\bar\phi(x)]
=\tilde k_{\tilde
f}[\phi-\bar\phi(x);\bar\phi(x)]\label{eq1.13}
\eea
and considering an infinitesimal field variation, 
\bea
d_V=\sum_{k=0}
d_V\phi^i_{\mu_1\dots\mu_k}
\frac{\partial^S}{\partial \phi^i_{\mu_1\dots\mu_k}}\ ,\label{vertdiff}
\eea
one has by construction 
$(d_Vk_{\tilde f})|_{\bar\phi(x),\varphi}=
\tilde k_{\tilde
f}[\varphi;\bar\phi(x)]$, 
and the asymptotic conservation law can be written from the point of
view of the full theory as 
\bea
d_H(d_V k_{\tilde
f})|_{\bar\phi(x),\varphi_s(x)}\longrightarrow 0.\label{1.13}
\eea
When the theory is asymptotically linear, we {\em define}
an asymptotically conserved $n-2$ form $k$ to be
an $n-2$ form for which 
this equation holds.
Two asymptotically conserved 
$n-2$ forms should be considered equivalent
if their
linearization around the given background agree asymptotically up to
the horizontal differential of an $n-3$ form when evaluated 
for asymptotic solutions, 
\bea
k_1
\sim k_2
\iff
\big(d_V(k_1
-k_2
)\big)|_{\bar\phi(x),\varphi_s(x)}
\longrightarrow
d_H\tilde l^{n-3}
\eea
where $\tilde l^{n-3}$ 
depends
linearly on the fields $\varphi$ and their derivatives.

\subsubsection{Bijective correspondence}

With these definitions and assumptions, we will show that every 
asymptotically conserved $n-2$ form is equivalent to an $n-2$ form  
$k_{\tilde f}$ obtained from (\ref{SUPERPOT1}) for some asymptotic
reducibility parameters $\tilde f^\alpha$, and
that there is a one-to-one correspondence
between equivalence classes $[\tilde f^\alpha]$ of asymptotic
reducibility parameters and equivalence classes
$[k]$ of asymptotically conserved $n-2$ forms. 
This is the analog of (the complete version of) Noether's first
theorem for the case of asymptotic symmetries.  
Furthermore, for irreducible gauge theories, there are no
nontrivial asymptotically 
conserved forms in degrees strictly smaller than $n-2$. 

\subsubsection{Charges\label{s134}}

Consider an $n-2$ dimensional compact 
manifold ${\cal C}^{n-2}$
without boundary, $\partial  {\cal C}^{n-2}=\emptyset$, 
that lies
in the asymptotic region and an asymptotically conserved $n-2$ form
$\tilde k_{\tilde f}$. 
The charge in the full theory is defined by
\bea
Q_{\tilde f}[\phi;\bar\phi(x)]
=\int_{{\cal C}^{n-2}}\tilde k_{\tilde
f}[\phi-\bar\phi(x);\bar\phi(x)]+N_{\tilde f},
\eea
where the normalization $N_{\tilde f}$ is the charge of the 
background $\bar\phi(x)$.
If we evaluate this integral for an asymptotic solution 
$\phi(x)=\bar\phi(x)+\varphi_s(x)$,
the charges are finite when
(\ref{iconvergencec}) holds, and 
we can then apply Stokes theorem because of
the conservation law \eqref{1.12} to prove 
asymptotic independence on
the choice of representatives for
the homology class $[{\cal C}^{n-2}]$ and for the equivalence class 
$[\tilde k_{\tilde
f}]$. 

\subsubsection{Algebra of asymptotic reducibility parameters 
\label{s135}}

Because we have assumed that $\delta_{f}
\phi^i=R^i_\alpha(f^\alpha)$ provide a generating set of non
trivial gauge symmetries, the
commutator algebra of the non trivial 
gauge symmetries closes on-shell in the sense
\bea
\delta_{f_1}R^i_\alpha(f_2^\alpha)-(1\longleftrightarrow 2)
\approx
R^i_\gamma(C^\gamma_{\alpha\beta}(f_1^\alpha,f_2^\beta)
+\delta_{f_1}f_2^\gamma-\delta_{f_2}f_1^\gamma),\label{1.18}
\eea
for some bidifferential operators
$C^\gamma_{\alpha\beta}$ ($\approx$ denotes
equality on-shell). Additional (sufficient) contraints on the asymptotic
reducibility parameters,  on the cubic vertex of the theory and on the
gauge symmetries of the linearized theory will be given
that guarantee that 
asymptotic reducibility
parameters $\tilde f^\alpha$ form a Lie algebra,
with bracket determined by the 
structure operators $C^\gamma_{\alpha\beta}$ evaluated at the background, 
\bea
[\tilde f_1,\tilde f_2]_M^\gamma=
C^{\gamma 0}_{\alpha\beta}(\tilde f_1^\alpha,\tilde f_2^\beta).
\eea
with $C^{\gamma 0}_{\alpha\beta}\equiv 
C^\gamma_{\alpha\beta}|_{\bar\phi(x)}$. 
This Lie algebra induces a well defined 
Lie algebra $\mathfrak g$ for the 
equivalence classes with bracket denoted by $[\ ,\ ]_G$:
\bea
[[\tilde f_1],[\tilde f_2]]_G=
[[\tilde f_1,\tilde f_2]_M].
\eea 

\subsubsection{Induced global symmetries}

Under approriate assumptions, 
the asymptotic reducibility parameters 
can be shown to determine asymptotic linear global
symmetries 
\bea
\delta^g_{\tilde f}\varphi^i=(d_V R^i_\alpha)|_{\bar\phi(x),\varphi}(\tilde
f^\alpha)\equiv  R^{i1}_\alpha(\tilde f^\alpha),
\eea
for the linearized theory,
in the sense that $\delta^g_{\tilde f}L^{\rm free}\asy d_H(\ )$. 

\subsubsection{Algebra of asymptotically
  conserved n--2 forms\label{s137}}

On the level of the equivalence classes of asymptotically 
conserved $(n-2)$-forms of the
linearized theory near the boundary, the Lie algebra $\mathfrak g$ 
of the 
equivalence classes of asymptotic 
reducibility parameters
can be represented asymptotically 
by a covariant Poisson bracket, which is defined
through the action of the associated global symmetry,
\bea
\{[\tilde k_{\tilde f_1}],[\tilde k_{\tilde f_2}]\}_F:=
[\delta^g_{\tilde f_1}\tilde k_{\tilde f_2}]
=
[\tilde k_{[\tilde f_1,\tilde
f_2]_M}].\label{i1.25}
\eea
The property $-[\delta^g_{\tilde f_2}\tilde k_{\tilde f_1}]=
[\tilde k_{[\tilde f_1,\tilde
f_2]_M}]$ implies that alternative equivalent expressions for 
the covariant Poisson bracket are $-[\delta^g_{\tilde
f_2}\tilde k_{\tilde f_1}]$ or $
\half ([\delta^g_{\tilde f_1}\tilde k_{\tilde f_2}]-[\delta^g_{\tilde
f_2}\tilde k_{\tilde f_1}])$.

\subsubsection{Algebra of charges and central extensions\label{s138}}

On the level of the charges of the full theory, the Lie algebra
$\mathfrak g$ can be represented by a covariant Poisson bracket that is
defined by applying 
an asymptotic symmetry $\delta_{\tilde f} \phi^i=
R^i_\alpha(\tilde f^\alpha)=R^i_{\tilde f}$. This representation 
may contain non trivial central
extensions. Explicitly,
\bea
\{Q_{\tilde f_1},Q_{\tilde f_2}\}_{CF}:=\delta_{\tilde f_1}Q_{\tilde f_2}
\sim
Q_{[\tilde
f_1,\tilde f_2]_M}-N_{[\tilde
f_1,\tilde f_2]_M}+K_{\tilde f_1,\tilde
f_2},
\eea
where $\sim$ is asymptotic equality when the charges are evaluated for
asymptotic solutions, the $N$'s are normalization constants, 
and the $K_{\tilde f_1,\tilde f_2}$ are central charges given by
\bea
K_{\tilde f_1,\tilde f_2}=Q_{\tilde f_2}
[R_{\tilde
f_1}|_{\bar\phi(x)};\bar\phi(x)]=
-Q_{\tilde f_1}[R_{\tilde
f_2}|_{\bar\phi(x)};\bar\phi(x)]. \label{centralcharge}
\eea
These $K$'s are Chevalley-Eilenberg $2$-cocycles 
on the Lie algebra $\mathfrak g$.

\subsection{Organization of the paper}

Our analysis of asymptotic symmetries and conservation laws
is guided by methods and results known in the context of
exact symmetries and conservation laws. Therefore
we first summarize in sections \ref{s2}
through \ref{s4} facts about exact symmetries 
and conservation laws. The results on
asymptotic symmetries and conservation laws are collected in section
\ref{s5} and illustrated in section \ref{s6} for 
the most prominent gauge theories.
The details of the analysis are presented
in section \ref{s7} and in the appendix.
Section 8 describes briefly the relation to other approaches
to asymptotic symmetries and conservation laws. 

\subsubsection{Section 2}

We review, besides well known facts on global 
symmetries and conserved currents, the results of 
\cite{Barnich:1995db,Barnich:2000zw} on the bijective 
correspondence between suitably defined equivalence classes of 
global symmetries and conserved currents in the context of gauge
theories, 
without using the 
cohomological tools related to the BRST formalism. 

\subsubsection{Section 3}

The bijective correspondence
between equivalence classes of reducibility parameters of gauge symmetries
and conserved $n-2$ forms \cite{Barnich:1995db,Barnich:2000zw} is
reviewed, 
independently of BRST
cohomological arguments.
Universal formulas for  
the conserved $n-2$ forms associated to reducibility 
parameters are given and the definition and properties of 
corresponding charges are recalled.
It is shown that there is a well defined Lie action of
equivalence classes of global symmetries on equivalence classes of
reducibility parameters and thus also on equivalence classes of
conserved $n-2$ forms. 

\subsubsection{Section 4}

It is shown how the expansion of an interacting
gauge theory around a solution allows one to associate global symmetries
to the (field
independent)  
reducibility parameters of the
linearized theory. The Lie action of these
symmetries is then used to define a Lie algebra 
of equivalence classes of field independent reducibility parameters.

\subsubsection{Section 5}

The section begins with a general discussion of the boundary
conditions followed by a discussion of asymptotic reducibility
parameters and asymptotically conserved $n-2$ forms and their 
algebra from the point of
view of the linearized theory near the boundary. The assumptions that
allow one to use and reexpress these results from the point of view of the
bulk theory are discussed next. 
Finally, some remarks on the associated boundary theory
are given. 

\subsubsection{Section 6}

It is shown how the familiar expression for the electric charge
in electrodynamics
arises from an asymptotically conserved $n-2$ form related to
an exact Killing vector of the background. 
Non abelian Yang-Mills theories are discussed next
and it is shown that and how our results reproduce those of
\cite{Abbott:1982jh}. Then we discuss in more detail
Einstein gravity in spacetime dimensions larger than $2$, with or without 
cosmological constant. 
We derive a general expression for the gravitational
asymptotically conserved $n-2$ forms which reproduces,
in the particular case that the reducibility parameters are
exact Killing vectors of the background, the
expressions given in \cite{Abbott:1982ff} and \cite{Anderson:1996sc}.
We also derive an explicit general expression for the
potential gravitational central charges which, to our knowledge,
is completely new, and illustrate the 
covariant theory of central charges in the
case of $3$-dimensional asymptotically anti-de Sitter gravity, where
the results of \cite{Brown:1986nw} obtained in the canonical framework
are recovered.  

\subsubsection{Section 7}

BRST cohomological methods are used
to reformulate, prove and partly generalize the statements of the
previous sections, with
subsection \ref{s7}.$x$ corresponding to section $x$ for $x=2,3,4,5$,
while in subsection \ref{s7}.1 we first recall the basic
ingredients of the BRST approach (antifields, 
ghost fields, Batalin-Vilkovisky master equation, antibracket,
BRST differential).
The cohomological formulation of Noether's first theorem
and of the relation between reducibility parameters and conserved
$n-2$ forms via descent equations \cite{Barnich:1995db,Barnich:2000zw} 
is briefly reviewed. 
The induced global symmetries, and the
Lie algebras discussed in the previous sections
are derived from the antibracket map.
In order to discuss asymptotic symmetries and
conservation laws, we define linear and exact linear
characteristc cohomology. When evaluated at a background this latter
cohomology is shown to provide the right framework to discuss asymptotic
reducibility parameters, asymptotically conserved $n-2$ forms and to
prove the bijective correspondence between the appropriate equivalence
classes. 

\subsubsection{Section 8}

The covariant theory of asymptotic symmetries and
conservation laws derived in the previous sections 
is related to the original Regge-Teitelboim
canonical approach \cite{Regge:1974zd}. 
The comparison with the covariantized
Regge-Teitelboim formalism \cite{Silva:1998ii}
recently proposed in the context of the
Lagrangian Noether method is direct. The main formulas that allow one to
connect our results to the covariant phase space method are given, and
finally we briefly compare 
the assumptions, techniques and results of our investigation 
to those of the original cohomological analysis in the context of 
the variational bicomplex \cite{Anderson:1996sc}.  

\subsubsection{Appendix}

In the appendix we first collect conventions and notation, especially
those concerning multiindices, then we give
compact expressions
for higher order Lie-Euler
operators and for the contracting homotopy of the horizontal complex,
and finally the proof of a central theorem of the paper.

\mysection{Global symmetries and conserved currents}\label{s2}

\subsection{Definitions}

Global symmetries are evolutionary vector fields with characteristic
$X^i$ such that their prolongation 
leaves the Lagrangian $L$ invariant up to a total divergence,
\bea
\delta_X L=\partial_\mu k^\mu.\label{1}
\eea
Conserved currents $j^\mu$ are currents whose divergence vanishes
when the Euler-Lagrange equations of motion hold,
\bea
\partial_\mu j^\mu=Y^{i(\nu)}\partial_{(\nu)}\frac{\delta L}{\delta
  \phi^i}.\label{2}
\eea
Note that in this context, the
Lagrangian only serves to define the dynamics through its
Euler-Lagrange derivatives and is defined up to a total
divergence because
\bea
\frac{\delta f}{\delta \phi^i}=0\Longleftrightarrow
f=\partial_\mu m^\mu,\label{3}
\eea
for some local functions $m^\mu$.
This ambiguity does not affect the definition of the global symmetries
because $[\delta_{ X},\partial_\mu]=0$. Note also that the spacetime points
$x^\mu$ are not transformed so that $\delta_X\phi^i$ corresponds to the
``variations of the fields at the same point''.

\subsection{From symmetries to conserved currents}

The current $V^\mu_i(Q^i,f)$ is defined through the equation 
\bea
\delta_Q f=
Q^i\frac{\delta f}{\delta\phi^i}+\partial_\mu V^\mu_i(Q^i,f)\label{fun},
\eea
for all $Q^i$. It then follows from \eqref{1} that 
$j^\mu=k^\mu-V^\mu_i(X^i,L)$ is a conserved current because it
satisfies 
\bea
\partial_\mu j^\mu=X^i\frac{\delta L}{\delta\phi^i}.
\eea

\subsection{From conserved currents to symmetries}

Using on the right hand side
of equation \eqref{2} repeatedly Leibniz' rule under the form 
$f\6_\mu g=\6_\mu(fg)-(\6_\mu f)g$, and bringing the terms 
$\6_\mu(fg)$ to the left hand side, one obtains
\bea
\partial_\mu j^{\prime\mu}=
(-\partial)_{(\nu)}Y^{i(\nu)}\vdd{L}{\phi^i}\ . 
\eea
Using the same notation as in (\ref{fun}), one
has
\[
(-\partial)_{(\nu)}Y^{i(\nu)}\vdd{L}{\phi^i}=
\delta_{X} L-\partial_\mu V^\mu_i(X^i,L),
\]
with 
\[
X^{i}=(-\partial)_{(\nu)}Y^{i(\nu)}.
\]
Combining these equations, one obtains that
equation \eqref{2} implies \eqref{1}, with
$X^{i}=(-\partial)_{(\nu)}Y^{i(\nu)}$.

\subsection{Bijective correspondence between equivalence classes}
\label{2.6}

In order to understand a complete version of Noether's first
theorem on the correspondence
between global symmetries and conserved currents in the context of
gauge theories,
it is crucial to define what are trivial
symmetries and currents and to consider equivalence classes of
symmetries respectively currents up to trivial ones.

Indeed, both correspondences between symmetries and currents 
are not uniquely defined because of the existence of 
identically conserved currents 
$
\partial_\mu j^\mu=0$,
and the existence of Noether identities,
$
N^{i(\mu)}\partial_{(\mu)}\delta L/\delta \phi^i=0$.
For identically conserved currents, 
the algebraic Poincar\'e lemma mentioned already
in the introduction guarantees that, at least locally,
\bea
\partial_\mu j^\mu=0\iff 
j^\mu=\partial_\nu k^{[\nu\mu]}+\delta^n_1 C,\label{ass1}
\eea
for some local functions $k^{[\nu\mu]}$ and 
some constant $C\in {\mathbb  R}$. This is equivalent to the statement
that the cohomology of $d_H$ vanishes in form degree $n-1$ 
(locally) except for
the constants in spacetime dimension $n=1$. 

The requirement that the operators $R^{i}_\alpha$ 
be a generating set of gauge 
symmetries means that every gauge symmetry, i.e.,
every symmetry of the Lagrangian whose characteristic $G^i$ depends 
linearly and
homogeneously on an arbitrary local function $f$ and its derivatives, 
$G^i(f)=G^{i(\mu)}\partial_{(\mu)}f$ with
$\delta_{G(f)}L=\partial_\mu k^\mu(f)$, can be written as 
\bea
G^i(f)=R^i_\alpha(Z^\alpha(f))+M^{+ji}(\vdd{L}{\phi^j},f),\label{gs}
\eea
for some operators $Z^\alpha=Z^{\alpha(\mu)}\partial_{(\mu)}$.
Here and in the following,
$M^{ji}(Q_j,g)$ and $M^{+ji}(Q_j,g)$ (with two arguments)
are regarded as differential operators 
acting on their second argument ($g$), and
$M^{+ji}(Q_j)$ (with only one argument)
as local functions equal to $M^{+ji}(Q_j,1)$,
\bea
M^{ji}(Q_j,\ \cdot\ )&=&
\partial_{(\nu)}Q_j\, M^{[j(\nu)i(\mu)]}\,\partial_{(\mu)}\, \cdot\ ,
\nonumber\\
M^{+ji}(Q_j,\ \cdot\ )&=&(-\partial)_{(\mu)}
(\ \cdot\ M^{[j(\nu)i(\mu)]}\,\partial_{(\nu)}Q_j),
\nonumber\\
M^{+ji}(Q_j)&=&(-\partial)_{(\mu)}
(M^{[j(\nu)i(\mu)]}\,\partial_{(\nu)}Q_j),
\label{Mdef}
\eea
where
\bea
M^{[j(\nu)i(\mu)]}=-M^{[i(\mu)j(\nu)]}.
\label{Mdef2}
\eea

Equivalently, every operator
$N^{i(\mu)}\partial_{(\mu)}$ defining a Noether identity 
can be written as
\bea
N^{i(\mu)}\partial_{(\mu)}\delta L/\delta \phi^i=0\iff 
N^{i(\mu)}\partial_{(\mu)}=Z^{+\alpha}\circ R^{+i}_\alpha+
M^{[j(\nu)i(\mu)]}\partial_{(\nu)}
\frac{\delta L}{\delta \phi^j}\partial_{(\mu)},\label{ass2}
\eea
for some operators $Z^{+\alpha}$ and some $M^{[j(\nu)i(\mu)]}$. 

Armed with these definitions, one 
can prove that there is a bijective correspondence
$[X^i]\longleftrightarrow [j^\mu]$ between
equivalence classes $[X^i]$ of global symmetries $X^i$ 
satisfying (\ref{1}), with
equivalence ($\sim$) defined by
\bea
X^i\sim X^i +R^i_\alpha(f^\alpha)+M^{+ji}(\vdd{L}{\phi^j}),
\eea
and equivalence classes
$[j^\mu]$ of conserved currents $j^\mu$ satisfying (\ref{2}),
with equivalence defined by
\bea
j^\mu\sim j^\mu+\partial_\nu k^{[\nu\mu]}
+l^{\mu i(\nu)}\partial_{(\nu)}\frac{\delta L}{\delta \phi^i}
+\delta^n_1 C,\quad C\in\mathbb{R}.
\eea
The proof using the Koszul-Tate resolution of the stationary surface
and descent equations techniques, originally given
in \cite{Barnich:1995db}, is briefly reviewed in section 
\ref{s7}.

Explicity, the correspondence is given by  
\bea
[X^i]&\longrightarrow& [k^\mu-V^\mu_i(X^i,L)],
\nonumber\\
{}[j^\mu]&\longrightarrow& [(-\partial)_{(\nu)}Y^{i(\nu)}].
\eea

{}From the point of view of equivalence classes of global symmetries, 
symmetries of the form $\delta_f\phi^i=R^i_\alpha(f^\alpha)$ (non
trivial gauge symmetries ) and of the form
$\delta_M\phi^i=M^{+ji}(\vdd{L}{\phi^j})$ (trivial gauge symmetries)
should thus be considered as trivial, while from the
point of view of equivalence classes of conserved currents, trivial
currents are  on-shell equal to the divergence of an arbitrary
superpotential.

One can furthermore show under
fairly general assumptions (linearizable, normal theories;
see theorems 6.5 and 6.6 in \cite{Barnich:2000zw},
and especially the remark after theorem 6.6 there) that global
symmetries whose characteristic $X^i$ vanishes when the
Euler-Lagrange equations of motion hold, can be assumed to be trivial 
\bea
\delta_X L=\partial_\mu k^\mu,\ X^i\approx 0\ \Longrightarrow\ X^i=
R^{i}_\alpha(f^\alpha)+M^{+ji}(\vdd{L}{\phi^j}),\label{6bis}
\eea
for some $f^\alpha$ and some $M^{[j(\nu)i(\mu)]}$.

In
the language of differential forms, the equivalence class $[j]$ 
is an element of the
characteristic cohomology 
in form degree $n-1$ associated with the
surface defined by the Euler-Lagrange equations of motion
\cite{Vinogradov:1978,Tsujishita:1982,Vinogradov:1984,%
Anderson1991,Bryant:1995},
\bea
H^{n-1}_{\rm char}=\{[j]; d_H j\approx 0, j\sim j +d_H k +t +\delta^n_1 C, 
t\approx
0\}, 
\eea
with $j,t$ in form degree $n-1$ and $k$ in form degree $n-2$.

\subsection{Application of Stokes theorem}

Stokes theorem implies that for a compact $n-1$ dimensional manifold
${\cal C}^{n-1}$ without boundary, $\partial{\cal C}^{n-1}=\emptyset$,
and a solution $\phi(x)$ of the equations of motion, the charge
\bea
Q([{\cal C}^{n-1}],[j])[\phi(x)]
=\int_{{\cal C}^{n-1}}j|_{\phi(x)}
\eea
does not dependent on the choice of representatives for the homology
class $[{\cal C}^{n-1}]$ or for the equivalence class $[j]$.

\subsection{Algebra}

The vector space of evolutionary vector fields is
an infinite dimensional
Lie algebra for the bracket defined by
\bea
[Q_1,Q_2]^i_L=\delta_{Q_1} Q^i_2-\delta_{Q_2} Q^i_1.
\eea
The vector space of global symmetries is 
an infinite dimensional  
Lie sub-algebra for this bracket.

Furthermore, the bracket of a trivial global symmetry with
any global symmetry is again trivial.
Indeed, $[R_\alpha(g^\alpha),X]_L^i$
defines a family of symmetries depending 
on the arbitrary local functions
$g^\alpha$ and, by the definition of a generating set of non trivial
gauge symmetries, it can thus be
written as in (\ref{gs}). Similarly, 
$[M^{+j}(\delta L/\delta\phi^j),X]_L^i$ defines a
global symmetry for any choice of functions 
$M^{[j(\nu)i(\mu)]}$, so that
it can again be written as in (\ref{gs}). (Moreover, 
$[M^{+j}(\delta L/\delta\phi^j),X]^i$ vanishes
on-shell because $\delta_{ X} (\delta
L/\delta\phi^i)\approx 0$, which implies by corollary 6.3 of 
\cite{Barnich:2000zw} that it can be assumed to be of the form of the
second term on the right hand side of \eqref{gs}.)

Hence, there is a well defined induced bracket for
the equivalence classes,
\bea
\big[[X_1],[X_2]\big]^i_L=\big[[X_1,X_2]_L^i\big].
\eea
The induced Lie algebra for the equivalence classes of
global symmetries is the relevant algebra from a physical point 
of view. 

There is also a well defined Lie bracket for equivalence classes of
conserved currents, the
Dickey bracket \cite{Dickey:1991xa}.  
This bracket has
the following equivalent expressions:
\bea
\big[[j_1],[j_2]\big]_D^\mu=[\delta_{X_1} j^\mu_2]=[-\delta_{ X_2}
j^\mu_1]
=[\frac{1}{2}
(\delta_{ X_1} j^\mu_2-\delta_{ X_2} j^\mu_1)]=
-[\omega^\mu(X_1,X_2)],\label{2.28}
\eea
where the presymplectic current $2$-form is defined by 
$\omega^\mu=d_V(V_i^\mu(d_V\phi^i,L))$. 
One can show \cite{Dickey:1991xa,Barnich:1996mr}
that the Lie algebras of equivalence classes of
global symmetries and equivalence classes of 
conserved currents are
isomorphic, 
\bea
\big[[X_1],[X_2]\big]_L^i\longleftrightarrow \big[[j_1],[j_2]
\big]_D^\mu.
\eea

\mysection{Reducibility parameters and conserved n--2
  forms}\label{s3}

\subsection{Reducibility parameters}

(Global) reducibility parameters are a collection of local functions 
$f^\alpha$ 
that satisfy the global reducibility identity
\bea
R^i_\alpha(f^\alpha)=M^{+ji}(\frac{\delta L}{\delta \phi^j}),\label{7}
\eea
for some skew symmetric functions
$M^{[j(\nu)i(\mu)]}$, cf.\ (\ref{Mdef}) and (\ref{Mdef2}).
The corresponding gauge transformations
leave thus all solutions of the
equations of motion invariant (``ineffective gauge transformations'').
Trivial reducibility parameters are given by 
functions $f^\alpha$ that
vanish on any solution of the equations of motion, $f^\alpha\approx
0$.
Indeed, such functions always define a global reducibility
identity, because if
$f^\alpha=k^{\alpha j(\nu)}\partial_{(\nu)}{\delta L}/{\delta
  \phi^j}$, then \eqref{7} holds with
$M^{[j(\nu)i(\mu)]}=-R^{+j(\nu)}_\alpha k^{\alpha i(\mu)}+
R^{+i(\mu)}_\alpha k^{\alpha j(\nu)}$, by using the Noether identity
(\ref{sec1ni}). We define equivalence classes 
$[f^\alpha]$ of reducibility parameters 
by identifying parameters
which coincide on-shell, 
\bea
f^\alpha\sim f^{\alpha\prime}\quad \Longleftrightarrow\quad
f^\alpha\approx f^{\alpha\prime}\quad\Longleftrightarrow\quad
f^\alpha-f^{\alpha\prime}=
k^{\alpha j(\nu)}\partial_{(\nu)}{\delta L}/{\delta \phi^j}.
\label{equiv_f}
\eea

\subsection{Conserved n--2 forms}

Conserved $n-2$ forms are defined by superpotentials whose divergence
vanishes when the Euler-Lagrange equations of motion hold,
\bea
\partial_\nu k^{[\nu\mu]}=J^{i\mu(\lambda)}
\partial_{(\lambda)}\vdd{L}{\phi^i}.\label{3.4}
\eea
Equivalence classes of conserved $n-2$ forms are defined by identifying
the superpotentials of $n-2$ forms that differ by 
the sum of a weakly vanishing superpotential and 
the divergence of an antisymmetric tensor with three indices, 
\bea
k^{[\mu\nu]}\sim k^{[\mu\nu]}+\partial_\sigma l^{[\sigma\mu\nu]}+
t^{[\mu\nu]i(\lambda)}\partial_{(\lambda)}\vdd{L}{\phi^i}
+\delta^n_2\epsilon^{\mu\nu}C.
\eea
In the language of differential forms, the equivalence classe $[k]$
is an element of the characteristic cohomology in form degree $n-2$
associated with the Euler-Lagrange equations of motion,
\bea
H^{n-2}_{\rm char}=\{[k]; d_H k\approx 0, k\sim k +d_H l +t+\delta^n_2
C, t\approx
0\}, 
\eea
with $k,t$ in form degree $n-2$ and $l$ in form degree $n-3$. 

\subsection{From reducibility parameters to conserved n--2 forms}

Contracting the reducibility identity \eqref{7} 
with $\delta L/\delta\phi^i$ and
using \eqref{sec1cur}, we obtain
\bea
\partial_\mu[
S_\alpha^{\mu i}(\vdd{L}{\phi^i},f^\alpha)]=M^{+ji}(\vdd{L}{\phi^j})\,
\vdd{L}{\phi^i}\ .\label{3.7bis}
\eea
By definition (\ref{Mdef}) of $M^{+ji}$ and the 
skew-symmetry (\ref{Mdef2}) of $M^{[j(\nu)i(\mu)]}$, one has
\bea
M^{+ji}(\vdd{L}{\phi^j})\,\vdd{L}{\phi^i}
&=&
(-\partial)_{(\mu)}
\Big[M^{[j(\nu)i(\mu)]}\,\partial_{(\nu)}\vdd{L}{\phi^j}\Big]
\vdd{L}{\phi^i}
\nonumber\\
&=&
-\6_\mu M^{\mu ji}(\vdd{L}{\phi^j},\vdd{L}{\phi^i})
+\underbrace{M^{[j(\nu)i(\mu)]}\, 
\partial_{(\nu)}\vdd{L}{\phi^j}\,
\partial_{(\mu)}\vdd{L}{\phi^i}}_{0}
\nonumber\\
&=&
-\6_\mu M^{\mu ji}(\vdd{L}{\phi^j},\vdd{L}{\phi^i})
\label{2.23}
\eea
where the second equality follows by using repeatedly Leibniz' rule
for the derivative $\partial_\mu$. 
Hence, the right hand side of \eqref{3.7bis}
is the divergence of a ``doubly'' weakly vanishing
Noether current $-M^{\mu ji}(\frac{\delta L}{\delta \phi^j},
\frac{\delta L}{\delta \phi^i})$
associated to the trivial global symmetry 
$\delta_{-M}\phi^i=-M^{+ji}(\vdd{L}{\phi^j})$ and
\eqref{3.7bis} reduces to the divergence identity
\bea
\partial_\mu J^\mu_f=0,
\quad
J^\mu_f=S_\alpha^{\mu i}(\vdd{L}{\phi^i},f^\alpha)+M^{\mu ji}(\frac{\delta
  L}{\delta \phi^j},\frac{\delta L}{\delta \phi^i}).\label{3.9}
\eea
Because of \eqref{ass1}, it follows that in dimensions $n\geq 2$,
there exists a superpotential $k^{[\mu\nu]}$ whose
divergence is conserved on-shell,
\bea
\partial_\nu k^{[\nu\mu]}_f=J^\mu_f\approx 0.\label{3.10}
\eea

\subsection{From conserved n--2 forms to reducibility
  parameters}\label{s3.5}

Given an $n-2$ form $k$ with associated superpotential satisfying
\eqref{3.4}, 
it follows, by contracting with $\partial_\mu$ and using skew-symmetry
of the indices of the superpotential, that the operators
$\partial_\mu \circ J^{i\mu}$, with
$J^{i\mu}=J^{i\mu(\lambda)}\partial_{(\lambda)}$
define a Noether identity. Because we assume that the 
$R^{+i}_\alpha$ define a generating set of non trivial Noether
identities \eqref{ass2}, these operators can be expressed as
\bea
\partial_\mu \circ J^{i\mu}=Z^{+\alpha}\circ
R^{i+}_\alpha
-M^{[j(\nu)i(\mu)]}\partial_{(\nu)}\frac{\delta L}{\delta
  \phi^j}\,\partial_{(\mu)}.\label{3.11}
\eea
The adjoint 
operator relations are
\bea
-J^{+i\mu}\circ \partial_\mu =R^{i}_\alpha\circ Z^\alpha-
M^{+ji}(\frac{\delta L}{\delta
  \phi^j},\ \cdot\ )
\label{adj_id}
\eea
with $M^{+ji}({\delta L}/{\delta \phi^j},\ \cdot\ )$ as in (\ref{Mdef}).
Applying (\ref{adj_id}) to 1, we find the relation (\ref{7}), with
the reducibility parameters given by
\bea
f^\alpha=Z^\alpha(1)=(-\partial)_{(\mu)}Z^{+\alpha(\mu)}.
\eea

\subsection{Operator currents}

So far, we have considered the problem from the point of view of the
reducibility parameters or the conserved $n-2$ forms. One can also make
the discussion from the point of view of the current that is
involved. Consider currents $J^{\mu i}(Q_i)=
J^{\mu i(\lambda)}\partial_{(\lambda)}Q_i$ that depend 
linearly and
homogeneously on arbitrary
local functions $Q_i$. We call such currents operator currents.
Let us consider operator currents that
either satisfy the condition 
that their divergence gives a Noether operator, 
\bea
\partial_\mu J^{\mu i}(Q_i)=N^i(Q_i),\quad
N^i(\vdd{L}{\phi^i})=0,\label{cond1}
\eea
or the condition that they are given by the divergence of a
superpotential upon replacing $Q_i$ by the Euler-Lagrange
derivatives of the Lagrangian, 
\bea
J^{\mu i}(\vdd{L}{\phi^i})=\partial_{\nu}k^{[\nu\mu]}.\label{cond2}
\eea
Equivalence classes of currents $J^{\mu i}(Q_i)$ satisfying either
condition \eqref{cond1} or \eqref{cond2} are obtained by identifying
currents that differ by a current defining itself a Noether identity
or the divergence of a superpotential operator,
\bea
J^{\mu i}(Q_i)\sim J^{\mu i}(Q_i)+N^{i\mu}(Q_i) +\partial_\nu 
[k^{[\nu\mu]i}(Q_i)],\
N^{i\mu}(\vdd{L}{\phi^i})=0. 
\eea
For every operator current satisfying \eqref{cond1}, 
one writes the
Noether operator $N^i$ as in the right hand side of \eqref{3.11} 
and finds again that one can associate reducibility parameters 
through $f^\alpha=(-\partial)_{(\mu)}Z^{+(\alpha)}$, 
while $\partial_\mu J^{\mu i}(\delta
L/\delta\phi^i)=0$ implies that equation \eqref{cond2} holds
for some local $k^{[\nu\mu]}$ (owing to \eqref{ass1}) which
gives thus a conserved $n-2$ form $k$.
Similarly, if the operator current satisfies \eqref{cond2},
the associated conserved $n-2$ form is again $k$ and associated
operator currents satisfying \eqref{cond1} as well as associated
reducibility parameters are constructed as in subsection \ref{s3.5}.

\subsection{Bijective correspondence between equivalence classes}

In order to prove directly that there is a bijective correspondence
between equivalence classes of reducibility parameters, equivalence
classes of conserved $n-2$ forms
and equivalence classes of operator currents satisfying \eqref{cond1}
or \eqref{cond2}, 
one has to use, in addition to
\eqref{ass1} and \eqref{ass2}, that, by the algebraic Poincar\'e lemma,
\bea
\partial_\nu k^{[\nu\mu]}=0\iff 
k^{[\mu\nu]}=\partial_\sigma
l^{[\sigma\mu\nu]}+\delta^n_2\epsilon^{\mu\nu} C,\label{ass3}
\eea
for some local functions $l^{[\sigma\mu\nu]}$ or (if $n=2$)
some constant $C\in {\mathbb R}$, and the
assumption that the generating set of non trivial gauge symmetries is 
irreducible in the sense that 
\bea
Z^{+\alpha}\circ R^{+i}_\alpha\approx 0\quad\Longrightarrow\quad
Z^{+\alpha}\approx 0.\label{ass4}
\eea
Because this equation has to hold as an operator equation, it is
equivalent to the adjoint operator 
equation, which can be written as 
\bea
R^{i}_\alpha(Z^\alpha(f))\approx 0\quad \forall f\quad\Longrightarrow
\quad Z^\alpha\approx 0. \label{locreduc}
\eea 
Irreducible gauge theories, to which the present investigation is
limited, are thus characterized by the absence of
``local'' reducibility identities, i.e., reducibility identities
defined by 
operators $Z^\alpha$ satisfying the
left hand side of \eqref{locreduc}
without being weakly zero themselves.
We stress the difference from
{\em global} reducibility identities which involve
particular local functions, the reducibility parameters, and which 
may very well exist in irreducible gauge theories.
This difference is analogous to the difference between global 
and gauge symmetries.

The direct proof of the bijective correpondences will not 
be given  
here, but we will briefly review in section \ref{s7}
the proof given in \cite{Barnich:1995db}
based on the Koszul-Tate resolution.

\subsection{Explicit computation of associated conserved n--2 form}
\label{s3.8}

\subsubsection{General expression}

In order to explicitly construct an expression for the superpotential 
$k^{[\mu\nu]}_f$ in \eqref{3.10}, associated to the
divergence free current $J^\mu_f$ of \eqref{3.9} built out of the data
of the reducibility identity defined by $f^\alpha$, one needs to use
the contracting homotopy that is involved in the algebraic Poincar\'e
lemma. 

For instance, one can use the formula for the homotopy operator
given in \cite{Andersonbook} (see also \cite{Olver:1993}, chapter 5 for an
expression with a different summation convention). The part concerning
the fields of this homotopy operator is given in appendix A, together
with the definition and relevant properties 
of the higher order Lie-Euler operators
$\delta/\delta\phi^i_{(\mu)}$. 
Explicitly, when applied to an $n-1$ form, one
finds from $d_H J_f=0$ that $J_f=-d_H k_f$, respectively
$J_f^\mu=\partial_\nu k^{[\nu\mu]}_f$, with 
\bea
k^{[\nu\mu]}_{f}=\int_0^1{dt}\
\Big(\partial_{(\lambda)}\big[\phi^i
(
\frac{|\lambda|+1}{|\lambda|+2}\frac{\delta J^\mu_f}{\delta
\phi^i_{(\lambda)\nu}}[x,t\phi])\big]
+ x^\nu J^\mu[t x,0] -\mu\longleftrightarrow \nu\Big)
.\label{2.31}
\eea

\subsubsection{Simplifications in the computation}

\begin{itemize}

\item If the current $J_f^\mu$ vanishes 
when the fields and their derivatives
are set to zero, 
one has $J_f^\mu[tx,0]=0$ in (\ref{2.31}).
This is for instance the case 
when the
Lagrangian contains only terms of degree $k\geq 2$ 
in the fields and their
derivatives. 

\item 
If the Euler-Lagrange equations are linear and
homogeneous in the fields and the generating set of non trivial
gauge transformations contains only field independent 
operators, any reducibility identity with field independent gauge
parameters $\bar f^\alpha=\bar f^\alpha(x)$
holds off-shell, because the left hand side
of (\ref{7}) does not depend on the fields, so that one can take
$M^{[j(\nu)i(\mu)]}=0$. It follows that $J^\mu_{\bar f}$ 
reduces to
\bea
J^\mu_{\bar f} = S_\alpha^{\mu i}
(\frac{\delta L}{\delta \phi^i},\bar f^\alpha)\equiv S^\mu_{\bar f},
\eea
which is then
also linear and homogeneous
in the fields (the field dependence coming only from the
Euler-Lagrange derivatives of $L$) and the integration over $t$ can be
evaluated trivially. Because in this case one has $J^\mu[x,0]=0$,
(\ref{2.31}) becomes
\bea
k^{[\nu\mu]}_{\bar f}=
\partial_{(\lambda)}\big[\frac{|\lambda|+1}
{|\lambda|+2}\phi^i\frac{\delta S^\mu_{\bar f}
}{\delta
\phi^i_{(\lambda)\nu}}-(\mu\leftrightarrow\nu)\big].\label{s1}
\eea

\item  If the current $S^\mu_{\bar f}$ 
contains at most second
order derivatives of the fields,
(\ref{s1}) reduces to
\bea
k^{[\nu\mu]}_{\bar f}=\frac{1}
{2}\phi^i\frac{\delta S^{\mu}_{\bar f}}{\delta
\phi^i_{\nu}}
+\frac{2}{3}\partial_\lambda(\phi^i\frac{\delta S^\mu_{\bar f}}{\delta
\phi^i_{\lambda\nu}})
-(\mu\leftrightarrow\nu).\label{simp}
\eea
The higher order Euler-operators are then explicitly given by
\bea
\frac{\delta f}{\delta \phi^i}&=&\frac{\partial^S f}{\partial
  \phi^i}-\partial_\lambda \frac{\partial^S f}{\partial \phi^i_\lambda}
+\partial_{\lambda}\partial_\rho
\frac{\partial^S f}{\partial \phi^i_{\lambda\rho}}\ ,\\
\frac{\delta f}{\delta \phi^i_\nu}&=&\frac{\partial^S f}{\partial
  \phi^i_\nu}-2\partial_\lambda
\frac{\partial^S f}{\partial \phi^i_{\lambda\nu}}\ ,\\
\frac{\delta f}{\delta \phi^i_{\lambda\nu}}&=&\frac{\partial^S f}{\partial
  \phi^i_{\lambda\nu}}\ ,
\eea
so that (\ref{simp}) becomes
\bea
k^{[\nu\mu]}_{\bar f}=\frac{1}
{2}\phi^i\frac{\partial^S S^{\mu}_{\bar f}}{\partial
\phi^i_{\nu}}
+[\frac{2}{3}
\phi^i_\lambda-\frac{1}{3}\phi^i
\partial_\lambda]
\frac{\partial^S S^\mu_{\bar f}}{\partial
\phi^i_{\lambda\nu}}
-(\mu\leftrightarrow\nu).\label{s2+}
\eea

\end{itemize}

\subsection{Application of Stokes theorem}

Stokes theorem implies that for an
$n-2$ dimensional compact manifold ${\cal
  C}^{n-2}$ without boundary, $\partial {\cal
  C}^{n-2}=\emptyset$, and a solution $\phi(x)$ of the equations of motion, 
the charge
\bea
Q([{\cal C}^{n-2}],[k])[\phi(x)]=\int_{{\cal C}^{n-2}}k|_{\phi(x)}
\eea
is independent of the choice of representatives both for the homology
class $[{\cal C}^{n-2}]$ and for the equivalence class  $[k]$.

\subsection{Algebra\label{s310}}

One can define a bilinear skew-symmetric operation among the
parameters of the non trivial gauge symmetries, i.e., the collection
of local functions $f^\alpha$ according to 
\bea
[f_1,f_2]_P^\gamma=C^\gamma_{\alpha\beta}(f^\alpha_1,f_2^\beta),\label{eq3.35}
\eea
where the structure operators 
$C^\gamma_{\alpha\beta}(f^\alpha_1,f_2^\beta)\equiv
C^{\gamma(\mu)(\nu)}_{\alpha\beta}\partial_{(\mu)}f^\alpha_1\partial_{(\nu)}
f_2^\beta$ depend on the choice of the
generating set according to \eqref{1.18}. 

The Jacobi identity for the Lie bracket $[\cdot,\cdot]_L$ 
of evolutionary vector fields
then implies a relation for the bracket of gauge parameters:
\bea
&[R_{f_3},[R_{f_1},R_{f_2}]_L]_L^i+{\rm
  cyclic}(1,2,3)=0\quad\Longrightarrow& \nonumber\\
&R^i_\rho([f_3,[f_1,f_2]_P+\delta_{f_1}f_2-\delta_{f_2}f_1]_P^\rho
+\delta_{f_3}[f_1,f_2]_P^\rho)+{\rm
  cyclic}(1,2,3)\approx 0.&
\eea
For irreducible gauge theories one deduces because of \eqref{locreduc} that
\bea
[f_3,[f_1,f_2]_P]_P^\rho+[f_3,\delta_{f_1}f_2-\delta_{f_2}f_1]_P^\rho
+\delta_{f_3}[f_1,f_2]^\rho_P+{\rm
  cyclic}(1,2,3)\approx 0.
\label{8.5}
\eea

We also note the commutation relation between a global symmetry $X^i$
and a gauge symmetry $R^i_\alpha(f^\alpha)$:
\bea
[X,R_f]^i_L=R^i_\alpha(X^{\alpha}_\beta(f^\beta)
+\delta_X f^\alpha)+
M^{+ji}_\alpha(\frac{\delta L}{\delta\phi^j},
f^\alpha),\label{glga}
\eea
for some operators $X^{\alpha}_\beta=
X^{\beta(\mu)}_\alpha\partial_{(\mu)}$,
and
$M^{+ji}_\alpha({\delta L}/{\delta\phi^j},\ \cdot\ )$
as in (\ref{Mdef}).

There is a well defined Lie action from equivalence classes of global
symmetries on equivalence classes of reducibility parameters. 
Indeed, by acting with a global symmetry
$\delta_X$ on $R^i_\alpha(f^\alpha)\approx 0$ and using the algebra
(\ref{glga}) together with $\delta_{f}\approx 0$,
we deduce that
$R^i_\alpha(X^{\alpha}_\beta(f^\beta)
+\delta_X f^\alpha)\approx 0$, so that the global symmetry defines 
a mapping
$(X^i,f^\alpha)\mapsto  X^{\alpha}_\beta(f^\beta)
+\delta_X f^\alpha$ which maps a collection of reducibility parameters
to another collection of reducibility parameters.
Furthermore, if $f^\alpha\approx 0$, the result vanishes weakly, so
that the mapping gives
trivial reducibility parameters and induces a well defined map for
equivalence classes of reducibility parameters.
Similarly, if
$X^i\approx 0$
the
left hand side of (\ref{glga}) defines, for all $f^\alpha$,
a weakly vanishing, and thus a trivial gauge symmetry,
and the irreducibility of the generating set then implies that
$X^{\alpha}_\beta(f^\beta)
+\delta_X f^\alpha\approx 0$ for all $f^\alpha$ and thus also in
particular for reducibility parameters $f^\alpha$.
Finally, if we choose
$X^i=R^i_\alpha(f_1^\alpha)$ with arbitrary local functions
$f_1^\alpha$, and $f^\alpha\equiv
f^\alpha_2$ reducibility parameters, 
the parameters in the reducibility identity 
$R^i_\alpha(X^{\alpha}_\beta(f^\beta)
+\delta_X f^\alpha)\approx 0$ depend on the arbitrary local function
$f_1$. This implies by the irreducibility of the generating set that 
$X^{\alpha}_\beta(f^\beta)
+\delta_X f^\alpha\approx 0$. 

Hence, the mapping
\bea
\left([X^i],[f^\alpha]\right)\mapsto \left[X^{\alpha}_\beta
(f^\beta)
+\delta_X f^\alpha\right]\label{eq3.39}
\eea
is well defined.

By the
isomorphism of equivalence classes of reducibility parameters and
equivalence classes of non constant 
conserved $n-2$ forms, 
it follows that this last space is an isomorphic Lie module.
Explicitly, the associated module action is defined by
\bea
\left([X^i],[k_{f}]\right)\mapsto [\delta_{X}
k_{f}].\label{eq3.40}
\eea
This will be proved by cohomological methods in section \ref{s7}.  
\bigskip

\noindent
{\bf Remark:}

\noindent
Suppose that $f^\alpha_1$ and $f^\alpha_2$ are
reducibility parameters,
$R^i_\alpha(f_1^\alpha)\approx 0$ and 
$R^i_\alpha(f_2^\alpha)\approx 0$.
Equation (\ref{1.18}) defining the algebra of the gauge transformations
then implies that $[f_1,f_2]^\alpha_P$ are also reducibility parameters
(because of $\delta_{f_1}\approx 0$, $\delta_{f_2}\approx 0$):
\bea
R^i_\gamma([f_1,f_2]_P^\gamma)\approx 0.
\label{2.64}
\eea
Furthermore,
the bracket of weakly vanishing gauge parameters with arbitrary gauge 
parameters is
again weakly vanishing (if, say,
$f_1^\alpha\approx 0$, then $[f_1,f_2]_P^\alpha\approx 0$, for all
$f_2^\alpha$), and
(\ref{8.5}) implies that the Jacobi identity holds for equivalence
classes of reducibility parameters.
Hence, the bracket $[\cdot,\cdot]_P$ induces a well defined Lie bracket among
equivalence classes of reducibility parameters:
\bea
\big[[f_1],[f_2]\big]_P^\gamma=\big[[f_1,f_2]_P^\gamma\big].
\eea
However, this algebra is always Abelian. Indeed, suppose that
$f_2^\alpha$ are reducibility parameters, while $f^1_\alpha$ are
arbitary local functions. Then \eqref{1.18} implies
\bea
R^i_\gamma([f_1,f_2]_P^\gamma+\delta_{f_1}f_2^\gamma)\approx 0
\eea
for arbitary $f_1^\alpha$, which implies by irreducibility 
\bea
[f_1,f_2]_P^\gamma+\delta_{f_1}f_2^\gamma\approx 0.
\eea
In the case where $f_1^\alpha$ are reducibility parameters,
$\delta_{f_1}\approx 0$, which gives the result. 

\mysection{Induced symmetries of the linearized theory}\label{s4}

\subsection{Gauge and global symmetries of the linearized theory}

Consider the change of variables $\phi=\bar\phi(x)+\varphi$, where
$\bar\phi(x)$ is a solution of the equations of motion,
\bea
\frac{\delta L}{\delta \phi^i}\Big|_{\bar\phi(x)}=0.
\eea
We have
\bea
L[\bar\phi(x)+\varphi]=L[\bar\phi(x)]+\partial_\mu
V^\mu_i(\varphi^i,L^1)+\sum_{n=2} L^n,\\
L^n=\frac{1}{n!}\frac{{\partial^S}^n L[\phi]}{\partial\phi^{i_1}_{(\mu_1)}
\dots
\partial\phi^{i_n}_{(\mu_n)}}\Big|_{\bar\phi(x)}\varphi^{i_1}_{(\mu_1)}\dots
\varphi^{i_n}_{(\mu_n)}.
\eea
[See Eq.\ (\ref{fun}) for the
notation $V^\mu_i(\varphi^i,L^1)$.]
In the following, we will assume that the operators
$R^{+i}_\alpha|_{\bar\phi(x)}$ provide an (irreducible) generating
set for the Noether identities of the ``free theory'' defined by
$L^{\rm free}[\varphi]\equiv  L^2$, i.e., that the
theory is ``linearizable'' \cite{Barnich:2000zw}
around the solution $\bar\phi(x)$.

Consider 
the equation expressing the fact that $R^i_{\alpha}(f^\alpha)$ are
symmetries of the Lagrangian for all local functions $f^\alpha$,
\bea
R^i_{\alpha}(f^\alpha)\frac{\delta L}{\delta\phi^i}=\partial_\mu
S^{\mu i}_\alpha(\frac{\delta L}{\delta\phi^i},f^\alpha).
\eea
Expanding in powers of the fields $\varphi^i$, we obtain to lowest non
trivial order,
\bea
 R^{i0}_{\alpha}( f^{\alpha 0})\frac{\delta 
  L^2}{\delta\varphi^i}
=\partial_\mu
 S^{\mu i0}_\alpha (\frac{\delta 
  L^2}{\delta\varphi^i}, f^{\alpha 0}),
\eea
which reflects the gauge invariance of $L^2$ under
the transformations $\delta^0_{ f} \varphi^i=
R^{i 0}_\alpha( f^\alpha)$. 
The next order gives 
\bea
\lefteqn{
 R^{i0}_{\alpha}( f^{\alpha 0})\frac{\delta 
  L^3}{\delta\varphi^i}+
\big( R^{i1}_{\alpha}( f^{\alpha 0})+
 R^{i0}_{\alpha}( f^{\alpha 1})\big)\frac{\delta 
  L^2}{\delta\varphi^i}
}
\nonumber\\
&=&\partial_\mu\Big(
 S^{\mu i0}_\alpha (\frac{\delta 
  L^3}{\delta\varphi^i}, f^{\alpha 0})+
 S^{\mu i1}_\alpha (\frac{\delta 
  L^2}{\delta\varphi^i}, f^{\alpha 0})+
 S^{\mu i0}_\alpha (\frac{\delta 
  L^2}{\delta\varphi^i}, f^{\alpha 1})\Big).
\label{4.7}
\eea
Suppose that $\4f^\alpha$ is 
the parameter of
a field independent reducibility identity of the free
theory, 
\bea
R^{i0}_{\alpha}( \4f^{\alpha})=0,\quad \4f^{\alpha}=\4f^{\alpha}(x).
\label{isometry}
\eea
[There can be
no equations of motion terms here because these are at least linear in the
fields.  If one would however consider reducibility parameters of the free
theory that depend on the fields, equations of motion terms are
relevant in general. The analysis can be extended to cover this
case. This is done in section \ref{s7} using cohomological methods.]
Specializing \eqref{4.7} by choosing parameters 
$f^{\alpha 0}=\4f^{\alpha}$ and
$f^{\alpha 1}=0$, it reads
\bea
R^{i1}_{\alpha}( \4f^{\alpha})
\frac{\delta L^2}{\delta\varphi^i}
=\partial_\mu\Big(
 S^{\mu i0}_\alpha (\frac{\delta L^3}{\delta\varphi^i}, \4f^{\alpha})+
 S^{\mu i1}_\alpha (\frac{\delta L^2}{\delta\varphi^i}, \4f^{\alpha})
\Big),
\eea
which means that $ R^{i1}_{\alpha}( \4f^{\alpha})$
defines a global symmetry
of the free theory. 

Hence, the parameters $ \4f^{\alpha}$ of a field
  independent reducibility identity of the free theory
  determine a
  global, linear, symmetry of the free theory given by 
$\delta^1_{\tilde f}\varphi^i=
 R^{i1}_{\alpha}( \4f^{\alpha})$.

\subsection{Algebra\label{s42}}

By expanding \eqref{1.18} in terms of $\varphi$ around
$\bar\phi(x)$, with $f_1^\alpha=\tilde f^\alpha$ 
field independent
reducibility parameters satisfying (\ref{isometry})
and $f_2^\alpha=g^\alpha$ arbitrary field
independent parameters,
one finds, to zeroth order, the action of a gauge symmetry of the
linearized theory on the global symmetry associated to the reducibility
parameters: 
\bea
[ R^{1}_{\tilde f}, R^{0}_{g}]_L
=-\delta^0_{g} R^{i 1}_{\beta}(\tilde f^{\beta})=
 R^{i 0}_{\gamma}( C^{\gamma 0}_{\alpha\beta}(\tilde
f^{\alpha},g^\beta)).
\label{algebra1}
\eea
If one chooses $g^\alpha=\tilde g^\alpha$ where
$\tilde g^\alpha$ are also field independent
reducibility parameters of the free theory 
($R^{i0}_{\4g}=0$), this gives
\bea
 R^{i 0}_{\gamma}( C^{\gamma 0}_{\alpha\beta}(\tilde
f^{\alpha},\tilde g^{\beta}))=0.
\label{algebra2}
\eea
The latter equation shows that 
the bracket
\bea
{}[\4f,\4g]^\gamma_M:= C^{\gamma 0}_{\alpha\beta}(\tilde
f^{\alpha},\tilde g^{\beta})\label{defbra}
\eea
gives again parameters of
a field independent reducibility identity, whenever 
$\4f^\alpha$ and $\4g^\alpha$ are such parameters.
Furthermore, \eqref{8.5} then guarantees that this bracket satisfies the
Jacobi identity so that the vector space of 
field independent reducibility parameters equipped with this bracket 
is a Lie algebra.
Note that this bracket may be non trivial and that there is no contradiction
with the analysis of section \ref{s310}. Indeed, 
the commutators $[\delta^0_{f_1},\delta^0_{f_2}]$ vanish
for all field independent parameters $f^\alpha_1$ and $f^\alpha_2$
(since $R^{i0}_\alpha$ is field independent), i.e., the structure
operators arising in these commutators vanish and
are not given by $C^{\gamma 0}_{\alpha\beta}$; hence the bracket
(\ref{defbra}) is {\em not} the counterpart in the linearized theory of 
the bracket (\ref{eq3.35}) in the full theory.

The Lie algebra with bracket (\ref{defbra})
can also be expressed in terms of a basis 
$\{\4f_A\}$ for the field independent reducibility parameters.
Such a basis is defined analogously to a basis
for Killing vector fields of a Riemannian metric:
each $\4f_A$ is a ``vector field'' with components $\4f_A^\alpha$ such
that (i) the vector fields $\4f_A$ are
linearly independent, and (ii) every vector field $\4f$ of
field independent
reducibility parameters $\4f^\alpha$ is a linear combination
$C^A\4f_A$ of 
the $\4f_A$ with constant coefficients $C^A$.
In particular, one thus has
\bea
 C^{\gamma 0}_{\alpha\beta}(\4f^{\alpha}_A,\4f^{\beta}_B)
=\C ABC \4f^{\gamma}_C
\label{algebra3}
\eea
for some constant coefficients $\C ABC$, which are 
the structure constants of the Lie algebra defined by $[\ ,\ ]_M$ 
in the basis $\{\4f_A\}$,
\bea
{}[\4f_A,\4f_B]^\alpha_M=\C ABC \4f^{\alpha}_C\ .
\label{algebra4}
\eea
Let us denote by $\delta_A$ the induced global symmetry
of the free theory
associated to $\4f_A$,
\[
\delta_A \varphi^i= R^{i1}_{\alpha}(\tilde f_A^{\alpha}).
\]
Note that some linear combinations of the 
subset of global symmetries $\{\delta_A\}$ might be trivial 
global symmetries. 
For $g^\alpha=\tilde g^\alpha$, (\ref{algebra1}) defines 
a Lie action of these global symmetries
on the field independent reducibility parameters.
Owing to
(\ref{algebra3}), this Lie module action is
\bea
(\delta_A,[\4f_B])\mapsto \C ABC [\4f_C].
\label{algebra5}
\eea

To first order in $\varphi$, Eq.\ \eqref{1.18} gives
the commutator algebra of the induced global symmetries,
\bea
[ R^1_{\tilde f},
 R^1_{\tilde g}]^i_L\approx^{\rm free} 
 R^{i1}_\gamma( C^{\gamma 0}_{\alpha\beta}(\tilde f^\alpha,
\tilde g^\beta))+
 R^{i0}_\gamma( C^{\gamma 1}_{\alpha\beta}(\tilde f^\alpha,
\tilde g^\beta)),
\label{algebra6}
\eea
where $\approx^{\rm free}$ means an equality when the equations of
motion of the free theory hold. 
As the second term on the right hand side is a 
trivial symmetry of the free theory (it is a gauge
transformation), one obtains, using (\ref{algebra3}),
\bea
[\delta_A,\delta_B]\sim\C ABC\delta_C
\label{algebra7}
\eea
where $\sim$ denotes equivalence in the 
free theory. Hence, the commutator algebra of the induced
global symmetries reflects, modulo trivial global
symmetries, the Lie algebra (\ref{algebra4})
associated to the reducibility parameters.

\mysection{Asymptotic symmetries and conservation laws}\label{s5}

\subsection{Boundary conditions}\label{Prerequisites}

Our aim is to capture general properties of symmetries, conservation
laws and their algebra in Lagrangian field theories, for different
models and  various choices of boundary conditions. Therefore we try
to avoid, as much as possible, too specific assumptions on the
boundary conditions. In fact, a detailed specification of the boundary
conditions cannot be done in a model independent manner.
For instance, a basic physical requirement on the boundary conditions 
could be that they contain certain solutions of the 
full equations of motion that are of physical interest and such a
requirement depends on the model and the particluar solutions under
investigation. 

What we want here are generic assumptions
in connection with the boundary conditions
that allow us to extend
the bijective correspondence between equivalence
classes of exact reducibility parameters 
and conserved $n-2$ forms
described in sections \ref{s3} 
to the asymptotic counterparts of these quantities.
Nevertheless we find it
useful to describe in the following a certain type of boundary conditions
and related assumptions that are sufficient for this bijective 
correspondence\footnote{They are not necessary, i.e., one may relax
them; a central requirement for our purpose is
the validity of the asymptotic acyclicity
properties described in section 5.4.}.
These conditions are adapted
from the Hamiltonian analysis of asymptotically anti-de Sitter gravity 
in $3$ and $4$ dimensions in \cite{Henneaux:1985tv,Brown:1986nw}. 
Three-dimensional asymptotically
anti-de Sitter gravity will be discussed in some detail
in section \ref{3d-adS}, to which we refer for a concrete
example. 

The conditions are formulated in terms of Landau's $O$-notation
and a corresponding ``asymptotic degree''
which characterize the behaviour of the functions of interest
near the boundary
[i.e., the behaviour of the fields, 
gauge parameters and local forms constructed of them;
the boundary need not be at (spatial) infinity]. 
We denote the asymptotic degree of a function $f$ by
$|f|$. The notation $f\asy O(g)$ and $f\asy o(g)$ mean $|f|\leq |g|$ and
$|f|< |g|$, respectively. For instance,
in three-dimensional asymptotically anti-de Sitter gravity,
the asymptotic degree of a function is its
leading power in the radial coordinate $r$ for $r\asy \infty$ so that
a function $f\asy r^m h(t,\winkel)$ has
asymptotic degree $|f|=m$. 

Moreover we assume that we can also 
assign a definite asymptotic degree to each of the
relevant differential operators,
independently of its arguments. For definiteness and
simplicity, let us
assume in particular 
that the derivatives $\6_\mu$ have asymptotic degree opposite
to the corresponding coordinates and differentials,
\bea
|x^\mu|=|dx^\mu|=-|\6_\mu|.
\label{x-degree}
\eea
This implicitly is an assumption on properties of the space of
functions in which the fields are assumed to live. For instance,
in three-dimensional asymptotically anti-de Sitter gravity, 
we assign asymptotic degree $-1$ to the derivative $\6_r$ with respect to
the radial coordinate meaning that
$f\asy O(r^m)\Longrightarrow \6_rf\asy O(r^{m-1})$
for $r\asy \infty$; this excludes in particular
functions with an oscillating dependence on
the radial coordinate
near the boundary,
such as outgoing or incoming waves.
Restrictions of this kind on the space of allowed functions
are quite commonly used in studies of asymptotic
quantities; for example, they were already used in
\cite{Sachs2} and also in \cite{Henneaux:1985tv,Brown:1986nw}.

The boundary conditions for 
the fields refer to a background $\5\phi^i(x)$
and the 
fields $\phi^i(x)$ are assumed to approach the background fields
near the boundary, i.e., $\phi^i(x)/\5\phi^i(x)\asy 1$ at some
rate. Accordingly, the
deviations $\varphi^i=\phi^i-\5\phi^i(x)$ of the fields from the
background, which are used as basic field variables near the boundary,
satisfy $\varphi^i(x)/\5\phi^i(x)\asy 0$. We denote the 
resulting asymptotic
behaviour of the fields $\varphi^i(x)$ by
\bea
\varphi^i(x)\equiv\phi^i(x)-\bar\phi^i(x)\asy O(\chi^i).\label{boundphi}
\eea 
In general, the fields $\varphi^i$ are not ``small''. Near the
boundary, however, 
they are small as compared to
the corresponding background fields so that
$\varphi^i/\5\phi^i\asy 0$. Nevertheless it may happen
that $\varphi^i$ does not approach zero at the boundary
if $\5\phi^i$ does not do so.
In the following, we assume that $\varphi^i(x)$ are generic
fields that satisfy 
the boundary conditions \eqref{boundphi}.

In the case of three
dimensional anti-de Sitter gravity for example, the metric deviations
$h_{\mu\nu}$ 
satisfying the boundary conditions 
are required to be of the form 
\bea
h_{\mu\nu}(x)\asy r^{m_{\mu\nu}}\tilde h_{\mu\nu}(t,\winkel)
+ o(r^{m_{\mu\nu}}),
\eea
with $\tilde h_{\mu\nu}(t,\winkel)$ arbitrary functions of
$t,\winkel$. 

When discussing the asymptotic behaviour of
a local form $\omega^p$, we will consider
the asymptotic degree of the differential form obtained
after evaluating the form $\omega^p$ for generic fields that 
satisfy the boundary conditions.
 
Equation \eqref{x-degree} implies that the differential $d_H$ and 
the associated contracting homotopy
$\rho_{H,\varphi}$ \eqref{phihomotopy} have 
vanishing asymptotic degree,
\bea
\omega^p\Big|_{\varphi(x)}\longrightarrow O(\chi^p)\quad
\Longrightarrow\quad
(d_H\omega^p)\Big|_{\varphi(x)}\asy O(\chi^p),\quad
(\rho_{H,\varphi}^p\omega^p)\Big|_{\varphi(x)}\asy O(\chi^p).
\label{asympacycd}
\eea
Furthermore, \eqref{x-degree} implies that
a field independent 
differential operator $Z=Z^{(\mu)}(x)\partial_{(\mu)}$ has the
same asymptotic degree as its adjoint
$Z^+=(-\partial)_{(\mu)} [Z^{(\mu)}(x)\cdot]$,
\bea
Z=Z^{(\mu)}(x)\partial_{(\mu)}\quad\Longrightarrow\quad
|Z|=|Z^+|.\label{adjoints}
\eea

Now, let $L[\phi^i;j_a(x)]$ be the Lagrangian of the model
under study. It may involve external sources $j_a(x)$ but
these and their derivatives are supposed to vanish in a neighborhood of
the boundary, so that near the boundary, the theory is described by
the source free Lagrangian $L[\phi^i;0]$. 
We shall assume here that the background is an exact solution of the 
field equations derived from $L[\phi^i;0]$. 

The boundary conditions may  
allow one to completely neglect a subset of the fields
$\varphi^i$ near the boundary (possibly after a field
redefinition), so that one may use a simplified Lagrangian there
(with less fields and less terms).
A typical example is the case where  ``matter fields'' 
in general relativity, electrodynamics or Yang-Mills theory decrease
sufficiently fast near the boundary so that there
one may use the pure Einstein, Maxwell or Yang-Mills
Lagrangian, respectively.

We denote by $O(\chi_i)$ the behaviour of the linearized
(source free) field
equations evaluated at $\varphi^i(x)$ and multiplied by the 
volume element,
\bea
\forall\varphi^i(x)\asy O(\chi^i):\quad
d^nx\,\frac{\delta
    L^{\rm free}}{\delta\varphi^i}\Big|_{\varphi(x)}\asy O(\chi_i),
\eea
with a $\chi_i$ of minimal asymptotic degree.
To determine that degree,
we use that the
linearized field equations take the form
\[
\frac{\delta
    L^{\rm free}}{\delta\varphi^i}=
D_{ij}\varphi^j
\]
where $D_{ij}=d_{ij}^{(\mu)}(x)\6_{(\mu)}$ are
differential operators involving the background fields and
their derivatives. The asymptotic degree of the function
$\chi_i$ is thus
\bea
|\chi_i|=\max_j\{|D_{ij}|+|\chi^j|+|d^nx|\},
\eea
where we use the convention $|0|=-\infty$, i.e.,
the vanishing of an operator $D_{ij}$ does not
affect $|\chi_i|$.
Furthermore, let us denote by $O(\chi_\alpha)$ the behaviour of the
linearized 
Noether identities when evaluated at a field 
that behaves like the
linearized equations of motion times the volume form,
\bea
\forall\psi_i\asy O(\chi_i):\quad
R^{+i0}_\alpha(\psi_i)\asy
O(\chi_\alpha),
\label{puregauge} 
\eea  
where
\bea
|\chi_\alpha|=\max_{i}\{|R^{i0}_\alpha|+|\chi_i|\}.
\eea

The left hand sides of the
(Euler-Lagrange) equations of motion of the full and the free 
theory and their total derivatives, $\partial_{(\mu)}\delta
L/\delta\phi^i$ and 
$\partial_{(\mu)}\delta L^{\rm free}/\delta\varphi^i$, have been assumed 
to satisfy important regularity conditions described for 
instance in  
\cite{Fisch:1990rp,Henneaux:1991rx} and spelled out in detail in 
the context of
Yang-Mills theories in \cite{Barnich:2000zw}. We assume that these
regularity conditions also hold asymptotically. By this we mean that
the 
leading order terms of 
$\partial_{(\mu)}\delta L^{\rm free}/\delta\varphi^id^nx$, after
substitution of generic fields that saturate the boundary conditions,
satisfy the mentioned regularity
conditions.

The Noether operators $R^{+i}_\alpha$ of the full theory were assumed
to form an irreducible generating set of Noether identitites, as
expressed by
\eqref{ass2} and \eqref{ass4}. Similarly, these Noether operators
evaluated at
the background $R^{+i0}_\alpha$ 
were assumed to form an irreducible generating set of
Noether operators for the linearized theory near the boundary. Now, we
require in addition that these properties also hold
asymptotically. More precisley, 
\bea
\forall\varphi^i(x):\
N^i(\frac{\delta L^{\mathrm{free}}}
{\delta\varphi^i})\Big|_{\varphi(x)}d^nx
\asy 0\ \Longrightarrow\
\exists \{Z^\alpha\}\ \forall\psi_i:\ 
N^i(\psi_i)\longrightarrow Z^{+\alpha}
(R^{+i0}_\alpha(\psi_i)),\quad
\label{asyacycdelta}
\eea
for field independent differential operators $N^i$,
and
\bea
\forall\psi_i:\ 
Z^{+\alpha} (R^{+i0}_\alpha(\psi_i))\asy 0\ \Longrightarrow\ 
\forall\psi_\alpha:\ Z^{+\alpha}(\psi_\alpha)\asy 0,
\label{asyacycdeltabis}
\eea
for field independent differential operators $Z^{+\alpha}$. 
Here and in the following, $\psi_i$ and $\psi_\alpha$
are generic fields satisfying the boundary conditions
\bea
\psi_i\asy O(\chi_i),\quad 
\psi_\alpha\asy O(\chi_\alpha).
\eea

\subsection{Analysis from the viewpoint of the linearized theory}

\subsubsection{Asymptotic solutions}

Asymptotic solutions are particular fields $\varphi_s(x)$
satisfying the boundary conditions
\eqref{boundphi} together with the condition
\bea
\frac{\delta
    L^{\mathrm{free}}}{\delta\varphi^i}\Big|_{\varphi_s(x)}d^nx\asy
  o(\chi_i).
\label{asysol}
\eea

\subsubsection{Asymptotic reducibility parameters}

{\bf Definition:}
Asymptotic reducibility parameters are
field independent gauge
parameters $\tilde f^\alpha$ satisfying the condition
\bea
\forall \psi_i:\quad
\psi_iR^{i0}_\alpha(\tilde f^\alpha)\asy 0.
\label{asympsym}
\eea

Because of (\ref{puregauge}),
this condition is automatically satisfied 
for parameters with asymptotic degrees
smaller than $-|\chi_\alpha|$. Such parameters
will thus be considered as trivial and called
``pure gauge''.
Equivalence classes of asymptotic 
reducibility parameters are defined 
by asymptotic reducibility parameters up to parameters that are pure gauge.
In particular, parameters that are pure gauge are thus equivalent
to zero ($\sim 0$),
\bea
\tilde f^\alpha\sim 0\quad
\Longleftrightarrow\quad \tilde f^\alpha\asy o(\chi^\alpha),
\label{puregauge2}
\eea
where $\chi^\alpha$ is a function with asymptotic degree
equal to $-|\chi_\alpha|$,
\bea
|\chi^\alpha|=-|\chi_\alpha|.
\label{puregauge?}
\eea

\subsubsection{Asymptotically conserved n--2 forms}

{\bf Definition:}
An asymptotically conserved $n-2$ form is 
an $n-2$ form
$\tilde k[\varphi]$ that depends linearly and homogeneously on
$\varphi^i_{(\mu)}$ 
such that 
\bea
\forall\varphi^i(x):\quad 
d_H\tilde k|_{\varphi(x)}\asy \tilde s^i(\frac{\delta L^{\rm
    free}}{\delta\varphi^i})|_{\varphi(x)},\label{cocyn-2}
\eea
with $\tilde s^i(Q_i)$ an $n-1$ form
that depends linearly and homogeneously
on $Q_i$ and its derivatives.

An asymptotically conserved $n-2$ form $\tilde k$ is trivial if 
\bea
\forall\varphi^i(x):\quad 
\tilde k|_{\varphi(x)}\asy \tilde t^i(\frac{\delta L^{\rm
    free}}{\delta\varphi^i})|_{\varphi(x)}+d_H
\tilde l|_{\varphi(x)},\label{cobn-2} 
\eea
with $\tilde t^i(Q_i)$ an $n-2$ form that depends linearly and homogeneously
on $Q_i$ and its derivatives.

\subsubsection{Bijective correspondence}\label{s524}

The $n-1$ form $s^i_\alpha(\psi_i,\tilde f^\alpha)$ is defined by 
\bea
\forall Q_i:\quad
d^nx\,Q_iR^{i0}_\alpha(\tilde
f^\alpha)=
d^nx\,R^{+i0}_\alpha(Q_i)\tilde f^\alpha+d_H s^i_{\alpha}(Q_i,\tilde
f^\alpha). 
\eea
For $Q_i={\delta L^{\rm free}}/{\delta\varphi^i}$, this relation
reduces to 
\bea
d_H \tilde s_{\tilde f}=d^nx\,\frac{\delta L^{\rm
    free}}{\delta\varphi^i}\,R^{i0}_\alpha(\tilde f^\alpha)\label{defin},
\eea 
with $\tilde s_{\tilde f}=s^i_\alpha({\delta L^{\rm
    free}}/{\delta\varphi^i},\tilde f^\alpha)$. 
Suppose that $\tilde f^\alpha$ are asymptotic reducibility parameters
so that \eqref{asympsym} holds. This implies 
\bea
\forall\varphi^i(x):\quad 
d_H \tilde s_{\tilde f}|_{\varphi(x)}\asy 0.\label{defin1}
\eea
Applying the contracting homotopy $\rho_{H,\varphi}$ to 
$\tilde s_{\tilde f}$ 
and using \eqref{defin1} together with \eqref{asympacycd}, it follows that 
\bea 
\forall\varphi^i(x):\quad 
\tilde s_{\tilde f}|_{\varphi(x)}\asy -d_H
\tilde k_{\tilde f}|_{\varphi(x)},
\eea
with $\tilde k_{\tilde f}=-\rho^{n-1}_{H,\varphi}\tilde s_{\tilde f}$.
Since $\tilde s_{\tilde f}$ depends linearly and homogeneously on
the ``left hand sides'' of the linearized field equations, 
the $n-2$ form $\tilde k_{\tilde f}$ is thus an asymptotically
conserved $n-2$ form. 

Conversely, for an asymptotically conserved $n-2$ form $\tilde k$,
application of $d_H$ to \eqref{cocyn-2} implies 
\bea
\forall\varphi^i(x):\quad 
d_H \tilde
s^i(\frac{\delta L^{\rm free}}{\delta\varphi^i})|_{\varphi(x)}\asy 0.
\eea
Hence $d_H \tilde
s^i(\cdot)\equiv d^nx N^i$ defines an asymptotic Noether
operator as in \eqref{asyacycdelta} which implies that
there are operators $Z^\alpha$ such that
$N^i(\psi_i)\asy Z^{+\alpha}
(R^{+i0}_\alpha(\psi_i))$. Setting $\psi_i=Q_i d^nx$,
we obtain $d^nx Z^{+\alpha}
(R^{+i0}_\alpha(Q_i))=d_H(\dots)+d^nx Q_i R^{i0}_\alpha(\tilde f^\alpha)$
with $\tilde f^\alpha=Z^\alpha(1)$. Furthermore
we have $d^nx N^i(Q_i)=d_H \tilde s^i(Q_i)$ by definition of $N^i$.
We thus obtain $\psi_i R^{i0}_\alpha(\tilde f^\alpha)=
d^nx Q_i R^{i0}_\alpha(\tilde f^\alpha)\asy
d_H\omega$ for some $(n-1)$-form $\omega$. Recall that
this holds for {\em all} $\psi_i$ with
$\psi_i\asy O(\chi_i)$. This is only possible
if both $\psi_i R^{i0}_\alpha(\tilde f^\alpha)\asy 0$ and
$d_H\omega\asy 0$. It follows that 
the $\tilde f^\alpha=Z^\alpha(1)$ satisfy
\eqref{asympsym} and are thus
asymptotic reducibility parameters.

We have thus shown that asymptotic reducibility parameters
correspond to asymptotically conserved $n-2$ forms and
vice versa. This correspondence extends to the
equivalence classes associated with these quantities. 
This will be proved in section \ref{s7} 
using cohomological
methods and is summarized by the following theorem.

\begin{theorem}
There is a bijective correspondence between the quotient space of
asymptotic reducibility parameters 
factored by pure gauge parameters 
on the one hand, and equivalence classes of asymptotically conserved $n-2$
forms on the other hand.  
\end{theorem}

\noindent {\bf Remark:}

Because of \eqref{asympacycd}, the asymptotic behaviour of
the forms $\tilde s_{\tilde f}$ and $\tilde k_{\tilde f}$ 
is determined by the asymptotic behaviour of the asymptotic
reducibility parameters according to
\bea
|\tilde s_{\tilde f}|_{\varphi(x)}|
\leq \max_{\alpha}\{|\tilde f^\alpha|+|\chi_\alpha|\},
\nonumber\\
|\tilde k_{\tilde f}|_{\varphi(x)}|\leq
\max_{\alpha}\{|\tilde f^\alpha|+|\chi_\alpha|\}.
\eea 
In particular, these forms vanish asymptotically for trivial asymptotic 
reducibility parameters because then one obtains
$|\tilde f^\alpha|+|\chi_\alpha|<-|\chi_\alpha|+|\chi_\alpha|=0$,
see \eqref{puregauge2} and (\ref{puregauge?}),
while they are asymptotically finite for asymptotic reducibility parameters
that satisfy
\bea
\forall\alpha:\quad |\tilde f^\alpha|\leq -|\chi_\alpha|.
\label{convergencec}
\eea
In this latter case, the horizontal differential of the asymptotically
conserved $n-2$ form $\tilde k_{\tilde f}$ vanishes asymptotically
when evaluated at an arbitrary asymptotic solution $\varphi_s(x)$,
\bea
d_H\tilde k_{\tilde f}|_{\varphi_s(x)}\asy 0.\label{e526}
\eea
Similarly, a trivial asymptotically conserved $n-2$ form,
evaluated at an arbitrary asymptotic solution $\varphi_s(x)$, is
asymptotically given by the horizontal differential of an $n-3$ form,
\bea
\tilde k_{\tilde f}\sim 0\quad\Longrightarrow\quad
\tilde k_{\tilde f}|_{\varphi_s(x)}\asy d_H l|_{\varphi_s(x)}.
\eea

\subsubsection{Asymptotic charges}

Consider an $n-2$ dimensional compact 
manifold ${\cal C}^{n-2}$
without boundary, $\partial  {\cal C}^{n-2}=\emptyset$, 
that lies
in the asymptotic region and an asymptotically conserved $n-2$ form
$\tilde k_{\tilde f}$. The associated 
charge in the linearized theory is defined by 
\bea
\4Q_{\tilde f}[\varphi;\bar\phi(x)]=
\int_{{\cal C}^{n-2}}\tilde k_{\tilde
f}[\varphi;\bar\phi(x)].
\eea 
If the condition \eqref{convergencec} holds, the charges are 
finite when evaluated at a field $\varphi(x)$
that satisfies the boundary conditions \eqref{boundphi}. If furthermore we 
evaluate the charge for a solution $\varphi_s(x)$ of the
linearized equations of motion, we can apply Stokes theorem because of
the conservation law \eqref{e526} to prove 
asymptotic independence of $\4Q_{\tilde f}$ on 
the choice of representatives for
the homology class $[{\cal C}^{n-2}]$ and for the equivalence class 
$[\tilde k_{\tilde f}]$. 

\subsubsection{Asymptotic algebra}

Let us suppose now, and in the following, that
the reducibility parameters $\tilde f^\alpha$ defined by
\eqref{asympsym} satisfy 
condition \eqref{convergencec}, i.e., that $\tilde f^\alpha\asy
O(\chi^\alpha)$, which  guarantees that the associated $n-1$ and
$n-2$ forms are asymptotically finite. Consider fields $\psi^\alpha$
satisfying the boundary conditions $\psi^\alpha\asy
O(\chi^\alpha)$. Suppose now that the additional constraints
\bea
\forall\psi_i,\varphi^i(x)
:&&
\psi_iR^{i1}_\alpha(\tilde f^\alpha)|_{\varphi(x)}\asy O(1),\label{constr1}\\
\forall\psi_\alpha,\psi^\alpha
:&&
\psi_\alpha C^{\alpha 0}_{\beta\gamma}
(\psi^\beta,\tilde f^\gamma)\asy
O(1),\label{constr2}\\ 
\forall \psi_i, \psi^\alpha:&& \frac{\delta}{\delta\varphi^j}
[\psi_i R^{i1}_\alpha(\psi^\alpha)]\asy O(\chi_j),\label{newass1}\\
\forall\varphi^i(x):&&\frac{\delta L^3}{\delta\varphi^i}
\Big|_{\varphi(x)} d^nx\asy
O(\chi_i)\label{newass}
\eea
hold
for asymptotic reducibility parameters $\tilde f^\alpha$ that
satisfy \eqref{convergencec}.  
Under these conditions, we have
\begin{theorem}
The vector space of asymptotic reducibility parameters 
forms a Lie algebra for the bracket
\eqref{defbra}. Furthermore, the bracket induced among 
equivalence classes of asymptotic reducibility parameters is well
defined,  
\bea
[[\tilde f],[\tilde g]]^\gamma_G=[[\tilde f,\tilde g]_M]^\gamma.
\eea
\end{theorem}
The space of equivalence classes of asymptotic reducibility parameters
equipped with the bracket $[\cdot,\cdot]_G$ defines the  
physically relevant Lie
algebra $\mathfrak{g}$. 
Again, the theorem will be proved in section
\ref{s7} by cohomological means.

If the additional constraints
\bea
\forall\varphi^i(x):&&
R^{i1}_\alpha(\tilde f^\alpha)\asy O(\chi^i),\label{newass2}\\
\forall\varphi^i(x):&&\Big[\frac{\delta}{\delta\varphi^j}[R^{i0}_\alpha(\tilde
f^\alpha)\frac{\delta L^3}{\delta\varphi^i}]\Big]_{\varphi(x)}d^nx\asy
O(\chi_j),\label{newass3}\\
\forall \psi^\alpha:&&R^{j0}_{\beta}(\tilde f^\beta)
\frac{\delta R^{i1}_\alpha(\psi^\alpha)}{\delta\varphi^j}\asy
O(\chi^i)\label{newass4}
\eea
hold for asymptotic reducibility parameters $\tilde f^\alpha$ that
satisfy \eqref{convergencec},
the Lie algebra $\mathfrak{g}$ of equivalence classes of asymptotic 
reducibility parameters
can be represented on the level of the equivalence classes of asymptotically 
conserved $(n-2)$-forms of the
linearized theory near the boundary
by a covariant Poisson bracket, which is defined
through the action of the associated ``global symmetry''\footnote{Strictly
  speaking, when $\tilde f^\alpha$ are asymptotic reducibility
  parameters, the variations $\delta^g_{\tilde
    f}\varphi^i=R^{i1}_\alpha(\tilde f^\alpha)$ are not global
  symmetries of the linearized theory, but it will be shown below that
  they induce symmetries of the
  equations of motion of the boundary theory.}:
\bea
\{[\tilde k_{\tilde f_1}],[\tilde k_{\tilde f_2}]\}_F:=
[\delta^g_{\tilde f_1}\tilde k_{\tilde f_2}]
=
[\tilde k_{[\tilde f_1,\tilde
f_2]_M}].\label{e537}
\eea
The property $-[\delta^g_{\tilde f_2}\tilde k_{\tilde f_1}]=
[\tilde k_{[\tilde f_1,\tilde
f_2]_M}]$ implies that alternative equivalent expressions for 
the covariant Poisson bracket are $-[\delta^g_{\tilde
f_2}\tilde k_{\tilde f_1}]$ or $
\half ([\delta^g_{\tilde f_1}\tilde k_{\tilde f_2}]-[\delta^g_{\tilde
f_2}\tilde k_{\tilde f_1}])$.

When evaluated for solutions of the linearized
equations of motion, the Lie algebra $\mathfrak g$ can also be represented
by a covariant Poisson bracket of the charges  
$\4Q_{\tilde f}$ of the free theory, defined in the same way:
\bea
\{\4Q_{\tilde f_1},\4Q_{\tilde f_2}\}_{CL}:=
\delta^g_{\tilde f_1}\4Q_{\tilde f_2}
\stackrel{\approx^{\rm free}}{\longrightarrow}
\4Q_{[\tilde f_1,\tilde
f_2]_M}.
\eea

That both of these representations also provide
representations of the Lie algebra $\mathfrak{g}$ follows from the
fact that the asymptotically conserved $n-2$ forms $\tilde k_{\tilde f}$ 
and the
associated charges $\tilde Q_{\tilde f}$ vanish asymptotically
whenever the $\tilde f^\alpha$ are pure gauge. 

The proof of these statements is postponed until
section \ref{s7}. 

\subsection{Analysis from the viewpoint of the bulk theory}\label{s53}

\subsubsection{Asymptotic linearity}

In order for the previous discussion of asymptotic reducibility
parameters and asymptotically conserved $n-2$ forms to correctly
describe these quantities from the point of view of the bulk theory,
additional assumptions on the Lagrangian, the gauge transformations 
and the boundary conditions are needed. They state 
that the theory is ``asymptotically linear''. 
By that we mean that 
in the vicinity of the boundary,
the full theory can be approximated by the linearized theory with
Lagrangian $L^{\rm free}$.

More precisely, this translates into the following 
requirements:  

(i) the only terms of the equations of motion
that are relevant near the boundary are the equations
of motion of the linear theory, 
\bea
[\frac{\delta L}{\delta \phi^i}-\frac{\delta L^{\rm free}}{\delta
  \varphi^i}]|_{\varphi(x)}d^nx \longrightarrow o(\chi_i),\label{lineom}
\eea

(ii) the generating set of non trivial Noether operators are 
appropriately described by the Noether operators of the linearized theory, 
\bea
\forall \psi_i\asy O(\chi_i):\ [R^{+i}_\alpha(\psi_i)-R^{+i0}_\alpha(\psi_i)]
|_{\varphi(x)}
\asy o(\chi_\alpha). \label{lingaugesym} 
\eea

(iii) the gauge transformation associated to asymptotic reducibility
parameters $\tilde f^\alpha$ are appropriately described by the
sum of the corresponding gauge transformation of the linearized
theory and the associated ``global'' symmetry,
\bea
\forall \tilde f^\alpha\ {\rm satisfying}\ \eqref{asympsym}:\
[R^{i}_\alpha(\tilde f^\alpha )-R^{i0}_\alpha(\tilde f^\alpha )-
R^{i1}_\alpha(\tilde f^\alpha )]
|_{\varphi(x)}\asy o(\chi^i).\label{expasymsym}
\eea

We shall also use the fact that the Euler-Lagrange derivatives
$\delta L^{\rm free}/\delta \varphi^i$ of the linearized Lagrangian
are the linearization of the Euler-Lagrange derivatives
${\delta L}/{\delta \phi^i}$ of the full Lagrangian,
\bea
(d_V\frac{\delta L}{\delta
  \phi^i})|_{\bar\phi(x),\varphi}
=\frac{\delta L^{\rm free}}{\delta \varphi^i}\ ,\label{id}
\eea
which holds for all $\bar\phi(x)$ and not only
for $\bar\phi(x)$ that are solutions of the equations of motion.
In (\ref{id}) and throughout this paper,
evaluation at $\bar\phi(x),\varphi$ is
obtained by replacing $\phi^i_{\mu_1\dots\mu_k}$ by ${\partial^k
\bar\phi^i(x)}/{\partial x^{\mu_1}\dots\partial x^{\mu_k}}$ and
the Grassmann odd variables
$d_V\phi^i_{\mu_1\dots\mu_k}$ by $\varphi^i_{\mu_1\dots\mu_k}$.

\subsubsection{Asymptotic solutions}

On account of \eqref{lineom}, from the point of view of the full
theory, asymptotic solutions can equivalently be 
defined by fields $\phi_s(x)$ that satisfy the boundary 
conditions \eqref{boundphi} together with the condition
\bea
\frac{\delta
    L}{\delta\phi^i}\Big|_{\phi_s(x)}d^nx\asy o(\chi_i).
\eea

\subsubsection{Asymptotic reducibility parameters}

{}From the point of view of the full theory, one can allow
for possibly field dependent gauge parameters $f^\alpha$. 
The condition for asymptotic reducibility parameters then becomes
\bea
\forall\psi_i\asy O(\chi_i):\ 
\psi_i R^i_\alpha(f^\alpha)|_{\bar\phi(x)}\asy 0,
\eea
while trivial asymptotic reducibility parameters correspond to
reducibility 
parameters $f^\alpha$ that fall off fast enough when evaluated at
the background,
\bea
f^\alpha|_{\bar\phi(x)}\asy o(\chi^\alpha).
\eea
The identification $f^\alpha|_{\bar\phi(x)}=\tilde f^\alpha$ shows
that there is no difference between the two points of view. 

\subsubsection{Asymptotic symmetries}

One can define asymptotic symmetries to be gauge 
transformations 
$\delta_f\phi^i=R^i_\alpha(f^\alpha)$ of the full theory with gauge
parameters that are asymptotic reducibility parameters. 
Trivial asymptotic symmetries are defined as asymptotic symmetries
that involve trivial reducibility parameters and equivalence classes
of asymptotic symmetries can as usual be defined by asymptotic
symmetries up to trivial ones. According to assumption \eqref{expasymsym}, 
the action of asymptotic symmetries near the boundary 
is determined by the action of the first two terms in their expansion:
\bea
\delta_{\4f}\phi^i=R^i_\alpha(\4f^\alpha)\asy
 R^{i 0}_\alpha( \4f^{\alpha})+R^{i 1}_\alpha( \4f^{\alpha})+o(\chi^i), 
\label{asygauge}
\eea

There is {\em no}
bijective correspondence between equivalence classes of 
asymptotic symmetries and asymptotic reducibility parameters. Indeed,
for instance  in the case of 
pure Maxwell theory, there is one exact reducibility parameter 
given by a constant gauge parameter, but the associated gauge transformation
vanishes. The reason why we will focus our attention on equivalence
classes of reducibility parameters and not on 
equivalence classes of asymptotic symmetries, is 
that the former and not the latter are in bijective
correspondence with equivalence classes of asymptotically conserved
$n-2$ forms. 

\subsubsection{Asymptotically conserved n--2 forms}

{}From the point of view of the full theory, an 
asymptotically conserved $n-2$ form $k$ is defined as an $n-2$ form 
whose linearization
at the background 
$\tilde k=(d_V k)_{\bar\phi(x),\varphi}$ satisfies
\eqref{cocyn-2}. Such a form is trivial if its linearization is,
i.e., if it satisfies \eqref{cobn-2}.  

An equivalent characterization of asymptotically conserved $n-2$ forms
and their relation to asymptotic reducibility parameters 
is the following. 
\begin{theorem}
Let $\Sigma$ be any $n-1$ dimensional hypersurface with boundary 
$\partial\Sigma$ and $\delta\phi^i=R^i_\alpha(f^\alpha)$ be a non
trivial gauge symmetry.  
The associated weakly vanishing ``Noether charge'' 
\bea
\int_\Sigma
S^{\mu i}_\alpha(\frac{\delta
L}{\delta\phi^i},f^\alpha)(d^{n-1}x)_\mu
\eea
can be improved
through the addition of 
a surface integral 
\bea
\oint_{\partial \Sigma}
k^{\mu\nu}_\alpha(f^\alpha) (d^{n-2}x)_{\mu\nu}
\eea
to a charge that is asymptotically extremal at $\bar\phi(x)$ for
arbitrary variations $d_V\phi^i$ (not restricted by any boundary
conditions) if and only if the 
$f^\alpha$ are asymptotic reducibility
parameters. 
For solutions of the equations of motions, the improved
Noether charge reduces to the surface integral whose
integrand is the associated asymptotically conserved $n-2$ form.
\end{theorem}
The proof of this theorem is given in section \ref{s7}.

\subsubsection{Algebra and central extensions for the full theory}
\label{fullalgebra}

Let
$Q_{\tilde f}$ be
the charge associated to a given collection of 
asymptotic reducibility parameters
$\tilde f^\alpha$,
\bea
Q_{\tilde f}[\phi;\bar\phi(x)]=\int_{{\cal C}^{n-2}}
\tilde k_{\tilde f}[\phi-\bar\phi(x);\bar\phi(x)] + N_{\tilde f}
\label{defcharge},
\eea
where the field independent normalization ``constant'' $N_{\tilde f}$ 
is the arbitrarily chosen charge of the background
and ${\cal C}^{n-2}$ denotes an $n-2$ dimensional compact and 
closed manifold
that lies in the asymptotic region. 

On the level of the
charges of the full theory, the Lie algebra $\mathfrak g$ of 
equivalence classes of asymptotic reducibility parameters is 
represented by acting with an asymptotic symmetry 
associated to one collection of reducibility parameters 
on the charge associated to another such collection,  
\bea
\{Q_{\tilde f_1},Q_{\tilde f_2}\}_{CF}:=\delta_{\tilde f_1}Q_{\tilde
  f_2}=\int_{{\cal C}^{n-2}}
\tilde k_{\tilde f_2}[R_{\tilde f_1};\bar\phi(x)].\label{5.19}
\eea 
Because of (\ref{asygauge}), only the first two terms in the expansion
of the asymptotic symmetries contribute near the boundary [since
we assume that (\ref{convergencec}) holds].
Central charges are contributions
to $\delta_{\4f_1}Q_{\4f_2}$ which have no counterpart
in the Lie algebra $\mathfrak g$
associated with the asymptotic reducibility parameters 
and with the charges in the
linearized theory. 
They arise from the first term on the
right hand side of (\ref{asygauge}), while the
``regular'' terms 
arise from the second term.
(We assume of course the validity of the assumptions of section 5.2.6,
that  guarantee that the algebra
of equivalence classes of asymptotic 
reducibility parameters is well defined and
can be represented by a Poisson algebra of the conserved charges 
for the free theory.)

\begin{theorem}\label{thmalgebra}
The covariant Poisson algebra 
of the charges defined by \eqref{5.19}
is given by
\bea
\{Q_{\tilde f_1},Q_{\tilde f_2}\}_{CF}
&\sim &
Q_{[\tilde f_1,\tilde f_2]_M}-N_{[\tilde f_1,\tilde f_2]_M}+
K_{\tilde f_1,\tilde f_2}\ ,
\label{asympcom}
\\
K_{\tilde f_1,\tilde f_2}&=&\int_{{\cal C}^{n-2}}
\tilde k_{\tilde f_2}[R^0_{\tilde f_1};\bar\phi(x)],
\label{centralcharge1}
\eea
where $\sim$ is asymptotic equality when the charges are evaluated 
for asymptotic solutions.

The $n-2$ forms $\tilde k_{f^\prime}[R^0_f;\bar\phi(x)]$ are
skew-symmetric, up to a $d_H$-exact $n-2$ form, under the exchange of 
arbitrary field independent gauge parameters $f,f^\prime$, 
\bea
\forall f^\alpha,f^{\alpha\prime}:\quad
k_{f}[R^0_{f'};\bar\phi(x)]=-k_{f'}[R^0_{f};\bar\phi(x)]+d_H(\dots).
\label{AsymmGeneral}
\eea 
This implies the
skew-symmetry of $K_{\tilde f_1,\tilde f_2}$ 
under exchange of $\tilde f_1^\alpha$ and $\tilde f_2^\alpha$,
and that $K_{\tilde f_1,\tilde f_2}$
are 2-cocycles 
on the Lie algebra of all
asymptotic reducibility parameters, 
\bea
& K_{\tilde f_1,\tilde f_2}=-K_{\tilde f_2,\tilde f_1}\ ,&
\label{Asymm}
\\[4pt]
& 
K_{[\tilde f_1,\tilde f_2]_M,\tilde f_3}
+K_{[\tilde f_3,\tilde f_1]_M,\tilde f_2}
+K_{[\tilde f_2,\tilde f_3]_M,\tilde f_1}= 0.
&
\label{Cocycle}
\eea
\end{theorem}

\noindent
The proof of this theorem is given in appendix \ref{appendixproof}.
We add a few comments:

\begin{itemize}

\item In general, the finiteness
of the charges (\ref{defcharge}) does not
imply the finiteness of
the central charges (\ref{centralcharge1}).
In particular, condition \eqref{convergencec}
which guarantees the existence of the charges $Q_{\4f}$ does
not guarantee the existence of the central charges, unless
the additional conditions
\bea
R^{i 0}_\alpha( \4f^{\alpha})\asy O(\chi^i) \label{BHboundcond}
\eea
on the asymptotic reducibility parameters are satisfied.
The reason is that $K_{\tilde f_1,\tilde f_2}$ arises
from $Q_{\tilde f_2}$ by substituting
$R^{i 0}_\alpha( \4f_1^{\alpha})$ for $\varphi^i$.
Furthermore, when
(\ref{convergencec}) holds, parameters
which satisfy
\bea
R^{i 0}_\alpha( \4f^{\alpha})\asy o(\chi^i) \label{BHtrivboundcond}
\eea
do not contribute to central charges. 
\eqref{BHboundcond} was the starting point of the analysis of
\cite{Brown:1986nw,Henneaux:1985tv}. In the case of
asymptotically
${\rm adS_3}$ gravity,
it implies the conditions (\ref{asympsym}), (\ref{convergencec}),
(\ref{constr1}), (\ref{constr2}).

\item Let $N$ and $K$ be the alternating linear maps 
on the Lie algebra of all asymptotic reducibility parameters defined by 
$N(\4f)=N_{\4f}$ and
$K(\4f_1,\4f_2)= K_{\4f_1,\4f_2}$, respectively.
The consistency condition (\ref{Cocycle}) can be written in terms of the 
Chevalley-Eilenberg differential $\delta^{CE}$ \cite{Chevalley} as 
$\delta^{CE}K=0$, while the term involving the normalization 
on the right hand side of the covariant Poisson bracket can be 
written as the coboundary $(\delta^{CE}N)(\4f_1,\4f_2)$.
The central charge 
$K_{\tilde f_1,\tilde f_2}$ can be removed from
(\ref{asympcom}) by a choice of normalization $N_{\tilde f}$ 
if there exists a
normalization $N_{\tilde f}$ such that
$K_{\tilde f_1,\tilde f_2}= N_{[\tilde f_1,\tilde f_2]_M}$, i.e., 
if the $2$ cocycle $K$ is a coboundary, $K=\delta^{CE} N$. 

\item The Lie algebra of physical interest is not the Lie 
algebra of all asymptotic reducibility parameters, but
the Lie algebra $\mathfrak g$ of equivalence classes of 
asymptotic reducibility parameters. 
Hence, nontrivial central charges are to 
be viewed as (representatives of) 
cohomology classes in degree $2$ of the Lie algebra $\mathfrak g$. 
This can be done consistently if
conditions \eqref{convergencec} and \eqref{BHboundcond} are satisfied,
provided trivial asymptotic reducibility parameters that 
satisfy \eqref{BHboundcond} automatically also satisfy 
\eqref{BHtrivboundcond}.
Then
the finite charges $\int_{{\cal C}^{n-2}}k_{\tilde f_2}[R_{\tilde f_1}^{0}]$
vanish whenever $\tilde f_1$ are trivial asymptotic
reducibility parameters,
so that $K_{\4f_1,\4f_2}$
really only depends on the
equivalence classes $[\tilde f_1]$, $[\tilde f_2]$.
Similarly, because the boundary conditions guarantee that the
charges $\int \tilde k_{\tilde f}[\phi-\bar\phi(x);\bar\phi(x)]$
vanish (asymptotically)
for trivial asymptotic reducibility parameters
(when evaluated at a solution satisfying the boundary condition),
the charge $Q_{\tilde f}$ only depends on
the equivalence class $[\4f]$ of the asymptotic reducibility
parameters.

\item Two particular important cases where the central charges
are necessarily trivial and can be absorbed by an appropriate
choice of
normalization are
the case that the Lie algebra cohomology of
equivalence classes of asymptotic reducibility parameters
in degree $2$ is trivial (``safe
algebras'', e.g., semi-simple finite dimensional
algebras), and the case that all
asymptotic reducibility parameters are equivalent
to exact Killing vectors of the background, because
$R^{i0}_{\tilde f}=0$ implies
$K_{\tilde f_1,\tilde f_2}=0$.
In the latter case, an appropriate choice is
to normalize the charges of the backgound to zero, $N_f=0$,
whereas for a semi-simple $\mathfrak g$,
the existence of an appropriate normalization
follows from $H^2(\mathfrak g)=0$. Furthermore,
because $H^1(\mathfrak g)=0$, the requirement that there should
be no central extension then completely fixes the
normalization of the background.

\end{itemize}

\subsubsection{Effective sources}

Usually, the charges (\ref{defcharge}) are
integrals over boundaries, i.e., $\cC^{n-2}=\6\Sigma$ is the boundary 
of an $(n-1)$-dimensional
region $\Sigma$ of spacetime.
One may then try to identify source terms
in $\Sigma$ and represent
the charges as $(n-1)$-dimensional
integrals over $\Sigma$ of the source terms.
For example, one may define ``source currents''
\bea
j^\mu_\mathrm{eff}:=
S^{\mu i}_\alpha (\frac{\delta L}{\delta \phi^i},\4f^\alpha)
-\6_\nu \4k_{\4f}^{[\nu\mu]}[\phi-\5\phi(x);\5\phi(x)].
\label{jeff}
\eea
This implies
\bea
Q_{\4f}-N_{\4f}
&\stackrel{(\ref{defcharge})}{=}&
\int_{\6\Sigma}
\4k_{\tilde f}[\phi-\5\phi(x);\bar\phi(x)]
\nonumber\\
&\stackrel{\mathrm{Stokes}}{=}&
\int_\Sigma d\4k_{\tilde f}[\phi-\5\phi(x);\bar\phi(x)]
\stackrel{(\ref{jeff})}{\approx}
\int_\Sigma (d^{n-1}x)_\mu\, j^\mu_\mathrm{eff}\ ,
\label{sourceint}
\eea
where $\approx$ denotes weak equality in the full theory. That is,
one has $Q_{\4f}=N_{\4f}+\int_\Sigma (d^{n-1}x)_\mu\, j^\mu_\mathrm{eff}$
for solutions of the field equations
satisfying the respective boundary conditions.
By construction, the currents $j^\mu_\mathrm{eff}$ thus 
yield the same value for the
charges upon integration and they are conserved, 
\[
\6_\mu j^\mu_\mathrm{eff}\approx 
-\6_\mu\6_\nu \4k_{\4f}^{[\nu\mu]}[\phi-\5\phi(x);\5\phi(x)]
=0.
\]

The motivation for the definition (\ref{jeff}) is that
the $j^\mu_\mathrm{eff}$ contain the terms in the
field equations that depend on external sources (if any), or, as in
\cite{Abbott:1982ff}, terms that are  
at least quadratic in the
fields $\varphi=\phi-\5\phi$. 

The difference between
our approach here and the one in \cite{Abbott:1982ff} is that we
concentrate first on the $n-2$ forms and then consider the effective
sources as derived quantities, instead of the other way around. The
advantage is that the procedure becomes constructive and ambiguities
or equivalences for various expressions of the charges can be
controlled. 

\subsection{Remarks on the boundary theory}\label{s54}

Suppose for definiteness that, in addition to the assumptions of section 
\ref{Prerequisites}, we are in the situation where we have coordinates 
$r,s^a$, (with
$s^a$ denoting for instance coordinates such as time or some angles) and
the boundary is at $r\asy\infty$ with boundary conditions
\bea
\varphi^i(x)=r^{m^i}\tilde\varphi^i(s)+o(r^{m^i}),
\eea
i.e., $|\chi^i|=m^i$. 
This means that, when evaluated at fields that satisfy the boundary
conditions, a linear local form is to leading order
a form that lives on the 
jet-bundle with base space coordinates $s^a$ and fiber coordinates
$\tilde \varphi^i$ and their derivatives with respect to 
$s^a$ with a parametrical
dependence on $r$. 

If we define 
\bea
\frac{\delta L^{\rm free}}{\delta\varphi^i}\Big|_{\varphi(x)}d^nx=
L^{as}_i+o(\chi_i),
\eea
so that $|L^{as}_i|=|\chi_i|$, 
the boundary theory that controls the leading order contributions of 
asymptotic solutions of the bulk
theory can be defined to be the linear theory for the fields
$\tilde \varphi^i$ with 
dynamics determined by the equations $\partial_{(a)} L^{as}_i=0$
[a priori, it is not
guaranteed that the equations $L^{as}_i=0$ derive from a variational
principle]. 

Asymptotic solutions are determined by exact solutions $\tilde\varphi^i(s)$
of the boundary theory,
\bea
L^{as}_i\Big|_{\tilde\varphi^i(s)}=0.
\eea
Denoting $m_i:=|\chi_i|$ and 
$m_\alpha:=|\chi_\alpha|$, we have
$\psi_i=r^{m_i}\tilde\psi_i(s)
+o(r^{m_i})$ and $\psi_\alpha=r^{m_\alpha}\tilde
\psi_\alpha(s)+o(r^{m_\alpha})$. 
One can decompose the generating set of
Noether operators of the linearized theory according to 
\bea
R^{+i0}_\alpha\psi_i=r^{m_\alpha}\tilde R^{+i0}_\alpha\tilde\psi_i(s) 
+o(r^{m_\alpha})
\eea
with $\tilde R^{+i0}_\alpha=\tilde R^{+i0(a)}_\alpha(s)\6_{(a)}$.
Generically
$\{\tilde R^{+i0}_\alpha\}$ will be a generating 
set of Noether operators of the boundary theory and
the asymptotic regularity conditions of section \ref{Prerequisites}
will imply standard regularity conditions for the boundary theory,
at least when the latter can be traced to
identities involving only field independent operators (as one
would expect for a linear theory).
Indeed, suppose that $\4N^i=r^{M-m_i}\4N^{i(a)}(s)\6_{(a)}$ is a field
independent Noether operator of the boundary theory, $\4N^i L^{as}_i=0$,
for some $M$.
Defining $N^i:=r^{-n'-M}\4N^i$ where $n'=|d^nx|$, we obtain
$d^nx N^i\delta L^\mathrm{free}/\delta \varphi^i\asy 0$.
\eqref{asyacycdelta} implies now
$r^{m_i}\4N^i\4\psi_i=r^M\4Z^{+\alpha}\4R^{+i0}_\alpha\4\psi_i$
for some operators $\4Z^\alpha=\4Z^{\alpha(a)}(s)\6_{(a)}$
and all $\4\psi_i$, i.e.,
$\4N^i=r^{M-m_i}\4Z^{+\alpha}\4R^{+i0}_\alpha$.

Suppose the functions $\tilde f^\alpha$ are asymptotic reducibility
parameters that satisfy condition
\eqref{convergencec} for finite charges, i.e.,
\bea
\tilde f^\alpha=\tilde f^\alpha_m+o(r^{-m_\alpha}),\quad
\tilde f^\alpha_m=r^{-m_\alpha}h^\alpha(s).
\eea
The leading order of the
definition \eqref{asympsym} of asymptotic reducibility parameters then
requires $\4f^\alpha_m$ to be exact reducibility parameters for the
operators $\tilde
R^{i0}_\alpha$ associated to the boundary theory, 
\bea
\tilde
R^{i0}_\alpha(\4f^\alpha_{m})=0.
\eea

If \eqref{constr1}-\eqref{newass2} hold
we obtain from \eqref{4.7}:
\bea
R^{i1}_\alpha(\tilde f^\alpha)\,
\frac{\delta L^\mathrm{free}}{\delta\varphi^i}\,d^nx
\asy d_H(\cdot),
\quad \Longrightarrow\quad 
\varphi^i\ \frac{\delta(\delta^1_{\tilde f}L^\mathrm{free})}
{\delta\varphi^i}\,d^nx \asy 0.
\eea 

Commuting the Euler-Lagrange derivative with the vector field 
$\delta^1_{\tilde f}=\partial_{(\mu)}
[R^{i1}_\alpha(\tilde
f^\alpha)]{\partial}/{\partial\varphi^i_{(\mu)}}$ then implies 
\bea
\varphi^j\Big[\delta^1_{\tilde f}\,
\frac{\delta L^\mathrm{free}}{\delta\varphi^j}
+(-\partial)_{(\mu)}[\frac{\partial R^{i1}_\alpha(\tilde
f^\alpha)}{\partial \varphi^j_{(\mu)}}\ \frac{\delta
L^\mathrm{free}}{\delta\varphi^i}]\Big] d^nx\asy 0.\label{asyeqmotsym}
\eea
Assuming that \eqref{newass2} holds, we may write
\bea
R^{i1}_\alpha(\tilde f^\alpha)=\tilde R^{i1}_\alpha(\tilde
f^\alpha_m)+o(\chi^i),
\label{copper1}
\eea
with $|\tilde R^{i1}_\alpha(\tilde
f^\alpha_m)|=|\chi^i|=m^i$.

By choosing $\varphi^i=r^{m^i}\tilde \varphi^i(s)$ and considering only
the leading order in \eqref{asyeqmotsym}, we obtain  
\bea
\tilde\delta^1_{\tilde f}L^{as}_j\approx^{\rm bd} 0,
\eea
where $\approx^{\rm bd}$ means equality when the equations of motions 
of the boundary theory hold,
and 
\bea
\tilde\delta^1_{\tilde f}=\partial_{(a)}[\tilde R^{i1}_\alpha(\tilde
f^\alpha_m)]\frac{\partial}{\partial \tilde\varphi^i_{(a)}}.
\eea
Hence, $\tilde\delta^1_{\tilde f}$ defines a symmetry of the equations
of motions of the boundary theory. 

For field independent gauge parameters, the contribution linear 
in the fields in the expansion of equation 
\eqref{1.18} gives 
\bea
\partial_{(\mu)}R^{j1}_\alpha(\tilde f_1^\alpha)\frac{\partial
  R^{i1}_{\beta}(\tilde
  f_2^\beta)}{\partial\varphi^j_{(\mu)}}
+\partial_{(\mu)}R^{j0}_\alpha(\tilde f_1^\alpha)\frac{\partial
  R^{i2}_{\beta}(\tilde
  f_2^\beta)}{\partial\varphi^j_{(\mu)}}-(1\longleftrightarrow
2)\approx^{\rm free}  R^{i1}_{\gamma}(C^{\gamma
  0}_{\alpha\beta}(\tilde f^\alpha_1,\tilde f^\beta_2).
\eea
Under the assumptions \eqref{newass2}, \eqref{expasymsym} and
\eqref{BHboundcond}, 
the leading order contribution to this equation gives
\bea
\tilde\delta^1_{\tilde f_1}\tilde R^{i1}_{\alpha}(\tilde f^\alpha_{2
  m})-(1\longrightarrow 2)\approx^{\rm bd} 
\tilde R^{i1}_{\alpha}([\tilde f_{1 m},\tilde f_{2 m}]_{\rm bd}),
\eea
while 
\bea
[\tilde f_1,\tilde f_2]^\alpha_M= [\tilde f_{1m},\tilde
f_{2m}]_{\rm bd}^\alpha+o(1/\chi_\alpha),
\eea
with $|[\tilde f_{1m},\tilde
f_{2m}]^\alpha|=|1/\chi_\alpha|=-m_\alpha$. 
Hence, on-shell for the boundary theory, the commutator algebra of the
equations of motion symmetries $\tilde \delta^1_ {\tilde f_m}$ 
represents the Lie algebra
$\mathfrak{g}$ of equivalence classes of asymptotic reducibility
parameters: 
\bea
[\tilde \delta^1_{\tilde f_{1 m}},\tilde \delta^1_{\tilde f_{2
    m}}]\approx^{\rm bd} \tilde \delta^1_{[\tilde f_{1m},\tilde
f_{2m}]_{\rm bd}}.
\eea

\mysection{Standard applications}\label{s6}

In this section, we illustrate and test the general results 
by applying them to the well studied cases of 
electrodynamics, Yang-Mills theory 
and Einstein gravity. We shall specify in each case
the superpotential of Eq.\ (\ref{superpot2}) 
and related quantities, such as asymptotic reducibility 
parameters and conserved charges.
In all cases treated here, the boundary conditions are imposed at the
boundary $\6\Sigma$ of a
spatial $(n-1)$-dimensional volume $\Sigma$ (not necessarily
at spatial infinity). 
For simplicity we shall assume that all ``matter fields'' fall off 
sufficiently fast to be negligible near $\6\Sigma$
and that
external sources vanish there. Accordingly, all
background matter fields vanish and the
background gauge or metric fields solve the source-free
Maxwell, Yang-Mills and Einstein equations,
respectively, possibly with
a cosmological constant
in the gravitational case. 
Furthermore, we shall mostly discuss the particular case of 
asymptotic reducibility parameters that are exact 
Killing vectors of the background
($R^{i0}_\alpha (\4f^\alpha)=0$) because
for these parameters the discussion can be made without
more specific assumptions on the boundary conditions.
The only exception where we consider precise boundary conditions and 
determine all 
the associated asymptotic reducibility parameters is the well-known 
example of three-dimensional
asymptotically anti-de Sitter gravity. The reason is of course that
this example gives rise to central extensions
in the algebra of conserved charges, and thus provides a
particularly nontrivial illustration of our general framework.

\subsection{Electrodynamics}

As a warm-up, we briefly discuss electrodynamics with
Lagrangian
$L=-\frac{1}{4}F^{\mu\nu}F_{\mu\nu} +\Lmat$ where
$F_{\mu\nu}=\6_\mu A_\nu-\6_\nu A_\mu$ are the
electromagnetic field strengths and
$\Lmat$ is a ``matter field Lagrangian''
of the standard type (such as
Dirac spinor fields minimally coupled to the gauge fields
via covariant derivatives), or contains terms with external
sources (such as $A_\mu j^\mu(x)$, with $\6_\mu j^\mu(x)=0$).
For the standard cases that $\Lmat$ contains only terms that are at least
quadratic in the matter fields and that
gauge transformations of the matter fields do not contain
derivatives of the gauge parameter,
the current in Eq.\ (\ref{smalls}) is 
$s^\mu_{\4f}=\4f\6_\nu f^{\nu\mu}$.
Here $\4f$ is an asymptotic reducibility parameter and
$f_{\mu\nu}=\6_\mu a_\nu-\6_\nu a_\mu$ is the
field strength of $a_\mu=A_\mu-\5A_\mu(x)$, with
$\5A_\mu(x)$ the background gauge fields.
The only asymptotic reducibility parameters
that are exact Killing vectors of the background
are constants, $\4f=c=\mathit{constant}$.
They yield $s^\mu_{c}=\6_\nu (c f^{\nu\mu})$ and
the corresponding superpotential (\ref{superpot2}) is
simply $\4k_c^{[\nu\mu]}=c f^{\nu\mu}$.
An asymptotically conserved $(n-2)$-form is thus
$c(d^{n-2}x)_{\nu\mu} (F^{\nu\mu}-\5F^{\nu\mu})$.
Owing to $d_V(F^{\nu\mu}-\5F^{\nu\mu})=d_VF^{\nu\mu}$,
a simpler (equivalent) choice is the $(n-2)$-form
\[
k_c[A]=c\, (d^{n-2}x)_{\nu\mu} F^{\nu\mu}.
\]
By integrating $k_c[A]$ over $\6\Sigma$, one
gets the corresponding conserved charge.
Notice that, actually, there is
a one-parameter family of conserved charges $Q_c$
para\-me\-tri\-zed by $c$. The charge is 
the ``generator'' $Q:=\6Q_c/\6c$ of this
family,\footnote{Here and in the following we use the notation
$d\sigma_i\equiv 2(d^{n-2}x)_{0i}$}
\[
Q=\int_{\6\Sigma}d\sigma_i\, F^{0i}.
\]
(\ref{jeff}) gives here $j^\mu_\mathrm{eff}
=\delta \Lmat/\delta A_\mu$, and (\ref{sourceint}) then implies 
\[
Q=-\int_{\6\Sigma} d\sigma_{i}\, F^{i0}=
-\int_\Sigma d\sigma\, \6_i F^{i0}\approx\int_\Sigma d\sigma\, j^0.
\]
Hence, when evaluated 
for solutions to the equations of 
motion, $Q$ agrees with
$\int_\Sigma d\sigma j^0$
where
$j^0=\delta \Lmat/\delta A_0$ is the charge density 
appearing in the Maxwell equation
$\6_i F^{i0}= -j^0$, so that the standard textbook expression for the 
electric charge is recovered. 

\subsection{Yang-Mills theory}

We consider a Lagrangian 
\bea
L=\frac{1}{4}\mathit{Tr}(F^{\mu\nu} F_{\mu\nu})+\Lmat,
\label{YMaction}
\eea
where $F_{\mu\nu}=\6_\mu A_\nu-\6_\nu A_\mu+[A_\mu,A_\nu]$ are the
nonabelian field strengths of the gauge
fields $A_\mu=A_\mu^a T_a$. We use here matrix notation and the 
conventions that $T_a$
are antihermitian representation matrices normalized
according to $\mathit{Tr}(T_a T_b)=-\delta_{ab}$.
Analogously to electrodynamics discussed before,
$\Lmat$ may contain matter fields or external sources
coupled to the gauge fields. Again, we assume that
all matter fields are negligible and all external sources
vanish near $\6\Sigma$. In particular, all
background matter fields vanish.

\subsubsection{Superpotentials}

Assuming a standard Lagrangian which contains
only terms that are at least quadratic in the matter fields,
one obtains
\bea
\frac{\delta L^\mathrm{free}}{\delta a_\mu^a}\,\delta^{ab}T_b=
\5D_\nu f^{\nu\mu}+[a_\nu, \5F^{\nu\mu}],
\label{linym}\eea
where 
\beann
a_\mu=A_\mu-\5A_\mu,\quad
f_{\mu\nu}=\5D_\mu a_\nu-\5D_\nu a_\mu,\quad 
\5D_\mu\,\cdot\, =\6_\mu +[\5A_\mu,\,\cdot\,].
\eeann
The currents
$s^\mu_{\4f}$ of Eq.\ (\ref{smalls}) are
\bea
s^\mu_{\4f}[a;\5A]=
-\mathit{Tr}(\4f\5D_\nu f^{\nu\mu}+\4f[a_\nu, \5F^{\nu\mu}]),
\label{ymsmalls}
\eea
where $\4f=\4f^a(x) T_a$ involves
the asymptotic reducibility parameters $\4f^a(x)$.
The latter are subject to (\ref{asympsym}) which requires in this case
\bea
\forall a_\mu^a\asy O(\chi_\mu^a):\quad
d^nx\, \mathit{Tr}(\5D_\nu f^{\nu\mu}\5D_\mu\4f
+[a_\nu, \5F^{\nu\mu}]\5D_\mu\4f)\asy 0,
\eea
where $\chi_\mu^a$ characterizes the boundary condition
for $a_\mu^a$.
According to (\ref{superpot2}), the associated superpotentials are given by 
\bea
\4k^{[\mu\nu]}_{\4f}[a;\5A]=-\mathit{Tr}\Big(\frac{3}{2}[\bar
A^\mu,a^\nu]\tilde f
+\frac{1}{2}a^\mu\partial^\nu \tilde f
+\tilde f\partial^\mu a^\nu -(\mu\leftrightarrow\nu)\Big).
\label{ymsuperpot}
\eea
Let us now discuss
asymptotic reducibility parameters that are
exact Killing vectors of the background,
\bea
\5D_\mu\4f=0.\label{kil}
\eea
For
parameters satisfying (\ref{kil}), we can simplify (\ref{ymsuperpot})
by substituting
$-[\5A^\nu,\4f]$ for $\partial^\nu\tilde f$. This yields
\bea
\4k^{[\mu\nu]}_{\4f}[a;\5A]\stackrel{(\ref{kil})}{=} 
-\mathit{Tr}(\4f f^{\mu\nu}),
\label{ymsuperpot1}\eea
which agrees with equation (5) of \cite{Abbott:1982jh}.
Equivalently, we can use (\ref{kil}) to substitute 
$-\partial^\nu\tilde f$ for $[\5A^\nu,\4f]$ in
(\ref{ymsuperpot}). Then we obtain,
using once again (\ref{kil}),
\bea
\4k^{[\mu\nu]}_{\4f}[a;\5A]\stackrel{(\ref{kil})}{=}
\6^\mu\cA^\nu-\6^\nu\cA^\mu,\quad \cA^\mu=-\mathit{Tr}(\4f a^\mu).
\label{ymsuperpot1a}\eea

{\bf Remark.}
Actually (\ref{ymsuperpot}) is
not restricted to the case that matter fields can be neglected near
$\6\Sigma$.
Rather, it even holds for solutions with possibly
non-negligible matter fields near $\6\Sigma$, assuming a Yang-Mills-matter 
Lagrangian of the standard type
(with matter fields that are fermions or scalar fields).
The reason is that, for
a standard Yang-Mills-matter system, the current $s^\mu_{\4f}$ involves
only the linearized field equations for the
gauge fields but not those for the matter fields because
the gauge transformations of standard matter fields do not 
involve derivatives of the gauge parameters. 
The matter field dependent terms in $s^\mu_{\4f}$ then
either do not contain derivatives at all (in the case of fermions)
or they contain precisely one derivative whose index coincides with
the index $\mu$ of $s^\mu_{\4f}$ (in the case of scalar fields).
As a consequence, they give no contributions to $\4k^{[\mu\nu]}_{\4f}$
at all,
as one easily reads off from (\ref{superpot2}). The only possible
effect that the matter fields may then have 
are extra conditions on the parameters $\4f$, 
but (\ref{ymsuperpot}) does not change.

\subsubsection{Asymptotically conserved n--2 forms}

Equation (\ref{ymsuperpot}) yields directly
asymptotically conserved $(n-2)$-forms given by
\bea
k_{\4f}[A;\5A(x)]= 
(d^{n-2}x)_{\mu\nu}\4k^{[\mu\nu]}_{\4f}[A-\5A;\5A].
\label{ymsuperpot2}\eea
When (\ref{kil}) holds,
there are somewhat more elegant, equivalent expressions for $k_{\4f}$
which do not explicitly depend on the background fields. The first
one corresponds to (\ref{ymsuperpot1a}) and reads
\bea
k'_{\4f}[A]\stackrel{(\ref{kil})}{=} 
-2(d^{n-2}x)_{\mu\nu}\6^\mu\mathit{Tr}(\4f A^\nu).
\label{ymsuperpot1b}\eea
Another one corresponds to (\ref{ymsuperpot1}) and reads
\bea
k''_{\4f}[A]\stackrel{(\ref{kil})}{=} 
-(d^{n-2}x)_{\mu\nu}\mathit{Tr}(\4f F^{\mu\nu}).
\label{ymsuperpot3}\eea
The  equivalence of all these
expressions is due to
\[
\Big[d_Vk''_{\4f}[A]\Big]_{\5A(x)}
\stackrel{(\ref{kil})}{=}d_Vk_{\4f}[A;\5A(x)]
\stackrel{(\ref{kil})}{=}d_Vk'_{\4f}[A] .
\]

\subsubsection{Example: asymptotically flat connections}

Let us finally consider asymptotically flat
connections as in \cite{Abbott:1982jh}, using a background
$\bar A_\mu= g^{-1}(x)\partial_\mu g(x)$.
The exact Killing vectors of such a background are easily found:
multiplying (\ref{kil}) from the left with $g(x)$ and from the right with
$g^{-1}(x)$ gives $\6_\mu[g(x)\4fg^{-1}(x)]=0$
and thus $g(x)\4fg^{-1}(x)=c^aT_a$ with 
constant parameters $c^a$. Hence, let us consider
\[
\4f=c^ag^{-1}(x)T_a g(x).
\] 
We define corresponding ``color'' charges by $Q_a:=\6Q_{\4f}/\6c^a$
where $Q_{\4f}$ is the integral over $\6\Sigma$ of
the asymptotically conserved $(n-2)$-forms. Using (\ref{ymsuperpot3}),
one obtains
\[
Q_a=
-\int_{\6\Sigma}d\sigma_i\, \mathit{Tr}
\Big[g^{-1}(x)T_a g(x) F^{0i}\Big].
\]
Because we have considered exact Killing vectors, 
there is no central extension in the covariant Poisson algebra of
the corresponding color charges:
\[
\{Q_a,Q_b\}_{CF}={f_{ab}}^cQ_c\ .
\]
This can be easily verified using
$\delta_{\4f}F^{0i}=[F^{0i},\4f]$. Here
${f_{ab}}^c$ are the structure constants of the
Lie algebra of the gauge group in the basis
associated to $\{T_a\}$,
$[T_a,T_b]={f_{ab}}^cT_c$. 

Of course, whether or not these charges vanish depends on the
behaviour of the gauge fields near the boundary. For instance,
the BPST instanton solution \cite{Belavin:1975fg}
of Yang-Mills theory in four-dimensional
Euclidean space
satisfies $A_\mu\asy g^{-1}(x)\partial_\mu g(x)$ 
at infinity but
yields only vanishing charges $Q_a$ because the field strengths
fall off too fast. The same is actually true for all Yang-Mills instantons
and related to the fact that they have
finite action as this requires that the field strengths
fall off faster than $1/r^2$.

\subsection{Einstein gravity}

We finally discuss 
Einstein gravity (without or with cosmological constant
$\Lambda$) in spacetime
dimensions $n\geq 3$ with Lagrangian
\[
L=\frac{1}{16\pi}\sqrt{-g}(R-2\Lambda)+\Lmat,
\]
where $R=g^{\mu\nu}{R_{\rho\mu\nu}}^\rho$,
${R_{\rho\mu\nu}}^\lambda=\6_\rho{\Gamma_{\mu\nu}}^\lambda+
{\Gamma_{\rho\sigma}}^\lambda{\Gamma_{\mu\nu}}^\sigma
-(\rho\leftrightarrow\mu)$. 
As before in the cases of electrodynamics and Yang-Mills theory,
$\Lmat$ may contain matter fields which are assumed to be negligible
near $\6\Sigma$, or external sources which vanish near $\6\Sigma$. We
introduce the standard notation $h_{\mu\nu}=h_{\nu\mu}$ for
the deviation of the metric fields from the background
metric $\bar g_{\mu\nu}(x)$ ($g_{\mu\nu}=h_{\mu\nu}+
\bar g_{\mu\nu}(x)$). The
background metric and its inverse are used to lower and raise 
world indices. In particular we thus use the notation
${h_\mu}^\nu=\5g^{\nu\rho}h_{\mu\rho}$ and $h^{\mu\nu}=\5g^{\mu\rho}
\5g^{\nu\sigma}h_{\rho\sigma}$. Furthermore, $h$ denotes
the trace of ${h_\mu}^\nu$, i.e., $h=\5g^{\mu\nu}h_{\mu\nu}$,
and $\5D_\mu$ are
background covariant derivatives, such as 
$\5D_\mu h^{\nu\rho}=\6_\mu h^{\nu\rho}+\5\Gamma_{\mu\lambda}{}^\nu
h^{\lambda\rho}+\5\Gamma_{\mu\lambda}{}^\rho h^{\nu\lambda}$,
$\bar D^\mu h=\6^\mu h=\5g^{\mu\nu}\6_\nu h$.
We write
the (full) equations of motion for the metric as
\bea
\cH^{\mu\nu}
+\frac {\sqrt{-\5g}}{2}\,T^{\mu\nu}_\mathrm{eff}=0,
\label{Einstein}
\eea
where $\cH^{\mu\nu}[h;\5g]$ is the linear
part of $\delta [(1/16\pi)\sqrt{-g}(R-2\Lambda)]/\delta g_{\mu\nu}$,
\bea
\lefteqn{
\cH^{\mu\nu}[h;\5g]:=
-\frac{1}{16\pi}\Big[d_V(
\sqrt{-g}R^{\mu\nu}-\frac 12\sqrt{-g} g^{\mu\nu}R+\sqrt{-g}g^{\mu\nu}\Lambda)
\Big]_{\bar g,h}}
\nonumber\\
&=&
\frac{\sqrt{-\bar g}}{32 \pi}\Big(-h\bar R^{\mu\nu}
+\frac{1}{2}h\bar R \bar g^{\mu\nu} + 2 h^{\mu\alpha}\bar R_\alpha{}^\nu
+2h^{\nu\beta}\bar R_\beta{}^\mu
-h^{\mu\nu}\bar R-h^{\alpha\beta}\bar R_{\alpha\beta}\bar g^{\mu\nu}
\nonumber\\
&&
+\bar D^\mu\bar D^\nu h+\bar D^\lambda\bar
D_\lambda h^{\mu\nu}
-2\bar D_\lambda \bar D^{(\mu}h^{\nu)\lambda} 
-\bar g^{\mu\nu}(\bar D^\lambda\bar D_\lambda h
- \bar D_\lambda\bar D_\rho h^{\rho\lambda})
\nonumber\\
&&
+2\Lambda h^{\mu\nu}-\Lambda\bar g^{\mu\nu}h\Big),
\label{Hmunu}\eea
and
$(1/2)\sqrt{-\5g} T^{\mu\nu}_\mathrm{eff}$ simply
collects all terms of $\delta L/\delta g_{\mu\nu}$
not contained in $\cH^{\mu\nu}$,
\[
T^{\mu\nu}_\mathrm{eff}:=\frac{2}{\sqrt{-\5g}}
\Big(\frac{\delta L}{\delta g_{\mu\nu}}
-\cH^{\mu\nu}
\Big).
\]
We also note that (\ref{Hmunu}) reduces to the following
expression when $\5g_{\mu\nu}$ solves the Einstein equations
$\5R_{\mu\nu}=2(n-2)^{-1}\Lambda \5g_{\mu\nu}$:
\bea
\cH^{\mu\nu}[h;\5g]=
\frac{\sqrt{-\bar g}}{32 \pi}\Big(
\frac{2\Lambda}{n-2}\,(2h^{\mu\nu}- \5g^{\mu\nu}h)
+\bar D^\mu\bar D^\nu h+\bar D^\lambda\bar D_\lambda h^{\mu\nu}
\nonumber\\
-2\bar D_\lambda \bar D^{(\mu}h^{\nu)\lambda} 
-\bar g^{\mu\nu}(\bar D^\lambda\bar D_\lambda h
- \bar D_\lambda\bar D_\rho h^{\rho\lambda})\Big).
\label{Hmunu2}\eea

\subsubsection{Superpotentials and asymptotically conserved n--2 forms}
\label{gravsp}

The assumption that the matter fields are negligible near the boundary
means more precisely that near the boundary they give no contribution to the
$(n-1)$-form constructed of
the current (\ref{smalls}). Then this $(n-1)$-form is
\bea
(d^{n-1}x)_\mu\,
s^\mu_\xi[h;\5g]\asy 
2(d^{n-1}x)_\mu{\cal H}^{\mu\nu}
\xi_\nu\ ,
\eea
where the use of $\asy$ (instead of $=$) indicates that terms with
matter fields (if any) have been neglected [in general, $s^\mu_\xi$
may contain terms with the linearized
equations of motion of matter fields, as the gauge transformations
of matter fields may contain derivatives of the gauge parameters].
$\xi_\nu=\xi_\nu(x)$ are asymptotic reducibility parameters satisfying
(\ref{asympsym}) which requires in this case
\bea
\forall h_{\mu\nu}\asy O(\chi_{\mu\nu}):\quad
d^nx\, {\cal H}^{\mu\nu}
\5D_\mu\xi_\nu\asy 0,
\label{xicondition}
\eea
where $\chi_{\mu\nu}$ characterizes the boundary condition
for $h_{\mu\nu}$.
Applying (\ref{superpot2}) to $2{\cal H}^{\mu\nu}\xi_\nu$, we find
for the superpotential $\4k^{[\nu\mu]}_\xi$:
\bea
\tilde k^{[\nu\mu]}_\xi[h;\5g]
&=&-\frac{\sqrt{-\bar g}}{16 \pi}\Big[\bar D^\nu
(h\xi^\mu)+\bar D_\sigma(h^{\mu\sigma}\xi^\nu)
+\bar D^\mu(h^{\nu\sigma}\xi_\sigma)\nonumber\\
&&
+\frac{3}{2}h\bar D^\mu\xi^\nu+\frac{3}{2}
h^{\sigma\mu}\bar D^\nu\xi_\sigma+\frac{3}{2}
h^{\nu\sigma}\bar D_\sigma\xi^\mu
-(\mu\leftrightarrow \nu)\Big].
\label{gsuperpot0}\eea
This expression can be more compactly written as
\bea
\tilde k^{[\nu\mu]}_{\xi} [h;\5g]
=
\frac{\sqrt{-\bar g}}{16\pi}\Big(\xi_\rho \5D_\sigma H^{\rho\sigma\nu\mu}
+\frac 12 H^{\rho\sigma\nu\mu}
\6_\rho\xi_\sigma
\Big),
\label{gsuperpot3}
\eea
where
$H^{\rho\sigma\mu\nu}[h;\5g]$ is
the following background tensor 
with the symmetries of the Riemann tensor:
\bea
H^{\mu\alpha\nu\beta}[h;\5g]&=&
-\7h^{\alpha\beta}\5g^{\mu\nu}
-\7h^{\mu\nu}\5g^{\alpha\beta}
+\7h^{\alpha\nu}\5g^{\mu\beta}
+\7h^{\mu\beta}\5g^{\alpha\nu},
\label{Hdef}\\
\7h_{\mu\nu}&=&
h_{\mu\nu}-\frac{1}{2}\bar g_{\mu\nu} h.
\label{hath}
\eea
The first term on the right hand side
of (\ref{gsuperpot3}) just collects all
terms in (\ref{gsuperpot0}) with background covariant derivatives
of $h_{\mu\nu}$, the second one the terms
with background covariant derivatives of
$\xi$ (we used that $\5D_{[\mu}\xi_{\nu]}=\6_{[\mu}\xi_{\nu]}$).
Equation (\ref{gsuperpot3}) generalizes superpotentials
that are familiar in the particular case of
asymptotically flat spacetimes to more general
asymptotics (asymptotically flat spacetimes will be briefly discussed
in the next subsection). It had been originally obtained for exact
Killing vectors of the background metric
in \cite{Abbott:1982ff}, equation (2.17).

Note that neither the terms in ${\cal H}^{\mu\nu}$ with
the background curvatures nor those with the cosmological constant
contribute to $\tilde k^{[\nu\mu]}_\xi$
because they do not contain derivatives of $h_{\mu\nu}$.
Hence, a cosmological constant affects the superpotential
only indirectly via its influence on the background.

{}From (\ref{gsuperpot0}) or (\ref{gsuperpot3}) one obtains
the asymptotically conserved $(n-2)$-form associated to the asymptotic
reducibility parameters
$\xi_\mu(x)$:
\bea
k_\xi[g;\5g]
=(d^{n-2}x)_{\nu\mu}\,\4k^{[\nu\mu]}_\xi
[g-\5g;\5g].
\label{gsuperpot2}\eea
Suppose now that the $\xi_\mu$ are exact Killing vectors of 
the background metric,
\bea
\5D_\mu\xi_\nu+\5D_\nu\xi_\mu=0.
\label{killing}
\eea
In this case, (\ref{gsuperpot2}) reduces to the
$(n-2)$-form given in equation (11) of \cite{Anderson:1996sc}:
\bea
k_\xi[h;\5g]
\stackrel{(\ref{killing})}{=}
\frac{\sqrt{-\bar g}}{16 \pi}\,(d^{n-2}x)_{\nu\mu}\,
\Big[\xi^\nu\bar D^\mu h
-\xi^\nu\bar D_\sigma h^{\mu\sigma}
+\xi_\sigma\bar D^\nu h^{\mu\sigma}
\nonumber\\
+\frac{1}{2}h\bar D^\nu\xi^\mu
-h^{\nu\sigma}\bar D_\sigma
\xi^\mu-(\mu\leftrightarrow \nu)\Big].
\label{gsuperpot}\eea
This is so because (\ref{gsuperpot2}) is given by
(\ref{gsuperpot}) plus the term
\bea
\frac{\sqrt{-\bar g}}{16 \pi}
\,(d^{n-2}x)_{\nu\mu}\,
h^{\rho\nu}(\5D^\mu\xi_\rho+\5D_\rho\xi^\mu).
\label{zusatz}
\eea
More generally, one may use the somewhat simpler
$(n-2)$-form (\ref{gsuperpot}) instead of (\ref{gsuperpot2})
whenever (\ref{zusatz}) vanishes asymptotically.

\subsubsection{Central charges}\label{gravcentralcharges}

According to section \ref{fullalgebra}, the central extensions
$K_{\xi',\xi}$ which can occur in the algebra 
of the gravitational conserved charges
arise from (\ref{gsuperpot2}) by substituting
there $\5D_\mu\xi'_\nu+\5D_\nu\xi'_\mu$ for $h_{\mu\nu}$ and then
integrating over $\6\Sigma$.
Performing this substitution in (\ref{gsuperpot0})
we obtain
\bea
\lefteqn{
\tilde k^{[\nu\mu]}_\xi[\5D_\lambda\xi'_\rho+\5D_\rho\xi'_\lambda;
\5g_{\lambda\rho}]
}
\nonumber\\
&=&
\frac{\sqrt{-\bar g}}{16 \pi}\Big[
\5D_\rho(\xi^\nu \5D^\mu \xi^{\prime\rho}
+\xi^\mu \5D^\rho \xi^{\prime\nu}
+\xi^\rho \5D^\nu \xi^{\prime\mu})
\nonumber\\
&&
-\5D_\rho\xi^\rho\, \5D^\nu\xi^{\prime\mu}
+\5D_\rho\xi^{\prime\rho}\, \5D^\nu\xi^{\mu}
+2\5D_\rho\xi^{\nu}\, \5D^\rho\xi^{\prime\mu}
\nonumber\\
&&
+\frac 12(\5D^\rho \xi^{\prime\nu}+\5D^\nu \xi^{\prime\rho})
(\5D^\mu \xi_\rho+\5D_\rho \xi^\mu)
-2\5R^{\mu\rho}\xi^\nu\xi'_\rho
+\5R^{\mu\nu\rho\sigma}\xi_\rho\xi'_\sigma\Big]
-(\mu\leftrightarrow \nu).\quad
\label{ccintegrand}
\eea
The terms in the first line on the right hand side
do not contribute to $K_{\xi',\xi}$
because they only contribute an exact form to its integrand,
\beann
(d^{n-2}x)_{\nu\mu}\sqrt{-\bar g}\,\5D_\rho 
(\xi^{[\nu} \5D^\mu \xi^{\prime\rho]})
&=&
(d^{n-2}x)_{\nu\mu}\6_\rho(\sqrt{-\bar g}\, 
\xi^{[\nu} \5D^\mu \xi^{\prime\rho]})
\\
&=&
d_H \Big[
(d^{n-3}x)_{\nu\mu\rho}\sqrt{-\bar g} \,
\xi^{\nu} \5D^\mu \xi^{\prime\rho}
\Big].
\eeann
When the background metric satisfies the
Einstein equations $\5R_{\mu\nu}=2(n-2)^{-1}\Lambda \5g_{\mu\nu}$,
we obtain for the
gravitational central charges:
\bea
K_{\xi',\xi}=\frac{1}{16 \pi}
\int_{\6\Sigma} (d^{n-2}x)_{\nu\mu}
\sqrt{-\bar g}
\Big[
-2\5D_\rho\xi^\rho\, \5D^\nu\xi^{\prime\mu}
+2\5D_\rho\xi^{\prime\rho}\, \5D^\nu\xi^{\mu}
\nonumber\\
+4\5D_\rho\xi^{\nu}\, \5D^\rho\xi^{\prime\mu}
+(\5D^\rho \xi^{\prime\nu}+\5D^\nu \xi^{\prime\rho})
(\5D^\mu \xi_\rho+\5D_\rho \xi^\mu)
\nonumber\\
+\frac{8\Lambda}{2-n}\,\xi^\nu\xi^{\prime\mu}
+2\5R^{\mu\nu\rho\sigma}\xi_\rho\xi'_\sigma
\Big].
\label{gcentralcharge}
\eea
This expression is manifestly skew symmetric
under exchange of $\xi$ and $\xi'$, owing to 
$(d^{n-2}x)_{\nu\mu}=-(d^{n-2}x)_{\mu\nu}$.
Remember, however, that it is not guaranteed to be finite unless
the charges $\int_{\6\Sigma}k_{\xi}[h;\5g]$
themselves are finite and
$\5D_\mu\xi_\nu+\5D_\nu\xi_\mu=O(\chi_{\mu\nu})=
\5D_\mu\xi'_\nu+\5D_\nu\xi'_\mu$ holds.

\subsubsection{Asymptotically
flat spacetimes}

We shall now briefly discuss the important case of
asymptotically flat spacetimes (with $\Lambda=0$). In particular
we shall show that
the superpotentials (\ref{gsuperpot0}) or, equivalently, (\ref{gsuperpot3}) 
reproduce
standard expressions for conserved quantities
in asymptotically flat spacetimes. We shall thus
use as background metric the Minkowski metric $\eta_{\mu\nu}$, 
so that $h_{\mu\nu}$ is the deviation of the $g_{\mu\nu}$
from the Minkowski metric,
\[
h_{\mu\nu}=g_{\mu\nu}-\eta_{\mu\nu}\ ,\quad
\5g_{\mu\nu}=\eta_{\mu\nu}=\mathrm{diag}(-1,+1,\dots,+1).
\]
The Einstein equations in the form (\ref{Einstein}) read now
\bea
\6_\rho\6_\sigma H^{\mu\rho\nu\sigma}=16\pi T^{\mu\nu}_\mathrm{eff},
\quad 
H^{\mu\rho\nu\sigma}=H^{\mu\rho\nu\sigma}[h;\eta],
\label{Elin}
\eea
with $H^{\mu\rho\nu\sigma}[h;\5g]$ as in
(\ref{Hdef}) [one has
$32\pi\cH^{\mu\nu}[h;\eta]
=-\6_\rho\6_\sigma H^{\mu\rho\nu\sigma}$].
The exact isometries of the
flat background are given by the Killing vector fields
$\xi_\lambda=c_\lambda$ and $\xi_\lambda=x^\rho (c_{\rho\lambda}
-c_{\lambda\rho})$
where $c_\lambda$ and $c_{\lambda\rho}$ are constant
parameters. For these $\xi$'s, (\ref{gsuperpot3}) reads,
respectively,
\bea
&&\tilde k^{[\nu\mu]}_{\xi_\lambda=c_\lambda} 
[h;\eta]
=
\frac{1}{16\pi}\,c_\rho \6_\sigma H^{\rho\sigma\nu\mu}\ ,
\\
&&\tilde k^{[\nu\mu]}_{\xi_\lambda=x^\rho (c_{\rho\lambda}-c_{\lambda\rho})} 
[h;\eta]
=
\frac{1}{16\pi}\,c_{\rho\sigma}[x^\rho\6_\lambda H^{\sigma\lambda\nu\mu}
-x^\sigma \6_\lambda H^{\rho\lambda\nu\mu}
+H^{\rho\sigma\nu\mu}].
\eea
Analogously to the procedure in electrodynamics and
Yang-Mills theory, we define
the associated charges through derivatives with respect to the
parameters $c_\mu$ and $c_{\mu\nu}$
of the integrated $(n-2)$-form (\ref{gsuperpot2}).
Denoting these charges by $P^\mu$ and $M^{\mu\nu}$, we obtain
\bea
P^\mu&:=&\frac{\6}{\6c_\mu}
\int_{\6\Sigma} d\sigma_i\, \tilde k^{[0 i]}_{\xi_\mu=c_\mu}
[h;\eta]=
\frac{1}{16\pi}\int_{\6\Sigma} d\sigma_i\, \6_\lambda H^{\mu\lambda 0 i},
\label{P}\\
M^{\mu\nu}&:=&\frac{\6}{\6c_{\mu\nu}}
\int_{\6\Sigma} d\sigma_i\, 
\tilde k^{[0 i]}_{\xi_\mu=x^\nu (c_{\nu\mu}-c_{\mu\nu})}
[h;\eta]
\nonumber\\
&=&\frac{1}{16\pi}\int_{\6\Sigma} d\sigma_i\, 
[x^\mu \6_\lambda H^{\nu\lambda 0 i}-
x^\nu \6_\lambda H^{\mu\lambda 0 i}+H^{\mu\nu 0 i}].
\label{M}\eea
Owing to $H^{\mu\nu 0 i}=H^{\mu i 0\nu}-H^{\nu i 0\mu}$,
these are precisely the expressions derived in
chapter 20 of \cite{Misner:1970aa} (but note that they are not
restricted to four dimensions).
$P^0$ gives the ADM mass formula \cite{ADM}
\bea
P^0=\frac{1}{16\pi}\int_{\6\Sigma} d\sigma_i\,
\eta^{ik}\eta^{jl}(\6_jg_{kl}-\6_kg_{jl}).
\eea

Equation (\ref{gsuperpot3})
can also be used to establish the relation
to the Landau-Lifshitz expressions \cite{LL} for the total momentum and 
angular momentum in asymptotically flat spacetimes.
Let us denote by $H^{\mu\rho\nu\sigma}_\mathrm{LL}$ the
Landau-Lifshitz weight-2-tensor density,
\bea
H^{\mu\rho\nu\sigma}_\mathrm{LL}=
-g(g^{\mu\nu}g^{\rho\sigma}-g^{\rho\nu}g^{\mu\sigma}).
\label{HLL}\eea
One has
\[
\Big[d_VH^{\mu\rho\nu\sigma}_\mathrm{LL}
\Big]_{\eta,h}
=H^{\mu\rho\nu\sigma}.
\]
This implies that
\bea
k'_\xi[g]
=\frac{1}{16 \pi}\,(d^{n-2}x)_{\nu\mu}
\Big(\xi_\rho \6_\sigma H^{\rho\sigma\nu\mu}_\mathrm{LL}
+\frac 12 H^{\rho\sigma\nu\mu}_\mathrm{LL}
\6_\rho\xi_\sigma\Big)
\label{LLform}
\eea
is equivalent to 
(\ref{gsuperpot2}) in asymptotically flat spacetimes
because of
\[
\Big[d_Vk'_\xi[g]
\Big]_{\eta,h}
= k_\xi[h;\eta].
\]
Note, however, that (\ref{LLform}) does in general not
vanish when evaluated for $g_{\mu\nu}=\eta_{\mu\nu}$,
in contrast to (\ref{gsuperpot2}); rather, 
for 
$g_{\mu\nu}=\eta_{\mu\nu}$ it equals
$(1/16\pi)(d^{n-2}x)_{\mu\nu}\6^\mu\xi^\nu$.
To obtain from (\ref{LLform}) equivalent asymptotically conserved
$(n-2)$-forms that vanish for $g_{\mu\nu}=\eta_{\mu\nu}$, 
one may simply subtract
$(1/16\pi)(d^{n-2}x)_{\mu\nu}\6^\mu\xi^\nu$ from
(\ref{LLform}).

Integrated over $\6\Sigma$,
(\ref{LLform}) reproduces the expressions
for the total momentum and angular momentum in
\S 96 of \cite{LL}. These expressions arise
from (\ref{P}) and (\ref{M}) by substituting 
$H_\mathrm{LL}^{\mu\nu 0 i}$ for
$H^{\mu\nu 0 i}$ everywhere in the integrands.

Analogously (\ref{gsuperpot2}) yields
the charges for the asymptotic
isometries of flat spacetimes
found in \cite{Bondi:1962px,Sachs:1962wk,Sachs2}, when
the parameters
of these isometries satisfy equation (\ref{xicondition}).

\subsubsection{Asymptotically 3d anti-de Sitter spacetimes with
central charges}\label{3d-adS}

The formulas derived 
in sections \ref{gravsp} and \ref{gravcentralcharges}
are valid in the presence of a non vanishing cosmological constant. 
That is why they can be used to rediscuss,
from a covariant point of view,
asymptotically anti-de Sitter spacetimes in
3-dimensional gravity.
The original Hamiltonian analysis in \cite{Brown:1986nw}, in addition
to its considerable intrinsic interest, was to illustrate that non trivial 
central extensions may occur in the classical algebra of the
canonical generators. In the same spirit, this model serves here as an
example for the covariant theory of such central extensions proposed
in section \ref{s5}.

The background metric is represented in
coordinates $\{x^\mu\}=\{t,r,\winkel\}$ 
as in section 4 of \cite{Brown:1986nw} by
\bea
(ds^2)_\mathrm{background}=-\frac{r^2}{\ell^2}\,dt^2
+\frac{\ell^2}{r^2}\,dr^2+r^2d\winkel^2,
\label{adSg}
\eea
where $\ell$ is a constant and $\winkel$ has
periodicity $2\pi$.
The nonvanishing components of the background
Christoffel connection are
\bea
\5\Gamma_{tt}{}^r=\frac{r^3}{\ell^4}\ ,\ 
\5\Gamma_{rr}{}^r=-\frac{1}{r}\ ,\ 
\5\Gamma_{\winkel\winkel}{}^r=-\frac{r^3}{\ell^2}\ ,\
\5\Gamma_{tr}{}^t=
\5\Gamma_{rt}{}^t=
\5\Gamma_{\winkel r}{}^\winkel=
\5\Gamma_{r\winkel}{}^\winkel=
\frac{1}{r}\ . 
\label{adSGamma}\eea
The background Ricci tensor and the cosmological
constant are
\bea
\5R_{\mu\nu}=2\Lambda\,\5g_{\mu\nu}\quad ,\quad
\Lambda=-1/\ell^2.
\label{adSRicci}\eea

As in section 4 of \cite{Brown:1986nw}, 
we study spacetimes which are
asymptotically anti-de Sitter in the sense
that the metric is $g_{\mu\nu}=\5g_{\mu\nu}+h_{\mu\nu}$
with boundary conditions
\bea
&h_{tt}\asy O(1),\quad
h_{rr}\asy O(r^{-4}),\quad
h_{\winkel\winkel}\asy O(1),&
\nonumber\\
&h_{tr}\asy O(r^{-3}),\quad 
h_{t\winkel}\asy O(1),\quad
h_{r\winkel}\asy O(r^{-3}).&
\label{adSbc}
\eea
The asymptotic behaviour is determined only by the dependence
on $r$ because the boundary conditions are imposed at
$r\asy \infty$. 
Using Eq.\ (\ref{Hmunu}), one obtains that
the boundary conditions (\ref{adSbc}) imply
\bea
&\cH^{tt}\asy O(r^{-3}),\quad
\cH^{rr}\asy O(r),\quad
\cH^{\winkel\winkel}\asy O(r^{-3}),&
\nonumber\\
&\cH^{tr}\asy O(r^{-2}),\quad 
\cH^{t\winkel}\asy O(r^{-3}),\quad
\cH^{r\winkel}\asy O(r^{-2}),&
\eea
where we restricted the
space of allowed functions to those which
satisfy $h_{\mu\nu}\asy O(r^m)\Rightarrow \6_r h_{\mu\nu}\asy O(r^{m-1})$
(and analogously for the derivatives of $h_{\mu\nu}$).
In particular we thus exclude oscillating functions
in the coordinate $r$, such as $r^m\sin(r)$.
As we must assign $O(r)$ to $d^3x$,
(\ref{xicondition}) imposes in this case
\bea
& 
\5D_t \xi_t \asy o(r^2),\quad
\5D_r \xi_r \asy o(r^{-2}),\quad
\5D_\winkel \xi_\winkel \asy o(r^2),
&
\nonumber\\
&
\5D_t \xi_r+\5D_r \xi_t \asy o(r),\quad
\5D_t \xi_\winkel+\5D_\winkel \xi_t \asy o(r^2),\quad
\5D_r \xi_\winkel+\5D_\winkel \xi_r \asy o(r).
&
\label{newAdScond}
\eea
The functions $\chi_\alpha$ in equation (\ref{puregauge})
are in this case $\chi_t=\chi_\winkel=1$, $\chi_r=1/r$.
Hence, trivial solutions to (\ref{newAdScond}) are:
\bea
\xi^\mu\sim 0\ \Longleftrightarrow\ 
\xi^t\asy 0,\ \xi^r\asy o(r),\ \xi^\winkel\asy 0.
\label{adS_3puregauge}
\eea
The general solution to the conditions (\ref{newAdScond}) 
in the space of functions satisfying
$\xi^\mu\asy O(r^m)\Rightarrow \6_r \xi^\mu\asy O(r^{m-1})$ is
\bea
\xi^t&\asy &\ell\, T(t,\winkel),
\nonumber\\
\xi^r&\asy &-r\6_\winkel\Phi(t,\winkel)
+o(r),
\nonumber\\
\xi^\winkel&\asy &\Phi(t,\winkel),
\label{adSxinew}
\eea
where $T(t,\winkel)$ and $\Phi(t,\winkel)$ are
functions of $t$ and $\winkel$ which are
$2\pi$-periodic in $\winkel$ and satisfy
\bea
\ell \6_t T(t,\winkel)=\6_\winkel \Phi(t,\winkel),\quad
\ell \6_t \Phi(t,\winkel)=\6_\winkel T(t,\winkel).
\label{adSxi2}
\eea
The general solution of these equations are
functions $T(t,\winkel)$ and $\Phi(t,\winkel)$ which are
superpositions of modes $f(nt/\ell)g(n\winkel)$
with $f,g\in\{\sin,\cos\}$, $n\in\mathbb Z$, see
\cite{Brown:1986nw} for details. We note that 
(\ref{adSxinew}) agrees to leading order
with the asymptotic Killing vector fields determined
in \cite{Brown:1986nw} from the conditions
$\cL_\xi\, \5g_{\mu\nu}\asy O(\chi_{\mu\nu})$. The 
latter conditions are stronger than (\ref{newAdScond})
and impose also constraints on
contributions to the $\xi$'s at subleading order (see remark
at the end of this section).
However, contributions to the $\xi$'s of subleading order
do not contribute to the charges obtained 
from (\ref{gsuperpot2}) because they are trivial, see (\ref{adS_3puregauge}). 
Furthermore, condition (\ref{convergencec}) is satisfied 
in this case and guarantees 
that the charges corresponding to (\ref{adSxinew})
are finite.
We choose $\6\Sigma$ the 
circle of radius $r$ for $r\rightarrow\infty$
(so that $dr=0$, $dt=0$ on $\6\Sigma$). The 
conserved charges are then
\bea
Q_\xi=
\frac 12\int_{\6\Sigma}dx^\rho\epsilon_{\mu\nu\rho}\,\4k^{[\mu\nu]}_\xi
[h,\5g]
=\lim_{r\rightarrow\infty}\int_0^{2\pi}d\winkel\,\4k^{[tr]}_\xi
[h,\5g],
\label{adSQ}
\eea
where $\4k^{[tr]}_\xi[h,\5g]$ 
is the $[tr]$-component of the
superpotential (\ref{gsuperpot0}) evaluated for
$h_{\mu\nu}=g_{\mu\nu}-\5g_{\mu\nu}(x)$
with the background metric (\ref{adSg}).
Explicitly one obtains
\bea
16\pi\4k^{[tr]}_\xi[h,\5g]
\asy-\xi^t(
\frac{r^4}{\ell^4}\,h_{rr}
+\frac{2}{\ell^2}\,h_{\winkel\winkel}
-\frac {r}{\ell^2}\,\6_r h_{\winkel\winkel}
)
-\xi^\winkel(2h_{t\winkel}-r\6_r h_{t\winkel}).
\label{ktr}
\eea
Notice that, indeed, the charges are finite and only the 
leading order terms
in (\ref{adSxinew}) contribute to them.
Equations (\ref{adSQ}) and (\ref{ktr}) may now be used to
compute explicitly the values of the charges for a given metric satisfying
the boundary conditions (\ref{adSbc}).
For example, let us consider the metric given in equation (4.2)
of \cite{Brown:1986nw}:
\[
ds^2=-(\frac{r^2}{\ell^2}+\alpha^2)\,dt^2
+2\alpha A\, dtd\winkel
+(\frac{r^2-A^2}{\ell^2}+\alpha^2)^{-1}dr^2+(r^2-A^2)\,d\winkel^2,
\]
where $A$ and $\alpha$ are constant parameters.
Evaluating (\ref{adSQ}) for $\xi=(\ell,0,0)$ (i.e.,
$T=1$, $\Phi=0$) and for $\xi=(0,0,-1)$ (i.e.,
$T=0$, $\Phi=-1$), respectively, one obtains
\beann
16\pi Q_{(\ell,0,0)}&=&
   2\pi\ell(\alpha^2+A^2/\ell^2),
\\
16\pi Q_{(0,0,-1)}&=&
   4\pi \alpha A,
\eeann
in agreement with
Eq.\ (4.12) of \cite{Brown:1986nw} (modulo conventions).

Let us finally discuss the algebra of the charges.
As we have pointed out, 
the existence of a well-defined algebra
generally may impose additional conditions on the asymptotic reducibility
parameters. Conditions which are sufficient 
for the existence of the algebra when (\ref{convergencec}) holds,
are given in
equations (\ref{constr1})--(\ref{newass}), 
(\ref{newass2})--(\ref{newass4}) and 
(\ref{BHboundcond}). In the present case, it turns out
that actually one only needs (\ref{newass2}) in order to get
a well-defined algebra; (\ref{newass2}) reads in this case
$\cL_\xi h_{\mu\nu}\asy O(\chi_{\mu\nu})$ and imposes
\bea
\xi^t&\asy &\ell\, T(t,\winkel)+O(r^{-2}),
\nonumber\\
\xi^r&\asy &-r\6_\winkel\Phi(t,\winkel)
+o(r),
\nonumber\\
\xi^\winkel&\asy &\Phi(t,\winkel)+O(r^{-2}),
\label{enough}
\eea
where the functions 
$T(t,\winkel)$ and $\Phi(t,\winkel)$ 
are still only subject to (\ref{adSxi2}).
(\ref{enough}) especially implies the existence 
(finiteness) of the central charges (\ref{gcentralcharge});
one obtains
\bea
K_{\xi_1,\xi_2} &=&\frac{1}{16\pi}
\lim_{r\rightarrow\infty}\int_0^{2\pi}d\winkel\,
\frac 2r\,(\6_\winkel\xi^r_1\,\6_\winkel\xi^t_2-
\6_\winkel\xi^r_2\,\6_\winkel\xi^t_1)
\nonumber\\
&=&\frac{2\ell}{16\pi}
\int_0^{2\pi}d\winkel\Big[
\6_\winkel T_1(t,\winkel)\,\6_\winkel^2\Phi_2(t,\winkel)
-\6_\winkel T_2(t,\winkel)\,\6_\winkel^2\Phi_1(t,\winkel)
\Big],
\label{BHcc}
\eea
which is the covariant expression for the central charge 
derived previously by different means in 
\cite{Terashima:2001gb}, equation (13).

Using a mode expansion of $\Phi(t,\winkel)$ and
$T(t,\winkel)$ as in \cite{Brown:1986nw},
it can be explicitly verified 
that the Poisson algebra (\ref{asympcom}) of the conserved charges
for parameters (\ref{enough})
coincides with the algebra of canonical generators
found in \cite{Brown:1986nw}.
As shown there, this algebra can be written as
the direct sum of two copies of the
Virasoro algebra.
\medskip

\noindent
{\bf Remarks:} 
\begin{itemize}
\item The final expression (\ref{BHcc}) for the central charges
involves solely the leading order terms in (\ref{enough})
which agree with those in (\ref{adSxinew}).
Nevertheless, (\ref{enough}) was used in the
computation, as we
dropped terms which vanish for $r\asy\infty$
on account of (\ref{enough}), but which
would in general diverge in this limit for parameters that satisfy
only the boundary conditions 
(\ref{adSxinew}). These terms are
\beann
\frac{1}{16\pi}
\lim_{r\rightarrow\infty}\int_0^{2\pi}d\winkel\,
\frac {r^3}{\ell^2}\Big[
\6_r\xi^t_1(\6_\winkel\xi_2^\winkel+\frac 1r\xi_2^r)
+\frac 12(\6_\winkel\xi^t_1-\ell^2\6_t\xi^\winkel_1)
\6_r\xi^\winkel_2-(1\leftrightarrow 2)
\Big].
\eeann
\item (\ref{enough}) is a weaker condition than 
(\ref{BHboundcond}) applied in the present case, as the latter 
imposes 
\bea
& 
\5D_t \xi_t \asy O(1),\
\5D_r \xi_r \asy O(r^{-4}),\
\5D_\winkel \xi_\winkel \asy O(1),
&
\nonumber\\
&
\5D_{(t} \xi_{r)} \asy O(r^{-3}),\
\5D_{(t} \xi_{\winkel)} \asy O(1),\
\5D_{(r} \xi_{\winkel)} \asy O(r^{-3}).
&
\label{oldAdScond}
\eea
The general solution of these conditions in the same space of functions
as above is
\bea
\xi^t&\asy &\ell\, T(t,\winkel)
+\frac{\ell^3}{2r^2}\,
\6^2_\winkel T(t,\winkel)
+O(1/r^4),
\nonumber\\
\xi^r&\asy &-r\6_\winkel\Phi(t,\winkel)
+O(1/r),
\nonumber\\
\xi^\winkel&\asy &\Phi(t,\winkel)
-\frac{\ell^2}{2r^2}\,\6^2_\winkel\Phi(t,\winkel)
+O(1/r^4).
\label{oldadSxi}
\eea
(\ref{oldAdScond}) are the conditions imposed
in \cite{Brown:1986nw}. The fact that (\ref{enough}) leads to the
same conclusions shows that these conditions can be relaxed.
This demonstrates that (\ref{BHboundcond}) is only  
a sufficient but not a necessary condition for finiteness of the
central charges as given in (\ref{centralcharge1}).
\end{itemize}

\mysection{Cohomological approach}\label{s7}

\subsection{Antifield BRST formalism}

\subsubsection{Koszul-Tate resolution}

The cohomological set-up used so far was given by the free variational
bicomplex (see e.g. \cite{Olver:1993,Anderson1991,Dickey:1991xa}),
i.e., horizontal and vertical form valued local functions with
horizontal differential $d_H=dx^\mu\partial_\mu$ and vertical
differential $d_V=d_V\phi^i_{(\mu)}\partial^S/\partial\phi^i_{(\mu)}$
and by the variational bicomplex pulled back to the surface defined by the
Euler-Lagrange equations of motion (and their total derivatives).

In the absence of vertical generators, 
one constructs a homological resolution of the horizontal complex
associated with the equations of motion by introducing ``antifields''
\cite{Zinn-Justin:1974mc,Batalin:1981jr}: 
for irreducible gauge theories, the
antifields are given by Grassmann odd generators $\phi^*_{i(\mu)}$
of antifield number $1$ and Grassmann even generators
$C^*_{\alpha(\mu)}$ of antifield number $2$. The so-called
Koszul-Tate differential 
\cite{Fisch:1990rp,Henneaux:1991rx} is defined by
\bea
\delta=\partial_{(\mu)}\frac{\delta L}{\delta \phi^i}
\frac{\partial^S}{\partial \phi^*_{i (\mu)}}+
\partial_{(\mu)}[R^{+i}_\alpha(\phi^*_i)]\frac{\partial^S}{\partial
  C^*_{\alpha (\mu)}}.
\eea
The cohomology of $\delta$ can then be shown to be trivial in the
space of horizontal forms
in the fields and the
antifields with strictly positive antifield number, $H_k(\delta)=0$
for $k\geq 1$, while $H_0(\delta)$ is given by equivalence classes of
forms $[\omega_0]$ in the original fields alone,
where two such forms have to be identified if they agree when
evaluated on every solution of the
equations of motion, $\omega_0\sim\omega_0^\prime$ if
$\omega_0-\omega_0^\prime\approx 0$.

\subsubsection{Antibracket, master equation and BRST differential}

In many problems involving the gauge symmetries of a classical
Lagrangian, and in particular for the discussion of the Lie algebra
associated with global reducibility identities, 
it is most convenient to 
extend the Koszul-Tate differential to the full BRST differential
of the antifield formalism. This differential is canonically generated
in the antibracket by the
so-called minimal solution of the master equation
\cite{Zinn-Justin:1974mc,Zinn-Justin:1989mi,%
Batalin:1981jr,Batalin:1983wj,Batalin:1983jr,Batalin:1984ss,%
Batalin:1985qj,Fisch:1990rp,Henneaux:1991rx} (for reviews, see 
e.g. \cite{Henneaux:1992ig,Gomis:1995he}).
This formulation is crucial
for the quantum theory, because canonical transformations in the
antibracket are used on the one hand to fix the gauge while
retaining the original
gauge invariance in the form of the gauge fixed BRST invariance, and,
on the other hand, to absorb trivial, BRST exact divergences. At
the classical level, a great advantage of the formalism is for
instance that different choices of field
parametrizations or of generating sets of gauge transformations are
again related by 
canonical transformations. 

The full antifield BRST formalism involves as additional fields not
only the antifields $\phi^*_i$ and $C^*_\alpha$, but also the ghosts 
$C^\alpha$ and their derivatives. They can be
understood as Grassmann odd gauge parameters. 
There is a well defined graded Lie bracket in $H^n(d_H)$, which is
induced by the local antibracket. 
With $\{\phi^A\}=\{\phi^i,C^\alpha\}$,
$\{\phi^*_A\}=\{\phi^*_i,C^*_\alpha\}$,
it is defined in terms of Euler-Lagrange 
left and right derivatives (indicated by
superscripts $L$ and $R$, respectively) through 
\bea
(f d^nx,g d^nx)=
\Big[\vddr{f}{\phi^A}\,\vddl{g}{\phi^*_A}-
\vddr{f}{\phi^*_A}\,\vddl{g}{\phi^A}\Big] d^nx.
\eea

To each equivalence class $[f d^nx]\in H^n(d_H)$, one can associate
a ``Hamiltonian'' vector field defined by 
\bea
\delta_{f d^n x}= 
\partial_{(\mu)}\vddr{f}{\phi^A}\,
\frac{{\partial^S}}{\partial\phi^*_{A(\mu)}}-
\partial_{(\mu)}\vddr{f}{\phi^*_A}\,
\frac{{\partial^S}}{\partial\phi^A_{(\mu)}}.
\eea
[The operators ${\partial^S}/\partial(\dots)$ are left derivatives.]
If $[\cdot,\cdot]$ denotes the graded commutator of vector fields, 
these vector fields satisfy
\bea
[\delta_{fd^nx},\partial_\mu]&=&0,\cr
[\delta_{fd^nx},d_H]&=&0,\cr
[\delta_{fd^nx},\delta_{gd^nx}]&=&\delta_{(fd^nx,gd^nx)}\label{vfcom}.
\eea
An algebraic proof of the last identity can be found for instance in
\cite{Barnich:1996mr}. 

The classical Lagrangian $L$ is extended to the Lagrangian $L_M$ that
solves the (classical) master equation 
\bea
(L_M\ d^nx,L_M\ d^nx)=d_H(\ ).
\eea
In an expansion according to the antifield number,
the Lagrangian $L_M$ reads 
\bea
L_M= L+\phi^*_i R^i_\alpha(C^\alpha)+\half C^*_\gamma
C^\gamma_{\alpha\beta}(C^\alpha,C^\beta)+\frac{1}{4}
M^{ij}_{\alpha\beta}(\phi^*_i,\phi^*_j,C^\alpha,C^\beta)\nonumber\\
+\mbox{terms
  of antifield number $\geq 3$},
\eea
where the structure operators $M^{ij}_{\alpha\beta}$ describe the
weakly vanishing terms in the commutators of gauge transformations. 
The full BRST differential involving the antifields is generated from
the solution of the master equation according to 
\bea
s=\delta_{L_Md^nx}.
\eea
In an expansion according to the antifield number, $s=\delta+\gamma
+s_1+ \dots$, it starts at
antifield number $-1$ with the Koszul-Tate differential $\delta$. The
component $\gamma$ at antifield number $0$ is the 
so-called
longitudinal differential 
along the gauge orbits \cite{Henneaux:1992ig}. 

The cohomology groups $H_k^n(\delta|d_H)$ can be shown
\cite{Barnich:1995db,Barnich:2000zw} to be 
isomorphic to the local BRST cohomological groups 
$H^{-k,n}(s|d_H)$, where the first superscript
denotes the ghost number obtained by assigning $1$ to
$C^\alpha$, 0 to $\phi^i,\6_\mu,x^\mu,dx^\mu$, 
$-1$ to $\phi^*_i$, and $-2$ to $C^*_\alpha$,
\bea
H_k^n(\delta|d_H)\simeq H^{-k,n}(s|d_H)\quad\mbox{for}\quad k>0.
\eea 
Owing to the properties of $H_k^n(\delta|d_H)$ summarized below,
this implies that
equivalence classes of 
global symmetries and of reducibility parameters, and at the
same time the associated characteristic cohomology, can be
described as local BRST cohomology classes. Explicitly,
representatives of $H_k^n(\delta|d_H)$ are completed by terms of
higher antifield number containing the ghosts $C^\alpha$ and their
derivatives to representatives of $H^{-k,n}(s|d_H)$, 
while conversely, representatives of $H^{-k,n}(s|d_H)$ determine
representatives of $H_k^n(\delta|d_H)$ by setting to zero the ghosts
$C^\alpha$ and their derivatives.  

One of the advantages of this description is that the 
antibracket map induces a graded Lie bracket (with grading +1) 
in local BRST cohomology, 
\bea
&(\cdot,\cdot)_M: H^{g_1,n}(s|d_H)\otimes
H^{g_2,n}(s|d_H)\longrightarrow H^{g_1+g_2+1,n}(s|d_H),&\nonumber\cr
&([\omega^{g_1,n}],[\eta^{g_2,n}])_M=[(\omega^{g_1,n},\eta^{g_2,n})].&
\eea
Alternative equivalent expressions for the antibracket map are%
\footnote{For simplicity we assume throughout this paper that
all fields $\phi^i$ are bosonic (Grassmann even). Then the Grassmann
parity of a local function of the fields, ghosts and antifields
equals its ghost number (modulo 2). This allows us to write
Grassmann parity dependent signs as in (\ref{nummer}) in terms of
ghost numbers.}
\bea
([\omega^{g_1,n}],[\eta^{g_2,n}])_M=[\delta_{\omega^{g_1,n}}\eta^{g_2,n}]
=-(-)^{(g_1+1)(g_2+1)}[\delta_{\eta^{g_2,n}}\omega^{g_1,n}].
\label{nummer}
\eea

This map will be used below to describe the Lie algebra of equivalence
classes of global symmetries, the Lie action of equivalence
classes of global symmetries on equivalence classes of reducibility
parameters, the global symmetries induced from reducibility parameters
and the Lie algebra of equivalence classes of reducibility
parameters.

\subsection{Global symmetries and conserved currents}

The equivalence classes of global symmetries can be identified with the
cohomology classes of the group $H_1^n(\delta|d_H)$, which admit
canonical representatives of the form $\omega_1^n=\phi^*_iX^id^nx$, 
while equivalence classes of conserved currents correspond
to cohomology classes of the group $H_0^{n-1}(d_H|\delta)$
with
representatives $\omega_0^{n-1}=j^\mu \,(d^{n-1}x)_\mu$. In this set-up, 
Noether's first theorem, in its complete(d) formulation
as in section
\ref{2.6}, is precisely the cohomological relation
\bea
H_1^n(\delta|d_H)\simeq H_0^{n-1}(d_H|\delta)/\delta^{n-1}_0 {\mathbb
  R},
\label{NOETHER}
\eea
which is a rather direct consequence of the 
properties of the cohomology of $\delta$ 
and the
fact that the cohomology of the horizontal differential
in the space of horizontal form valued local functions in the fields
and antifields is given (locally)
by $H^k(d_H)=\delta^k_0{\mathbb R}$ for $k\leq
n-1$ (algebraic Poincar\'e lemma).

For completeness, let us note that because of (\ref{3}),
$H^n(d_H)$ is given by the equivalence classes $[\omega^n]$ of $n$-forms
having the same Euler-Lagrange derivatives with respect to the fields
and the antifields,
$\omega^n\sim \tilde \omega^n$ iff $\frac{\delta}{\delta Z^A}
(\omega^n-\tilde \omega^n)=0$, for all
$Z^A\in\{\phi^i,\phi^*_i,C^*_\alpha\}$.

We also note that Noether's first theorem holds in exactly the same
form for reducible gauge theories, although one needs to introduce
additional antifields of antifield number higher than $2$  and extend
the definition of $\delta$ on these additional antifields in such a
way that the cohomology of $\delta$ remains trivial in strictly
positive antifield number and unchanged in antifield number $0$.

\subsection{Reducibility parameters and conserved n--2 forms}\label{s73}

\subsubsection{Characteristic cohomology and $H^n(\delta|d_H)$.}

The cohomology group $H_0^{n-k}(d_H|\delta)$ is also called the
characteristic cohomology
in form degree $n-k$ and is represented by conserved
$(n-k)$ forms. The characteristic cohomology
is the cohomology of the horizontal complex
associated to the (Euler-Lagrange) equations of motion \cite{Vinogradov:1978,%
Tsujishita:1982,Vinogradov:1984,Anderson1991,Bryant:1995}.
Its representatives are local forms which are $d_H$-closed
on-shell modulo local forms which are $d_H$-exact on-shell;
in other words: the representatives are conserved local forms.
Using the cohomology of $d_H$ and of
$\delta$, one can prove 
\cite{Barnich:1995db,Barnich:2000zw}:
\bea
H_0^{n-k}(d_H|\delta)/\delta^n_k{\mathbb R}\simeq
H_{k}^{n}(\delta|d_H)\quad\mbox{for}\quad 1\leq k\leq n-1.
\label{NOETHERBBH}
\eea
Note that this generalizes (\ref{NOETHER}) to $k\geq 1$
and might therefore be regarded as a generalization of
Noethers first theorem.
For $k=2$, it encodes the bijective correspondence
between conserved $(n-2)$ forms and global reducibility
identities as we shall explain in more detail below.
For irreducible gauge theories, one can show under fairly general
assumptions (linearizable, normal) \cite{Barnich:1995db,Barnich:2000zw} 
that there
is no characteristic
cohomology in form degree strictly smaller than $n-2$
except for the constant 0-forms, i.e., that
$H_0^{n-k}(d_H|\delta)=\delta^n_k{\mathbb R}$ and
$H_{k}^{n}(\delta|d_H)=0$ for $k>2$. More generally,
one can show for reducible
(linearizable, normal) gauge
theories of reducibility order $r$ ($r=-1$ for
models without nontrivial gauge symmetry, $r=0$ for
irreducible gauge theories, etc):
$H_0^{n-k}(d_H|\delta)=\delta^n_k{\mathbb R}$ and
$H_{k}^{n}(\delta|d_H)=0$ for $k>r+2$ \cite{Barnich:1995db}.

\subsubsection{Descent equations}

Let us now explain in more detail that and how
the isomorphism (\ref{NOETHERBBH}) yields the
bijective correspondence
between conserved $(n-2)$ forms and global reducibility
identities. The isomorphism (\ref{NOETHERBBH})
is based on so-called descent equations for $\delta$ and $d_H$.
For $k=2$, these descent equations
relate, in intermediate steps, $H_0^{n-2}(d_H|\delta)$
to $H_1^{n-1}(\delta|d_H)$ and $H_1^{n-1}(\delta|d_H)$
to $H_{2}^{n}(\delta|d_H)$,
\bea
H_0^{n-2}(d_H|\delta)/\delta^{n-2}_0{\mathbb R}\simeq
H_1^{n-1}(\delta|d_H)\simeq
H_{2}^{n}(\delta|d_H).\label{7.4}
\eea
[Analogous intermediate steps are behind
(\ref{NOETHERBBH}) for $k>2$ \cite{Barnich:1995db,Barnich:2000zw}.]

$H_0^{n-2}(d_H|\delta)/\delta^{n-2}_0{\mathbb R}\simeq
H_1^{n-1}(\delta|d_H)$
is explicitly given by associating to any
class $[\omega_0^{n-2}]\in H_0^{n-2}(d_H|\delta)$ (except for the
constants in $2$ spacetime dimensions),
a class
$[\omega_1^{n-1}]\in H_1^{n-1}(\delta|d_H)$.
$H_1^{n-1}(\delta|d_H)\simeq
H_{2}^{n}(\delta|d_H)$ is explicitly given by associating to any
class $[\omega_1^{n-1}]\in H_1^{n-1}(\delta|d_H)$
a class $[\omega_2^n]\in H_{2}^{n}(\delta|d_H)$.
The representatives satisfy the chain of descent equations
\bea
\delta\omega_2^n+d_H\omega_1^{n-1}=0,\label{9}\\
\delta\omega_1^{n-1}+d_H\omega_0^{n-2}=0.\label{10}
\eea

Under the same general assumptions as above, one can show
\cite{Barnich:1995db,Barnich:2000zw} that
$H_{2}^{n}(\delta|d_H)$ is isomorphic to the space of equivalence classes
of global reducibility identities up to trivial ones.

For a collection of
functions $f^\alpha$  and $M^{[j(\nu)i(\mu)]}$ that satisfy (\ref{7}),
the forms
$\omega^n_2,\omega^{n-1}_1,\omega^{n-2}_0$ that satisfy
the descent equations (\ref{9}) and (\ref{10}) can be constructed as
follows. 
The form $\omega^n_2$ is given by
\bea
\omega^n_2=[f^\alpha
C^*_\alpha-\frac{1}{2}\phi^*_{j(\nu)}\phi^*_{i(\mu)}
M^{[j(\nu)i(\mu)]}]d^nx.\label{o2}
\eea
The form 
\bea
\omega^{n-1}_1=-[S_\alpha^{\mu i}(\phi^*_i,f^\alpha)+
M^{\mu ji}(\frac{\delta L}{\delta \phi^j},\phi^*_i)](d^{n-1}x)_\mu,
\eea
is a particular solution to the first of the descent equations
(\ref{9}) because of (\ref{7}).
Finally, 
a particular solution to the
second of the descent equations (\ref{10}) is given by
\bea
\omega^{n-2}_0=-k^{[\mu\nu]}_{f}
(d^{n-2}x)_{\mu\nu}\ ,
\eea
with $k^{[\mu\nu]}_{f}$ given by \eqref{2.31}. 

The advantage of this cohomological formulation is that the
ambiguities in the solutions of the descent equations 
are automatically taken care of by
the triviality of the
cohomology of $d_H$ and $\delta$ in the appropriate degrees: 
the
various forms are all defined only up to the addition of $d_H$ and
$\delta$ exact terms. This leads to the isomorphisms \eqref{7.4},
which states the bijective correspondence between the equivalence
classes of reducibility parameters, conserved $n-2$ forms (up to the
constant form in $n=2$) and operator currents satisfying \eqref{cond1}
or \eqref{cond2}. 

\subsubsection{Lie algebra and action of global symetries 
from antibracket map}\label{s733}

In ghost number $g=-1$, the antibracket map 
describes the Lie algebra of equivalence classes of global symmetries,
up to a shift in the grading, and an overall minus sign.
Indeed, if the cocycle 
\bea
\omega^{-1,n}_X=(\phi^*_i X^i-C^*_\alpha
X^\alpha_\beta(C^\beta)+\dots )d^nx 
\eea
describes the global symmetry
with characteristic $X^i$, we have 
\bea
([\omega^{-1,n}_{X_1}],[\omega^{-1,n}_{X_2}])_M
=[\omega^{-1,n}_{[X_2,X_1]_L}].
\eea
Furthermore, due to the graded Jacobi identity for $(\cdot,\cdot)_M$,
\bea
([\omega^{-1,n}_{X_1}],([\omega^{-1,n}_{X_2}],[\omega^{*,n}])_M)_M
-([\omega^{-1,n}_{X_2}],([\omega^{-1,n}_{X_1}],[\omega^{*,n}])_M)_M
\nonumber\\
=([\omega^{-1,n}_{[X_2,X_1]_L}],[\omega^{*,n}])_M,
\eea
there is a well defined Lie action of equivalence 
classes of global symmetries
on local BRST cohomology classes. In particular, if the cocycle
\bea
 \omega^{-2,n}_f=(C^*_\alpha f^\alpha-
\frac{1}{2}\phi^*_{j(\nu)}\phi^*_{i(\mu)}M^{[j(\nu)i(\mu)]}+C^*_\alpha
k^{\alpha i}_\beta(\phi^*_i,C^\beta)+\dots )d^nx, 
\eea
describes the reducibility parameters $f^\alpha$, we can 
choose 
\bea
([\omega^{-1,n}_{X}],[\omega^{-2,n}_{f}])_M
=[\omega^{-2,n}_{-(X,f)}],
\eea
with $(X,f)^\alpha=\delta_X f^\alpha+X^\alpha_\beta(f^\beta)$, in
agreement with \eqref{eq3.39}. 

Owing to
\bea
H^{n-2}_{\rm char}/\delta^n_2{\mathbb R}\simeq H^n_2(\delta|d_H)\simeq
H^{-2,n}(s|d_H),
\eea
there is an isomorphic Lie action of equivalence classes of global
symmetries on equivalence classes of conserved $n-2$ forms. 

The proof that this Lie action is given by \eqref{eq3.40} proceeds as
follows. The vector field $\delta_{\omega^{-1,n}_{X}}$ anticommutes not
  only with $d_H$ but also with $s$. Indeed, \eqref{vfcom} implies
  that $[\delta_{\omega^{*,n}},s]=0$ for every $s$
  modulo $d_H$ cocycle $\omega^{*,n}$.  Hence, by appying
  $\delta_{\omega^{-1,n}_{X}}$ to the descent equations 
\bea
s\omega^{-2,n}_f+d_H\omega^{-1,n-1}_f=0,\cr
s\omega^{-1,n-1}_f+d_H\omega^{0,n-2}_f=0,
\eea
one can move $\delta_{\omega^{-1,n}_{X}}$ past $s$ and $d_H$. The
  result then follows from the fact that 
$[\delta_{\omega^{-1,n}_{X}}\omega^{-2,n}_f]=
([\omega^{-1,n}_{X}],[\omega^{-2,n}_f])_M$
and that 
$[\delta_{\omega^{-1,n}_{X}}\omega^{0,n-2}_f]=[-\delta_X
k^{\mu\nu}(d^{n-2}x)_{\mu\nu}+\dots]$. 

Let us also show by cohomological means that 
the bracket $[\cdot,\cdot]_P$ induced by
\eqref{eq3.35} in the space of equivalence classes of reducibility
parameters is trivial. Indeed, consider the trivial global symmetry
$X^i=R^i_\alpha(f^\alpha_1)$ for an arbitrary local function
$f^\alpha_1$ described by the cocycle 
\bea
\omega^{-1,n}_{R_{f_1}}&=&s (C^*_\alpha f^\alpha_1 d^nx) + d_H(\ )\nonumber\\
&=&[\phi^*_i
R^i_\alpha(f^\alpha_1)+C^*_\alpha\partial_{(\mu)}R^i_\alpha(C^\alpha)
\frac{\partial^S f^\alpha_1}{\partial\phi^i_{(\mu)}} +C^*_\alpha
C^\alpha_{\beta\gamma}(C^\beta,f^\gamma_1)+\dots]d^nx . 
\eea
For a given cocycle $\omega^{-2,n}_{f_2}$, we have
\bea
([\omega^{-1,n}_{R_{f_1}}],[\omega^{-2,n}_{f_2}])_M=[0]=
[\omega^{-2,n}_{-(R_{f_1},f_2)}], 
\eea
which implies the triviality of the resulting reducibility parameters,
\bea
(R_{f_1},f_2)^\alpha=\delta_{f_1} f^\alpha_2 -\delta_{f_2}f^\alpha_1+
C^\alpha_{\beta\gamma}(f^\beta_1,f^\gamma_2)\approx 0,
\eea
and also the triviality of the conserved $n-2$ form obtained by applying
$\delta_{f_1}$ to a given conserved $n-2$ form. In particular, if
$f^\alpha_1$ are reducibility parameters, 
we get 
\bea
C^\alpha_{\beta\gamma}(f^\beta_1,f^\gamma_2)\approx 0,
\eea
which proves the triviality of the bracket induced by
$[\cdot,\cdot]_P$ among equivalence classes of reducibility
parameters. 

\subsection{Induced symmetries and associated algebra}

\subsubsection{Global symmetries out of reducibility parameters}

Let $[\omega^{0,n}]\in H^{0,n}(s|d_H)$. 
Because $(\cdot,\cdot)_M:
H^{0,n}(s|d)\otimes H^{-2,n}(s|d)\longrightarrow H^{-1,n}(s|d_H)$, the
antibracket map with a given $[\omega^{0,n}]$ provides a way to induce a
possibly non trivial global symmetry from a set of reducibility
parameters,
\bea
([\omega^{0,n}],[\omega^{-2,n}_f])_M=[\omega^{-1,n}_f].\label{indu}
\eea

Because $(\cdot,\cdot)_M:
H^{-1,n}(s|d)\otimes H^{-2,n}(s|d)\longrightarrow H^{-2,n}(s|d_H)$,
there is an action of global symmetries, and in particular of induced
global symmetries,  on reducibility parameters.
Using the graded Jacobi identity for the antibracket $(\cdot,\cdot)$,
it follows that 
\bea
(\omega^{-1,n}_{f_1},\omega^{-2,n}_{f_2})=
-(\omega^{-1,n}_{f_2},\omega^{-2,n}_{f_1})+(\omega^{0,n},
(\omega^{-2,n}_{f_1},\omega^{-2,n}_{f_2}))+d_H(\ ).
\eea
By assumption, we are dealing with irreducible gauge theories which do not 
admit non trivial characteristic cohomology in 
ghost number $-3$, 
\bea
H^{n-3}_{\rm char}/\delta^n_3{\mathbb R}\simeq H^n_3(\delta|d_H)\simeq
H^{-3,n}(s|d_H)\simeq 0,
\eea
so that the $s$ modulo $d_H$ cocycle 
$(\omega^{-2,n}_{f_1},\omega^{-2,n}_{f_2})$
is trivial, $(\omega^{-2,n}_{f_1},\omega^{-2,n}_{f_2})=s(\ )+ d_H(\
)$. It follows that the
action of induced global symmetries on reducibility parameters is
skew-symmetric (modulo trivial terms), 
\bea
([\omega^{-1,n}_{f_1}],[\omega^{-2,n}_{f_2}])_M=
-([\omega^{-1,n}_{f_2}],[\omega^{-2,n}_{f_1}])_M.\label{skew}
\eea
Explicitly, for 
\bea
\omega^{0,n}=(v_0+\phi^*_i
v^{i}_\alpha(C^\alpha)+\frac{1}{2}C^*_\gamma
v^\gamma_{\alpha\beta}(C^\alpha,C^\beta) +\dots)d^nx ,
\eea
with $v_0$ a local $\phi$-dependent function and
$v^{i}_\alpha$, $v^\gamma_{\alpha\beta}$ local $\phi$-dependent operators,
\eqref{indu} holds with
\bea
\omega^{-1,n}_f&=&\Big(\phi^*_i
[v^{i}_\alpha(f^\alpha)-M^{+ji}(\vddr{v_0}{\phi^j})]
\nonumber\\
&&+C^*_\alpha[k^{\alpha  i}_\beta(\vddr{v_0}{\phi^j},C^\beta)
+\partial_{(\mu)} v^i_\beta(C^\beta)\ddl{\phi^i_{(\mu)}}{f^\alpha}
+v^\alpha_{\beta\gamma}(C^\beta,f^\gamma)]+\dots\Big)d^nx. \quad
\eea
 
{\bf Remark:}
We note that 
if the reducibility parameters give rise to a reducibility
identity off-shell, $M^{[j(\nu)i(\mu)]}=0$, the induced global
symmetry can only be non trivial if the dependence
 on the antifields of antifield number 1
of $\omega^{0,n}$ is non trivial.

{}From the point of
view of a free theory, the existence of such an element $\omega^{0,n}$
is a necessary
condition for the existence of an interacting theory with a non
trivial deformation $R^{i1}_\alpha$ of the generating set of gauge
symmetries.
More precisely, it is a necessary condition
for the existence of a first order deformation. 

Starting from an interacting gauge theory that deforms
the gauge transformations of the linearized theory
in a non trivial way through terms linear in the fields, 
the cubic part $\omega^n_0=L_M^3\ d^n x$ 
that arises in an expansion in the number of fields and
antifields of the solution $L_M$ of the
master equation around a solution of the
classical equations of motion, $L_M\ d^nx =L_M^2\ d^nx+ d_H(\ ) +
L_M^3\ d^nx +\dots$, is automatically a cocycle (modulo $d_H$) for the BRST
differential of the linearized theory $s^{\rm free}$ generated by
$L_M^2$, with a non trivial
dependence on the antifields.

Explicitly, for $ L_M^3\ d^n x=(L^3+\varphi^*_i
R^{i1}_\alpha(C^\alpha)+ \frac{1}{2}C^*_\gamma v^{\gamma
  0}_{\alpha\beta}(C^\alpha,C^\beta))d^nx $, the induced symmetry for field
independent reducibility parameters $f^\alpha$ is given by 
$\omega^{-1,n}_f=(\varphi^*_i
R^{i1}_\alpha(f^\alpha)+C^*_\alpha v^{\alpha
  0}_{\beta\gamma}(C^\beta,f^\gamma))d^nx $. 

\subsubsection{Lie algebra of reducibility parameters}

According to the previous section, for a given element $[\omega^{0,n}]\in
H^{0,n}(s|d_H)$, there exists
a bilinear operation $[f_1,f_2]^\alpha$ between
reducibility parameters $f_1^\alpha,f_2^\alpha$
defined so that  
\bea
[\omega^{-2,n}_{[f_2,f_1]}]=
([\omega^{-1,n}_{f_1}],[\omega^{-2,n}_{f_2}])_M,
\eea
which induces a skew-symmetric bracket among equivalence
classes of reducibility parameters,
\bea
[[f_1],[f_2]]_M=[[f_1,f_2]].
\eea
Explicitly, one can choose 
\bea
[f_1,f_2]^\alpha=v^\alpha_{\beta\gamma}(f^\beta_1,f^\gamma_2)+\partial_{(\mu)}
v^i_\beta(f^\beta_1)\frac{{\partial^S}
  f^\alpha_2}{\partial\phi^i_{(\mu)}}-
\partial_{(\mu)}v^i_\beta(f^\beta_2)\frac{{\partial^S}
  f^\alpha_1}{\partial\phi^i_{(\mu)}}\nonumber\\
-k_{1\beta}^{\alpha i}(\vddr{v_0}{\phi^i},f^\beta_2)-\partial_{(\mu)}
  M^{+ji}_1(\vddr{v_0}{\phi^i})\frac{{\partial^S}
  f^\alpha_2}{\partial\phi^i_{(\mu)}}.
\eea
The graded Jacobi identity for $(\cdot,\cdot)$ implies that 
\bea
(\omega^{-1,n}_{f_1},(\omega^{-1,n}_{f_2},\omega^{-2,n}_{f_3}))=
(\omega^{-1,n}_{f_2},(\omega^{-1,n}_{f_1},\omega^{-2,n}_{f_3})
+((\omega^{-1,n}_{f_1},\omega^{-1,n}_{f_2}),\omega^{-2,n}_{f_3})+d_H(\ ).
\label{jac}
\eea
Suppose the element $[\omega^{0,n}]$ satisfies the condition
\bea
([\omega^{0,n}],[\omega^{0,n}])_M=[0] \iff
(\omega^{0,n},\omega^{0,n})+s (\ )+d_H (\ )=0.\label{integra}
\eea
The graded Jacobi identity for
$(\cdot,\cdot)$ then implies 
\bea
(([\omega^{-1,n}_{f_1}],[\omega^{-1,n}_{f_2}])_M,[\omega^{-2,n}_{f_3}])_M=
([\omega^{-1,n}_{f_3}],([\omega^{-1,n}_{f_2}],[\omega^{-2,n}_{f_1}])_M)_M.
\label{perm}
\eea
By using \eqref{jac}, \eqref{skew} and \eqref{perm} to transform the
first term, it follows that
\bea
([\omega^{-1,n}_{f_1}],([\omega^{-1,n}_{f_2}],[\omega^{-2,n}_{f_3}])_M)_M
+ {\rm cyclic}\ (1,2,3)=0.
\eea
This implies that the bracket $[\cdot,\cdot]_M$ among equivalence
classes of reducibility parameters satisfies the Jacobi identity. We
denote the Lie algebra of equivalence classes of reducibility
parameters equipped with the bracket  $[\cdot,\cdot]_M$ by
$\mathfrak{g}$.

One can introduce a basis $\{f_A^\alpha\}$ 
in the space of equivalence classes of
reducibility parameters. Such a basis has the property that all
reducibility parameters can be expressed as a linear combination of
the basis, up to trivial reducibility parameters,
\bea
f^\alpha\approx k^A f_A^\alpha,
\eea
and that the basis vectors are independent in the sense that
\bea
k^A f_A^\alpha\approx  0 \Longrightarrow k^A=0.
\eea 
A corresponding basis of $H^{-2,n}(s|d_H)$ is then given by
$\{\omega^{-2,n}_{f_A}\}$ and satisfies
\bea
\omega^{-2,n}=k^A\omega^{-2,n}_{f_A}+s(\ )+ d_H(\ ),\label{gen}
\eea
for any $s$ modulo $d_H$ cocycle $\omega^{-2,n}$
and 
\bea
k^A\omega^{-2,n}_{f_A}=s(\ )+ d_H(\ )\Longrightarrow k^A=0.\label{ntr}
\eea
Induced global symmetries associated to the basis can be defined by 
\bea
\omega^{-1,n}_{f_A}=(\omega^{0,n},\omega^{-2,n}_{f_A})+s(\ )+d_H(\ ).
\eea
We note that, in general, the set $\{\omega^{-1,n}_{f_A}\}$ 
is not a basis of $H^{-1,n}(s|d_H)$
because the induced global symmetries do
not necessarily span all the non trivial global symmetries and some
linear combinations of the induced global symmetries can be trivial
global symmetries. 

By reasonings similar to the above using in addition \eqref{gen} and
\eqref{ntr}, it follows from these definitions that 
\begin{itemize}
\item
the module action
of the induced global symmetries can be described by skew-symmetric
structure constants $C^C_{AB}$,
\bea
(\omega^{-1,n}_{f_A},\omega^{-2,n}_{f_B})=C^C_{AB}\omega^{-2,n}_{f_C}+s(\
)+d_H(\ );
\eea
\item the algebra of the induced global symmetries involves the
same structure constants,
\bea
(\omega^{-1,n}_{f_A},\omega^{-1,n}_{f_B})=C^C_{AB}\omega^{-1,n}_{f_C}+s(\
)+d_H(\ );
\eea
\item the structure constants satisfy the Jacobi identity provided 
$\omega^{0,n}$ satisfies \eqref{integra},
\bea
C^D_{AE}C^E_{BC}+{\rm cyclic}\ (A,B,C)=0.
\eea
\end{itemize}

{\bf Remarks:} (a) From the point of view of a free theory, the condition
\eqref{integra} is a necessary condition for the first order
deformation to be extendable to a second order deformation. 

(b) The cocycle 
$\omega^n_0=L_M^3\ d^n x$ of $H^{0,n}(s^{\rm free}|d_H)$,
obtained from the interacting theory
by expanding $L_M$, automatically satisfies \eqref{integra} in
the linearized theory, because the expansion of the master equation to
order $4$ reads
\bea
\half(L^3_M d^nx, L^3_M d^nx)+ s^{\rm free} L^4_M d^nx +d_H(\ )=0.
\eea

\subsection{Asymptotic symmetries and conservation laws}

\subsubsection{Linear characteristic cohomology}

Our starting point for understanding asymptotic symmetries,
conservation laws and their interplay is the
approach of reference \cite{Anderson:1996sc}
where asymptotic conservation laws have been 
studied in the context of the variational bicomplex, including
the vertical
generators $d_V\phi^i_{(\mu)}=d\phi^i_{(\mu)}-\phi^i_{(\mu)\nu}dx^\nu$
and the vertical differential
$d_V=d_V\phi^i_{(\mu)}{\partial^S}/{\partial\phi^i_{(\mu)}}$.
The vertical differential corresponds to an ``infinitesimal field
variation'' (independent of the variation of the base space) with these
variations and their derivatives being Grassmann odd. The concept
``vanish on all solutions of the equations of motion'' ($\approx 0$)
includes the equations $\delta L/\delta \phi^i=0$ and
$d_V(\delta L/\delta \phi^i)=0$ for the fields $\phi^i, d_V\phi^i$. 

Linear characteristic cohomology is defined in terms of vertical
1-forms and horizontal $n-k$ forms through the cocycle condition
\bea
d_H \omega_0^{n-k,1}+\omega_0^{n-k+1,1
  i(\mu)}\partial_{(\mu)}\frac{\delta L}{\delta \phi^i}+
\omega_0^{n-k+1,0
  i(\mu)}\partial_{(\mu)}d_V\frac{\delta L}{\delta \phi^i}=0,\label{3.31}
\eea
and the coboundary condition
\bea
\omega_0^{n-k,1}=d_H\eta^{n-k-1,1}_0+\eta_0^{n-k,1
  i(\mu)}\partial_{(\mu)}\frac{\delta L}{\delta \phi^i}+
\eta_0^{n-k,0
  i(\mu)}\partial_{(\mu)}d_V\frac{\delta L}{\delta \phi^i}.\label{3.32}
\eea
The Koszul-Tate resolution is extended to
 the full variational bicomplex associated with
all the equations of motion
\cite{Barnich:1995db,Barnich:1996mr} through the addition of the
additional vertical generators $d_V \phi^*_{i (\mu)},d_V C^*_{\alpha
  (\mu)}$ and the definition
$\delta_T=\delta+\delta_V$, where
\bea
\delta_V=-\partial_{(\mu)}d_V\frac{\delta L}{\delta \phi^i}
\frac{{\partial^S}}{\partial d_V \phi^*_{i (\mu)}}-
\partial_{(\mu)}d_V\left[R^{+i}_\alpha(\phi^*_i)\right]
\frac{{\partial^S}}{\partial
  d_V C^*_{\alpha (\mu)}}.
\eea
The cocycle and coboundary conditions (\ref{3.31}) and (\ref{3.32})
can then be written as
\bea
d_H \omega_0^{n-k,1}+\delta_T\omega_1^{n-k+1,1}=0,\label{lincoc}\\
\omega_0^{n-k,1}=d_H\eta^{n-k-1,1}_0+\delta_T\eta_1^{n-k,1}.\label{lincob}
\eea
As in the case without vertical generators, one uses descent
equation techniques to show for instance the isomorphism
$H_0^{n-k,1}(d_H|\delta_T)\simeq H_k^{n,1}(\delta_T|d_H)$. Note that in
this case, the isomorphism holds exactly in all spacetime dimensions
and not only up to constants, because of the presence of the vertical
generators.

The following technical lemma is a direct generalization of 
theorems 6.5 and 6.6
of \cite{Barnich:2000zw}.
\begin{lemma}[Trivial linear characteristic cohomology]
For linearizable, normal gauge theories,

(i) if $k\geq 3$ and the theory is irreducible, or

(ii) if $k=2$ and $N_{d_V
  C^*_\alpha}(\omega_2^{n,1})=0=N_{C^*_\alpha}(\omega_2^{n,1})$, or

(iii) if $k=1$ and $\omega_1^{n,1}\approx 0$,

\noindent
then
\bea
\delta_T\omega^{n,1}_k+d_H\omega^{n-1,1}_k=0\Longrightarrow \omega^{n,1}_k=
\delta_T\eta^{n,1}_{k+1}+d_H\eta^{n-1,1}_k.
\eea
\end{lemma}

\subsubsection{Exact linear characteristic cohomology}

Exact linear characteristic cohomology is defined through elements
$\omega_0^{n-k,0}$ such that $\omega_0^{n-k,1}=d_V\omega_0^{n-k,0}$
satisfies the cocycle condition (\ref{lincoc}).
For such a
representative, the second term of the cocycle condition
(\ref{lincoc}) can also be assumed to be $d_V$-exact.
This can be understood as a consequence of the fact that
\bea
H_k^*(\delta_T|d_V)=0,\ {\rm for}\ k>0,\label{delmover}
\eea
which itself follows from the fact that the contracting
homotopy, which allows one to prove that $H_k^*(\delta_T)=0$ for $k>0$,
anticommutes with $d_V$. Hence, exact linear characteristic
cohomology
is defined through the cocycle condition
\bea
d_H d_V\omega_0^{n-k,0}+\delta_Td_V\omega_1^{n-k+1,0}=0\label{spec}
\eea
for the form $\omega_0^{n-k,0}$.
The cocycle $\omega_0^{n-k,0}$ is trivial as an
element of exact linear characteristic cohomology if
\bea
d_V\omega_0^{n-k,0}=d_H\eta^{n-k-1,1}_0+\delta_T\eta^{n-k,1}_1. \label{cobel}
\eea

We shall now show that standard characteristic cohomology
and exact linear characteristic cohomology are isomorphic, except
for the presence of the constants in the former.
Because $\{d_V,d_H\}=0=\{d_V,\delta_T\}$, there is a well defined map
from standard characteristic cohomology to exact
linear characteristic cohomology: $[\omega_0^{n-k,0}]\longrightarrow
[\omega_0^{n-k,0}]$. 
The kernel of this map is given by ${\mathbb R}$ in form degree
$0$.
Indeed, the kernel is defined by a cocycle $\omega^{n-k,0}_0$ of
standard characteristic cohomology such that (\ref{cobel}) holds.
Let $\omega^{k,l}=b^{k,l}+\tilde \omega^{k,l}$, where $b^{k,l}$ is
obtained from $\omega^{k,l}$ by setting to zero all the fields,
antifields, their derivatives and their vertical
derivatives. By applying the contracting homotopy 
$\rho_V (\cdot)=\int_0^1 dt/t\ [\partial_{(\mu)}Z^a\partial/
\partial d_V
Z^a_{(\mu)}(\cdot)][tZ,td_VZ,x,dx]$ of $d_V$ to
(\ref{cobel}), we get
\bea
\tilde \omega_0^{n-k,0}=-d_H\rho_V
\eta^{n-k-1,1}_0-\delta_T\rho_V \eta^{n-k,1}_1
+\{\rho_V, \delta_T\}\eta^{n-k,1}_1.\label{vextra}
\eea
Furthermore,
\bea
\{\rho_V,\delta_T\}=\partial_{(\mu)}[2\frac{\delta L}{\delta\phi^i}
-\frac{\delta N_\phi(L)}{\delta\phi^i}]\frac{\partial}{\partial
d_V\phi^*_{i(\mu)}}-\partial_{(\mu)}[N_\phi(R^{+j}_\alpha)(\phi^*_j)]
\frac{\partial}{\partial
d_V C^*_{\alpha(\mu)}}.\label{extra}
\eea
This shows that
for linear theories one has $\{\rho_V,\delta_T\}=0$, so that
$\tilde
\omega_0^{n-k,0}$ is a trivial characteristic cohomology class.  For
linearizable theories, an induction on the homogeneity in the fields
$Z,d_VZ$ with $\{Z\}=\{\phi,\phi^*,C^*\}$, 
allows one to prove the same result in the space of formal
power
series in $Z,d_V Z$, which is extended to the case of linearizable,
normal theories, to spaces involving a finite number of derivatives
as
in section 6 of \cite{Barnich:2000zw}. The part $b^{n-k,0}$ of
$\omega_0^{n-k,0}$ satisfies $d_Hb^{n-k,0}=0$, implying
$b^{n-k,0}=d_H b^{n-k-1,0}+\delta^{n-k}_0 k$, $k\in {\mathbb R}$,
which gives the result.

The map from standard to exact linear characteristic cohomology is
surjective. Indeed, if
$\omega^{n-k,0}_0=b^{n-k,0}+\tilde\omega^{n-k,0}_0$, the part
$b^{n-k,0}$ is always trivial in exact linear characteristic cohomology
because $d_Vb^{n-k,0}=0$ satisfies (\ref{cobel}) with
$\eta^{n-k-1,0}_0=0=\eta^{n-k,1}_1$, while $\tilde\omega^{n-k,0}_0$
corresponds to a cocycle of standard characteristic cohomology. This
follows by using the free cohomology of $d_V$ for the cocycle
condition (\ref{spec}) with $\omega^{n-k,0}_0$ replaced by
$\tilde\omega^{n-k,0}_0$ and $\omega^{n-k+1,0}_1$ by
$\tilde\omega^{n-k+1,1}_0$.

Hence, except for the constants, exact and standard characteristic
cohomology are indeed isomorphic, and nothing is gained by considering exact
linear characteristic
cohomology. However, this changes if one evaluates at a
fixed background. 

\subsubsection{Koszul-Tate resolution of the linearized theory } 

In this and the following subsubsections, the Lagrangian $L$ is
the source free Lagrangian relevant near the boundary and  
$\bar\phi(x)$ is a solution of the associated field equations. 
The differential $\delta_T|_{\bar\phi(x)}$ is nilpotent because for a 
solution $\bar\phi(x)$ of the field equations relevant near the boundary,
$R^{+i}_\alpha(\delta L/\delta\phi^i)=0$ implies
$R^{+i}_\alpha|_{\bar\phi(x)}((d_V\delta
L/\delta\phi^i)|_{\bar\phi(x)})=0$, while
$\delta_T|_{\bar\phi(x),\phi^*=0}\equiv\delta^{\rm free}$ with
\bea
\delta^{\rm free}
=-\partial_{(\mu)}(d_V\frac{\delta L}{\delta \phi^i})|_{\bar\phi(x)}
\frac{\partial^S}{\partial d_V \phi^*_{i (\mu)}}-
\partial_{(\mu)}\left[R^{+i}_\alpha|_{\bar\phi(x)}(d_V\phi^*_i)
\right]
\frac{\partial^S}{\partial
  d_V C^*_{\alpha (\mu)}},
\eea
is acyclic in positive vertical antifield number
because, up to an overall shift of grading,  
it is the Koszul-Tate differential associated with the free
theory valid near the boundary.
Furthermore, the identity
\bea
\delta^{\rm free}(d_V\omega)|_{\bar\phi(x),\phi^*=0}
=-(d_V\delta_T\omega)|_{\bar\phi(x),\phi^*=0},\label{3.16}
\eea
for all $\omega$, will allow us to relate expressions constructed in
the free theory near the boundary to expressions constructed
in the full theory.

\subsubsection{Boundary
  conditions and asymptotic acyclicity} \label{s754}

In the following, the asymptotic behaviour of forms 
is understood after evaluation for fields and
antifields that satisfy the asymptotic behaviour $\varphi^i(x)\asy
O(\chi^i)$, $d^nx \phi^*_i\asy O(\chi_i)$, $d^nx C^*_\alpha\asy
O(\chi_\alpha)$ with the $\chi_i$, $\chi_\alpha$ 
as defined in section \ref{Prerequisites}.

Our aim is to extend the results derived
in \cite{Barnich:1995db,Barnich:2000zw} and reviewed in sections
\ref{s3} and \ref{s73} on
exact reducibility parameters and conserved $n-2$ forms 
to their asymptotic counterparts. 
The analysis in \cite{Barnich:1995db,Barnich:2000zw} is
based on acyclicity properties of $d_H$ and $\delta$.
Therefore these results can be extended to the 
asymptotic context when the differentials 
$d_H$ and $\delta^{\rm free}$ have analogous
``{\em asymptotic acyclicity properties}''.
More precisely, what one needs is
\bea
\left\{\begin{array}{lcl}
d_H\omega^k\longrightarrow 0 &\Longleftrightarrow&
\omega^k\longrightarrow d_H\eta^{k-1}\quad {\rm for}\ 0<k<n,\\
\omega^n\longrightarrow d_H\eta^{n-1}&\Longleftrightarrow&
\forall d_VZ^A:\ 
\mbox{$\displaystyle{d_VZ^A
\frac{\delta \omega^n}{\delta d_V Z^A}\longrightarrow 0
}$},
\end{array}\right.
\label{7.66}
\eea
and
\bea
\left\{\begin{array}{llcl}
& \delta^{\rm free}\omega_k\longrightarrow 0 &\Longleftrightarrow&
\omega_k\longrightarrow \delta^{\rm free}\eta_{k+1}\quad {\rm for}\
k\geq 1,\\
\forall \omega_0\asy O(1):\ &\omega_0\asy\delta^{\rm free} \eta_1
&\Longleftrightarrow &\forall\varphi_s(x):\ 
\omega_0|_{\varphi_s(x)}\asy 0,
\end{array}\right.
\label{7.67}
\eea
on forms which are homogeneous and linear in the variables $\{d_V Z^A\}= 
\{d_V\phi^i,d_V\phi^*_i, d_VC^*_\alpha\}$ and their derivatives,
with coefficients that are ordinary differential forms
made up of $x^\mu$ and $dx^\mu$
[$\varphi_s(x)$ are asymptotic solutions as in
(\ref{asysol})]\footnote{As the last conditions in
  \eqref{7.66} and \eqref{7.67} show, what we call acyclicity
  properties includes not only absence of non trivial cohomology in
  appropriate degrees. In addition it requires 
  that the horizontal complex provides an algebraic
  resolution of equivalence classes of local $n$ forms with 
  asymptotically identical
  Euler-Lagrange derivatives, while the Koszul-Tate complex provides 
  asymptotically a
  resolution of the horizontal forms pulled back to the surface defined by
  the linearized equations of motion.}.
Therefore, we shall assume that
the boundary conditions are such that 
(\ref{7.66}) and (\ref{7.67}) hold. At first glance, this
may appear to be a strong assumption. However, as we tried to show by
a detailed analysis of exact reducibility parameters and
conserved $n-2$ forms, it is actually quite natural. 

The validity of the first part of 
(\ref{7.66}) is related to properties of the
contracting homotopy associated to $d_H$: equation (\ref{cc1}) gives,
for a $k$-form $\omega^k[x,d_VZ]$ depending
linearly on the $d_VZ^A$ and their derivatives:
\bea
\omega^k[x,d_VZ]=
\rho^{k+1}_{H,d_VZ}(d_H\omega^k)+d_H(\rho^k_{H,d_VZ}\omega^k),
\label{homoexample}
\eea
with homotopy
operators $\rho_{H,d_VZ}$ as in (\ref{phihomotopy}). The homotopy
operators remove a differential $dx^\nu$ and one
derivative $\6_\nu$ of one of the fields, and
redistribute the other derivatives over the fields and the 
coefficient functions. 
Hence,
whenever the fields, as functions of $x^\mu$, and the 
coefficient functions 
are sufficiently well-behaved (as discussed and illustrated in more
details in sections \ref{Prerequisites}, \ref{s54} and
\ref{3d-adS}), asymptotic acyclicity of $d_H$ will indeed hold. 
Similarly, the second part of (\ref{7.66}) is related to the identity
\bea
\omega^n[x,d_VZ]=d_VZ^A
\frac{\delta \omega^n}{\delta d_V Z^A}+d_H(\rho^n_{H,d_VZ}\omega^n).
\eea
This identity evidently provides
the implication $\Longleftarrow$ of 
the second part of (\ref{7.66}); it also
gives $\Longrightarrow$ whenever the following
(very reasonable) implication holds
\bea
\forall d_VZ^A:\
d_VZ^A
\frac{\delta \omega^n}{\delta d_V Z^A}\asy d_H\eta^{\prime
  n-1}\Longrightarrow d_VZ^A
\frac{\delta \omega^n}{\delta d_V Z^A}\asy 0.
\eea

The first condition in \eqref{7.67} is trivially satisfied for $k\geq
3$ because we consider irreducible gauge theories and forms that are
linear and homogeneous in the fields and antifields
[in irreducible gauge theories, there are no such forms because
there are no antifields with antifield number
$\geq 3$]. For $k=1,2$,
this condition is equivalent to \eqref{asyacycdelta} and
(\ref{asyacycdeltabis}), respectively. 
The second condition in \eqref{7.67} is a
consequence of the asymptotic regularity conditions discussed in
sections \ref{Prerequisites} and \ref{s54}. 

\subsubsection{Definitions and bijective correspondence}

Let $1\leq k\leq n$. 
The form $\omega_0^{n-k,0}$ is an asymptotic conservation law of
order $n-k$, relative to the fixed background
$\bar \phi^i(x)$, if the cocycle condition (\ref{spec})
holds asymptotically when evaluated at $\bar \phi(x)$,
\bea
d_H (d_V\omega_0^{n-k,0})|_{\bar\phi(x)}+
\delta^{\rm free}(d_V\omega_1^{n-k+1,0})|_{\bar\phi(x)}\longrightarrow 0.
\label{ascons}
\eea
The asymptotic conservation law $\omega_0^{n-k,1}$ at $\bar
\phi(x)$ is trivial 
if
\bea
(d_V\omega_0^{n-k,0})|_{\bar\phi(x)
}
\longrightarrow  d_H\tilde\eta^{n-k-1,1}_0+
\delta^{\rm free}(d_V\eta_1^{n-k,0})|_{\bar\phi(x)
},
\label{trcons}
\eea
where $\tilde\eta^{n-k-1,1}_0$ is a local form involving
linearly only the
fields $d_V\phi^i$.
Asymptotic characteristic cohomology is defined as the set of
equivalence classes of asymptotic conservation laws up to trivial
ones.

The form $\omega_k^{n,0}$ is an asymptotic degree $k$
symmetry for the fixed background $\bar\phi(x)$ if
\bea
\delta^{\rm free}(d_V\omega_k^{n,0})|_{\bar\phi(x),\phi^*=0}+
d_H(d_V\eta_{k-1}^{n-1,0})|_{\bar\phi(x),\phi^*=0}\longrightarrow 0.
\eea
The degree $k$ asymptotic symmetry at $\bar\phi(x)$ is trivial if
\bea
(d_V\omega_k^{n,0})|_{\bar\phi(x),\phi^*=0}\longrightarrow \delta^{\rm free}
\tilde\eta_{k+1}^{n,1}+
d_H(d_V\eta_{k}^{n-1,0})|_{\bar\phi(x),\phi^*=0},\label{trsym}
\eea
where $\tilde\eta_{k+1}^{n,1}$ involves linearly only the
vertical derivatives of the antifields.
Equivalence classes of asymptotic degree $k$ symmetries are
defined as asymptotic degree $k$ symmetries modulo trivial ones.

Because the forms are linear and homogeneous in the fields and antifields 
 and the theory is irreducible, only the cases $k=1,2$ can give
 non trivial cohomology. Furthermore, the asymptotic acyclicity
 properties assumed in subsubsection \ref{s754} allow one to prove the
 bijective correspondence between equivalence classes of 
asymptotic degree $k$ symmetries and of degree $n-k$ conservation laws
exactly as done in \cite{Barnich:1995db,Barnich:2000zw} in the exact case.

\subsubsection{Asymptotic global symmetries and 
  asymptotically conserved currents}

The cocycle condition both for the asymptotic global (i.e., degree
$1$) symmetries and
the asymptotically conserved $n-1$ forms at $\bar\phi(x)$ reads:
\bea
\delta^{\rm free}(d_V\omega^{n,0}_1)|_{\bar\phi(x),\phi^*=0}+
d_H(d_V\omega^{n-1,0}_0)|_{\bar\phi(x),\phi^*=0}\longrightarrow
0\label{cocdeg1}.
\eea
The general form of an asymptotic global symmetry is
\bea
\omega^{n,0}_1=\phi^*_{i(\mu)}Q^{+i(\mu)}
d^nx=(\phi^*_iQ^{i} +\partial_\mu T^{i\mu}(\phi^*_i))d^nx,
\eea
where we used ``integrations by parts'':
we applied repeatedly Leibniz' rule
$(\6f)g=-f(\6g)+\6(fg)$ and collected the
terms $\6(fg)$ in $\6_\mu T^{i\mu}(\phi^*_i)$.
The second term in $\omega^{n,0}_1$ corresponds
to the trivial asymptotic symmetry $d_H
d_V(T^{i\mu}(\phi^*_i))|_{\bar\phi(x),\phi^*=0}(d^{n-1}x)_\mu$ and
can be absorbed
by the trivial asymptotically conserved $n-1$ form  $\delta^{\rm free}
d_V(T^{i\mu}(\phi^*_i))|_{\bar\phi(x),\phi^*=0}(d^{n-1}x)_\mu$.
Accordingly, we can assume, $\omega^{n,0}_1=\phi^*_{i}Q^{i} d^nx$.
With $\omega^{n-1,0}_0=j^\mu (d^{n-1}x)_\mu$, the cocycle condition
(\ref{cocdeg1})
is explicitly given by
\bea
-(d_V\frac{\delta L}{\delta \phi^i})|_{\bar\phi(x)}Q^{i}|_{\bar\phi(x)}d^nx
+\partial_\mu (d_Vj^{\mu})|_{\bar\phi(x)}d^nx\longrightarrow 0,
\eea
which means
\bea
\frac{\delta L^{\rm free}}{\delta \varphi^i}Q^{i}|_{\bar\phi(x)}d^nx
\longrightarrow
\partial_\mu (d_Vj^{\mu})|_{\bar\phi(x),\varphi}d^nx.
\eea
The coboundary condition (\ref{trsym}) for $k=1$, with
\bea
\tilde \eta^{n,1}_2=-d_VC^*_{\alpha(\mu)}\tilde
f^{\alpha(\mu)}d^nx,
\eea
and $\tilde f^{\alpha(\mu)}$
depending
on $x$ alone, implies that the asymptotic symmetry defined by
$Q^{i}$ is trivial at $\bar\phi(x)$ if and only if
\bea
\forall \psi_i\asy
O(\chi_i):\ Q^{i}|_{\bar\phi(x)}\psi_i\longrightarrow
\psi_iR^i_\alpha|_{\bar\phi(x)}(\tilde f^{\alpha}), 
\eea
with $\tilde f^{\alpha}=(-\partial)_{(\mu)}\tilde f^{\alpha(\mu)}$. 
This follows by ``integrations by parts'' and using the fact that the
coboundary condition holds for all $d_V\phi^*_i$ satisfying the
boundary conditions. The coboundary condition for asymptotically
conserved $n-1$ forms states that such a form is trivial if it is
asymptotically $d_H$-exact up to a form
that is proportional to the field equations of the linearized theory.

In order to be able to interpret
asymptotic global symmetries and asymptotically conserved $n-1$ forms
from the point of view of the bulk theory, we use relation
(\ref{3.16}) to rewrite the cocycle condition \eqref{cocdeg1}
as 
\bea
-d_V(\delta_T \omega^{n,0}_1)|_{\bar\phi(x),\phi^*=0}+
d_H(d_V\omega^{n-1,0}_0)|_{\bar\phi(x),\phi^*=0}\longrightarrow 0,
\eea
or explicitly,
\bea
-d_V(\frac{\delta L}{\delta \phi^i}Q^{i})|_{\bar\phi(x)}d^nx+
\partial_\mu (d_Vj^{\mu})|_{\bar\phi(x)}d^nx\longrightarrow 0.
\eea
In other words, consider the weakly vanishing charge $\int_M
{\delta L}/{\delta \phi^i}Q^{i}$ associated to the transformation 
$\delta\phi^i=Q^{i}$. The necessary and sufficient condition that
allows one to improve this charge by the substraction of a surface
integral $\oint_{\partial
  M} j$ to a charge that is asymptotically extremal at $\bar\phi(x)$ 
for arbitary
variations $d_V\phi^i$ not restricted by any boundary conditions is
the requirement that $Q^i$ defines an asymptotic global symmetry. For
solutions of the equations of motion, the improved charge reduces to
the surface integral whose integrand is the asymptotically conserved
$n-1$ form. 

\subsubsection{Asymptotic reducibility parameters and
  conserved n--2 forms}\label{3.7}

The cocycle condition for an asymptotic degree $2$ symmetry at
$\bar\phi(x)$ is
\bea
\delta^{\rm free}(d_V\omega^{n,0}_2)|_{\bar\phi(x),\phi^*=0}
+d_H(d_V\omega^{n-1,0}_1)|_{\bar\phi(x),\phi^*=0}\longrightarrow 0.
\label{asdes1}
\eea
Applying $\delta^{\rm free}$ and using asymptotic acyclicity of $d_H$ gives
\bea
\delta^{\rm free}(d_V\omega^{n-1,0}_1)|_{\bar\phi(x),\phi^*=0}
+d_H\tilde\omega^{n-2,1}_0\longrightarrow 0.\label{asdes2}
\eea
The general form of an asymptotic degree 2
symmetry is
\bea
\omega^{n,0}_2=[f^\alpha
C^*_\alpha-\frac{1}{2}\phi^*_{j(\nu)}\phi^*_{i(\mu)}
M^{[j(\nu)i(\mu)]}]d^nx,
\eea
where again ``integrations
by parts'' have been done to reduce the first term to one not involving the
derivatives of $C^*_\alpha$.
With $\omega^{n-1,0}_1=T^{\mu i}(\phi^*_i){(d^{n-1}x)_\mu}$, the cocycle
condition gives
\bea
-R^{+i}_\alpha|_{\bar\phi(x)}(d_V\phi^*_i)f^\alpha|_{\bar\phi(x)}d^nx
+\partial_\mu
T^{\mu i}|_{\bar\phi(x)}(d_V\phi^*_i)d^nx\longrightarrow 0.
\eea
By using the fact that this equation has to hold for arbitrary
$d_V\phi^*_i$ satisfying the boundary conditions, one finds that
asymptotic degree 2 symmetries are determined by asymptotic
reducibility parameters $\tilde f^\alpha=f^\alpha|_{\bar\phi(x)}$
satisfying \eqref{asympsym}.
According to (\ref{1.3}), $\omega^{n-1,0}_1$ can be taken to be
$S^{\mu i}_\alpha(\phi^*_i,\4f^\alpha){(d^{n-1}x)_\mu}$.

The equation defining the corresponding
asymptotically conserved $n-2$ form $k^{[\nu\mu],0}$ can be taken
to be
\bea
\delta^{\rm free} (d_VS^{\mu i}_\alpha(\phi^*_i,
f^\alpha))|_{\bar\phi(x),\phi^*=0}{(d^{n-1}x)_\mu}+
\partial_\nu(d_Vk^{[\nu\mu],0}_\alpha(f^\alpha))|_{\bar\phi(x)}
{(d^{n-1}x)_\mu}
\longrightarrow 0,\label{3.38}
\eea
which gives explicitly
\bea
S^{\mu i}_\alpha(\frac{\delta L^{\rm
    free}}{\delta\varphi^i},
f^\alpha))|_{\bar\phi(x)}{(d^{n-1}x)_\mu}
+\partial_\nu(d_Vk^{[\nu\mu],0}_\alpha(f^\alpha))|_{\bar\phi(x),\varphi}
{(d^{n-1}x)_\mu}
\longrightarrow 0.
\eea
How the expression $d_Vk^{[\nu\mu],0}_\alpha(f^\alpha))$ can be
explicitly constructed
out of $s^\mu_f=
S^{\mu i}_\alpha({\delta L^{\rm
    free}}/{\delta\varphi^i},
f^\alpha))|_{\bar\phi(x)}$ has been explained in section \ref{s3.8}. 

The coboundary condition for the asymptotic symmetry
at
$\bar\phi(x)$ is
$(d_V\omega^{n,0}_2)|_{\bar\phi(x),\phi^*=0}\longrightarrow \delta^{\rm
  free}\tilde\eta^{n,1}_3+
d_H(d_V\eta^{n-1,0}_2))|_{\bar\phi(x),\phi^*=0}$. Because there is no
$\tilde\eta^{n,1}_3$ in irreducible gauge theories, 
and integrations by parts have already been used to remove
all derivatives from $C^*_\alpha$, the asymptotic degree
$2$ symmetries are trivial iff the associated asymptotic reducibility
parameters are trivial as defined in \eqref{puregauge2}. 

Hence, equivalence classes of asymptotic degree 2 symmetries are in
bijective correspondence to equivalence classes of asymptotic
reducibility parameters and theorem 1 follows from the bijective
correspondence between asymptotic degree 2 symmetries and degree $n-2$
conservation laws. 

According to (\ref{3.16}), equation (\ref{3.38}) can also be written as
\bea
-(d_V\delta_T S^{\mu i}_\alpha(\phi^*_i,
f^\alpha))|_{\bar\phi(x)}{(d^{n-1}x)_\mu}+
\partial_\nu(d_Vk^{[\nu\mu],0}_\alpha(f^\alpha))|_{\bar\phi(x)}
{(d^{n-1}x)_\mu}
\longrightarrow 0,
\eea
which gives
\bea
-(d_VS^{\mu i}_\alpha(\frac{\delta L}{\delta\phi^i},
f^\alpha))|_{\bar\phi(x)}{(d^{n-1}x)_\mu}+\partial_\nu(d_V
k^{[\nu\mu],0}_\alpha(f^\alpha))|_{\bar\phi(x)}{(d^{n-1}x)_\mu}
\longrightarrow 0.\label{3.40}
\eea
This equivalent formulation of the cocycle condition for asymptotic
degree $2$ symmeties and degree $n-2$ conservation laws proves theorem
3 of subsection \ref{s53}.

\subsubsection{Asymptotic algebra}

As we have seen above, the asymptotic degree $2$ symmetry can be
identified with the $n$ form $\omega^{-2,n}_{\tilde f}=\tilde f^\alpha
C^*_\alpha d^nx$ of the free theory, with $\tilde f^\alpha$ asymptotic
reducibility parameters. Assuming here and everywhere below 
that condition \eqref{convergencec} holds, the
asymptotic behaviour of this $n$-form is
$\omega^{-2,n}_{\tilde f}\asy O(1)$. Furthermore we have $s^{\rm free}
\omega^{-2,n}_{\tilde f}=\delta^{\rm free}\omega^{-2,n}_{\tilde f}
\asy d_H(\ )$. 

Let us define the asymptotic behaviour of the
ghosts $C^\alpha$ to be the same as that of the asymptotic 
reducibility parameters: $C^\alpha\asy
O(\chi^\alpha)$. 
The cubic
vertex  $\omega^{0,n}=[
L^3+\phi^*_iR^{i1}_\alpha(C^\alpha)+ 1/2 C^*_\alpha C^{\alpha
  0}_{\beta\gamma}(C^\beta,C^\gamma)]d^nx$ induced by the full theory
satisfies $s^{\rm free}
\omega^{0,n}=d_H(\ )$. Furthermore, assumptions 
\eqref{constr1} and (\ref{constr2})
then guarantee that
the form representing the induced ``global symmetry'' is
at most of order $1$
\beann
&\omega^{-1,n}_{\tilde f}=\delta_{\omega^{0,n}}\omega^{-2,n}_{\tilde
  f}+d_H(\cdot)=[ \phi^*_iR^{i1}_\alpha(\tilde
f^\alpha)+C^*_\alpha C^{\alpha 0}_{\beta\gamma}(C^\beta,\tilde f^\gamma)]d^nx
\asy O(1),
\eeann
with
\bea
\delta_{\omega^{0,n}}=
\partial_{(\mu)}[
R^{+i1}_\alpha(\phi^*_i)+
C^{+\alpha 0}_{\beta\gamma}(C^*_\alpha,C^\beta)]
\frac{\partial^S}{\partial
  C^*_{\gamma(\mu)}}-\partial_{(\mu)}
[\frac{1}{2}
C^{\alpha 0}_{\beta\gamma}(C^\beta,C^\gamma)]\frac{\partial^S}{\partial
  C^\alpha_{(\mu)}}\nonumber\\
+\partial_{(\mu)}[\frac{\delta}{\delta\varphi^j}(L^3+
\phi^*_iR^{i1}_\alpha(C^\alpha))]
\frac{\partial^S}{\partial
  \phi^*_{j(\mu)}}+\partial_{(\mu)}[R^{i1}_\alpha(C^\alpha)]
\frac{\partial^S}{\partial
  \varphi^i_{(\mu)}},
\eea
and 
$[s^{\rm free}, \delta_{\omega^{0,n}}]=0$.
When $\delta_{\omega^{0,n}}$ is applied to 
$s^{\rm free}\omega^{-2,n}_{\tilde f}$,
only the part that acts on the antifields $\phi^*_i$ is involved, 
and conditions \eqref{newass1} and \eqref{newass} guarantee that this
action does not increase the asymptotic degree. It follows that
$\omega^{-1,n}_{\tilde f}$ is 
asymptotically a BRST cocycle modulo $d_H$ 
\bea
s^{\rm free}\omega^{-1,n}_{\tilde f}
=\delta_{\omega^{0,n}}
\underbrace{s^{\rm free}\omega^{-2,n}_{\tilde f}}_{\asy d_H()}
\asy d_H(\cdot).\label{coblo}
\eea
Similarly, 
\beann
\omega^{-2,n}_{[\tilde f_2,\tilde f_1]_M}=
\delta_{\omega^{-1,n}_{\tilde f_1}}\omega^{-2,n}_{\tilde
  f_2}+ d_H(\cdot)
=d^nx C^*_\alpha[\tilde f_2,\tilde f_1]_M^\alpha
\asy O(1),
\eeann
so that $[\tilde f_1,\tilde
  f_2]_M^\alpha$ satisfies condition
  (\ref{convergencec}). 
Explicitly, 
\bea
\delta_{\omega_{\tilde f_1}^{-1,n}}=
-\partial_{(\mu)}[C^{+\alpha 0}_{\beta\gamma}(C^*_\alpha,\tilde
f^\beta_1)]\frac{\partial^S}{\partial
  C^*_{\gamma(\mu)}}-\partial_{(\mu)}
[C^{\alpha 0}_{\beta\gamma}(C^\beta,\tilde
f^\gamma_1)]\frac{\partial^S}{\partial
  C^\alpha_{(\mu)}}\nonumber\\
+\partial_{(\mu)}[\frac{\delta}{\delta\varphi^j}(
\phi^*_iR^{i1}_\alpha(\tilde f^\alpha_1))]
\frac{\partial^S}{\partial
  \phi^*_{j(\mu)}}-\partial_{(\mu)}[R^{i1}_\alpha(\tilde f^\alpha_1)]
\frac{\partial^S}{\partial
  \varphi^i_{(\mu)}}.
\eea
Now, $[s^{\rm free},\delta_{\omega_{\tilde f_1}^{-1,n}}]=
\delta_{s^{\rm free}\omega_{\tilde f_1}^{-1,n}}$ with 
\beann
s^{\rm free}\omega_{\tilde f_1}^{-1,n}=[
\frac{\delta L^2}{\delta \varphi^i}R^{i1}_\alpha(\tilde
f^\alpha_1)-\phi^*_i
\partial_{(\mu)}R^{j0}_\beta(C^\beta)\frac{\partial R^{i1}_\alpha(\tilde
f_1^\alpha)}{\partial \varphi^j_{(\mu)}}+R^{+i0}_\alpha(\phi^*_i)
C^{\alpha 0}_{\beta\gamma}(C^\beta,\tilde f_1^\gamma)]d^nx.
\eeann
In the action of $\delta_{s^{\rm free}\omega_{\tilde f_1}^{-1,n}}$ on 
$\omega^{-2,n}_{\tilde
  f_2}$ only the ghost dependent part of $s^{\rm
  free}\omega_{\tilde f_1}^{-1,n}$ is involved, and this part does
not change the asymptotic behaviour because the ghost dependent terms
are at most of order $1$. It follows from \eqref{coblo}
that $\delta_{s^{\rm free}\omega_{\tilde f_1}^{-1,n}}\omega^{-2,n}_{\tilde
  f_2}\asy 0$. 
Since again the part of $\delta_{\omega_{\tilde f_1}^{-1,n}}$ that
acts on $s^{\rm free}\omega^{-2,n}_{\tilde
  f_2}$ only involves the antifields $\phi^*_i$ and is at most of
order $1$, 
\beann
&s^{\rm free}\omega^{-2,n}_{[\tilde f_2,\tilde f_1]_M} \asy d_H(\cdot).&
\eeann
This implies that $\psi_iR^{i0}_\alpha([\tilde f_1,\tilde
  f_2]_M^\alpha)\asy 0$ for all $\psi_i$, meaning that $[\tilde f_1,\tilde
  f_2]_M^\alpha$ are asymptotic reducibility parameters. 
Hence, asymptotic reducibility
  parameters form a Lie algebra for the bracket $[\cdot,\cdot]_M$. 
Finally, 
  when the $\tilde f_2^\alpha$ are pure gauge, the
  associated form $\omega^{-2,n}_{\tilde f_2}$
vanishes asymptotically and so do the forms $\omega^{-1,n}_{\tilde
  f_2}$ and $\omega^{-2,n}_{[\tilde f_2,\tilde f_1]_M}$, 
  which implies that $[\tilde f_2,\tilde f_1]^\alpha_M$ are pure gauge.   
The skew symmetry \eqref{skew} of the bracket induced in
  cohomology shows that the same conclusion holds if the $\tilde
  f_1^\alpha$ are pure gauge. Hence, there is a well defined induced Lie
  algebra for the quotient space of asymptotic reducibility
  parameters modulo pure gauge parameters. This completes the proof of
  theorem 2.  

In subsubsection \ref{s733}, we have applied 
$\delta_{\omega_X^{-1,n}}$ to the set of descent equations
that relates the $n$ forms representing the (exact) reducibility 
parameters to
the associated conserved $n-2$ forms in order to
investigate the action of the associated global symmetry 
on equivalence classes of conserved $n-2$ forms. In order to use the
same reasoning here, we use the fact that the asymptotic behaviour
of the operator $\delta_{\omega^{-1,n}_{\tilde f_1}}$
is at most of order 1 due to the additional assumption
\eqref{newass2}. Furthermore, since (\ref{4.7}) and (\ref{1.18}) imply
\bea
s^{\rm free}\omega_{\tilde f_1}^{-1,n}=[
-\frac{\delta L^3}{\delta \varphi^i}R^{i0}_\alpha(\tilde
f^\alpha_1)+\phi^*_i
\partial_{(\mu)}R^{j0}_\beta(\tilde f_1^\beta)
\frac{\partial R^{i1}_\alpha(C^\alpha)}{\partial \varphi^j_{(\mu)}}
]d^nx+d_H(\cdot),
\eea
the additional assumptions \eqref{newass3} and \eqref{newass4}
guarantee that $[s^{\rm free},\delta_{\omega_{\tilde f_1}^{-1,n}}]=
\delta_{s^{\rm free}\omega_{\tilde f_1}^{-1,n}}\asy 0$. 
This implies that $\delta_{\omega_{\tilde f_1}^{-1,n}}$ can be
applied to the asymptotic descent equations \eqref{asdes1} and
\eqref{asdes2} and allows one to prove the statements on the
representation of the Lie algebra of equivalence classes of asymptotic
reducibility parameters in exactly the same way as in the case of
equivalence classes of exact reducibility parameters.

\mysection{Relation to other approaches}\label{s8}

\subsection{Hamiltonian approach}

A systematic approach to asymptotic conservation laws,
especially in the
context of general relativity, was given in
\cite{Regge:1974zd} in the context of the Hamiltonian formalism. In
order to make contact with this approach,
we will apply the covariant Lagrangian results obtained in the
previous sections to the action in first order
Hamiltonian form,
\bea
S=\int dt\int d^{n-1}x\ (\frac{1}{2}\dot z^A \sigma_{AB}z^B
-h-\gamma_a\lambda^a ),
\label{actfirstorder}
\eea
where we assume for simplicity that 
the constraints $\gamma_a$ are first class,
irreducible and time independent and 
that $\sigma_{AB}$ is the symplectic matrix with 
$\sigma^{AB}\sigma_{BC}=\delta^A_C$.
In the following, we will use a local Poisson bracket with spatial
Euler-Lagrange derivatives for spatial $n-1$ forms, 
\bea
\{f\, d^{n-1}x,g\, d^{n-1}x\}=\frac{\tilde \delta f}{\delta z^A}\sigma^{AB}
\frac{\tilde \delta g}{\delta z^B}\, d^{n-1}x.
\eea
If $\tilde d_H$ denotes the spatial exterior derivative, this Poisson bracket
is well defined in the space $H^{n-1}(\tilde d_H)$, and thus does not
depend on ``boundary terms'' that are added to improve spatial
functionals. Similiarily, the Hamiltonian vector field associated to
an $n-1$ form 
\bea
\tilde \delta_{fd^{n-1}x}=\partial_{(i)}\frac{\tilde\delta f}{\delta
  z^A}\sigma^{AB}\frac{\partial^S}{\partial z^B_{(i)}}
\eea
only depends on the class $[fd^{n-1}x]\in H^{n-1}(\tilde d_H)$.
If we denote
\bea
\hat\gamma_a=\gamma_a\, d^{n-1}x,\ \ \hat h_E=h\, d^{n-1}x+\lambda^a
\hat\gamma_a, 
\eea
an irreducible generating set of gauge transformations for
\eqref{actfirstorder} (see
e.g. \cite{Henneaux:1992ig} chapter 3) is given by
\bea
\delta_f z^A&=&\tilde\delta_{\hat\gamma_af^a}z^A=\frac{\tilde\delta
  (\gamma_a f^a)}{\delta
  z^B}\sigma^{BA},\\
\delta_f \lambda^a&=&\frac{D}{Dt} f^a +\tilde\delta_{\hat h_E}f^a
+C_{bc}^{a}(\lambda^b,f^c)+V_b^{a}(f^b),
\eea
where the gauge parameters $f^a$ may depend on $x^\mu$, the Lagrange
multipliers and their derivatives as well as the canonical variables
and their spatial derivatives and 
\bea
&&\frac{D}{Dt}=\frac{\partial}{\partial t}+\dot\lambda^a\frac{\partial}{
\partial \lambda^a}+\ddot\lambda^a\frac{\partial}{
\partial \dot\lambda^a}
+ \dots,\\
&&\{\hat \gamma_a \xi^a_1,\hat \gamma_b \xi^b_2\}
=\hat \gamma_c C^c_{ab}(\xi^a_1,\xi^b_2)+\tilde d_H(\ ),\\ 
&&\{h\,d^{n-1}x\ ,\hat \gamma_a \xi^a\} =\hat \gamma_b V^b_a(\xi^a)
+\tilde d_H(\ ),
\eea
with $\xi^a=\xi^a(x)$. 

According to section \ref{s3}, if one is interested in equivalence classes 
of exact reducibility parameters, 
the remaining
ambiguity in the $f^a$ is that two $f^a$ that agree on the
constraint surface defined by $\gamma_a$ and their spatial derivatives
have to be identified. Let $\approx^\prime 0$ denote terms that vanish
on this surface. The condition for exact reducibility parameters then
reads
\bea
&&\frac{\tilde\delta (\gamma_a f^a)}{\delta z^A}\approx^\prime 0,\\
&&\frac{D}{Dt} f^a+\tilde\delta_{\hat h_E}f^a
+C_{bc}^{a}(\lambda^b,f^c)+V_b^{a}(f^b)\approx^\prime 0.
\eea
This last condition means that one can
assume that $f^a$ is independent of the Lagrange multipliers or any of
its derivatives. This is so because the highest
order time derivative of a Lagrange multiplier needs to be multiplied
by a term that vanishes weakly, so it can be be absorbed. This can be
repeated until all the dependence on the Lagrange
multipliers has been absorbed, so that $f^a=f^a[x,z]$ and 
$D/Dt$ reduces to $\partial/\partial t$ in the above condition. 

If $\psi_A$ are arbitrary
fields with the asymptotic behaviour of the Hamiltonian evolution
equations times the volume form and $\psi_a$ arbitrary fields that
behave like the constraints times the volume form,
the conditions that determine the asymptotic reducibility parameters
$\tilde f^a(x)$ for a given background solution $\bar z^A(x)$ are
\bea
&&\psi_A\frac{\tilde\delta (\gamma_a \tilde f^a)}{\delta z^A}\Big|_{\bar
  z(x)
}\longrightarrow 0,\label{5.24}\\
&&\psi_a\Big[\frac{\partial\tilde f^a }{\partial t} 
+V_b^{a0}(\tilde f^b)\Big]
\longrightarrow 0\label{5.25}.
\eea
The first condition
implies that the Euler-Lagrange derivatives with
respect to the canonical coordinates of
the constraints contracted with the asymptotic reducibility parameters
have to vanish asymptotically when evaluated at
the background, while the second condition fixes the asymptotic time
behaviour of the asymptotic reducibility parameters. 

In order to construct the associated conservation laws, note that in
the first order case
\bea
R^i_\alpha(f^\alpha)\frac{\delta^L L}{\delta\phi^i}&=&
\frac{\tilde \delta (\gamma_cf^c)}{\delta z^C}\sigma^{CA}(-\sigma_{AB}\dot z^B-
\frac{\delta^L h}{\delta z^A}-\frac{\delta^L \lambda^a\gamma_a}{\delta
  z^A}) \nonumber\\
&&-(\frac{D}{Dt} f^a+\tilde\delta_{\hat h_E}f^a
+C_{bc}^{a}(\lambda^b,f^c)+V_b^{a}(f^b))\gamma_a
\nonumber\\
&=&-\frac{d}{dt}(f^a\gamma_a)-\partial_k \Big(
V^k_A(\frac{\delta h_E}{\delta z^C}
\sigma^{CA},\gamma_cf^c)+j^{kb}_c(\gamma_b,f^c)\big),
\eea
where the current $j^{kb}_c(Q_b,f^c)$ is determined in terms
of the Hamiltonian structure operators through the formula
\bea
\partial _k
j^{kb}_c(Q_b,f^c)&=&Q_aV^a_b(f^b)-f^bV^{+a}_b(Q_a)\nonumber\\
&&+Q_aC^a_{b,c}(\lambda^b,f^c)-C^{+a}_{bc}(\gamma_a,\lambda^b)f^c
\eea
for all $Q_b,f^c$.
Hence 
\bea
S^{0i}_\alpha(\frac{\delta L}{\delta\phi^i},f^\alpha)&\equiv& -\gamma_af^a,\\
S^{ki}_\alpha(\frac{\delta L}{\delta\phi^i},f^\alpha)&\equiv&
-V^k_A(\frac{\delta h_E}{\delta z^C}
\sigma^{CA},\gamma_cf^c)-j^{kb}_c(\gamma_b,f^c).
\eea
By applying theorem 3 for the surface $\Sigma$ defined by $t={\rm
  cste}$, 
we have {\em proved} that, if the gauge parameters $f^a$ are 
asymptotic reducibility parameters,  
the integrated constraints
\bea
G_a[f^a]=\int d^{n-1}x\ \gamma_a f^a
\eea
can be be improved through the subtraction of a surface integral
\bea
\oint_{\partial\Sigma} k^{0i}_\alpha(f^\alpha)(d^{n-2}x)_i
\eea
to a charge that is asymptotically extremal at the background 
$\bar z(x)$ for arbitrary variations of the canonical variables 
not restricted by any boundary conditions. The integrand of the
surface integral
is determined the time components of the associated 
conserved $n-2$ form $
k^{\mu\nu}_\alpha(f^\alpha)(d^{n-2}x)_{\mu\nu}$. 
This  provides an a posteriori justification of the 
Hamiltonian procedure of \cite{Regge:1974zd}. 

\subsection{Lagrangian Noether method}

Theorem 3 or its non integrated formulation in
\eqref{3.40} can be understood either as 
a precise formulation, 
a generalization or a justification 
of the Lagrangian Noether method of references 
\cite{Julia:1998ys,Silva:1998ii,Henneaux:1999ct,Julia:2000er,Julia:2000sy}.

\subsection{Covariant phase space approach}

Let us first recall the two main formulas from
the calculus of variations.
The first variational formula is simply
\bea
d_V L=d_V\phi^i\frac{\delta L}{\delta \phi^i}
+\partial_\mu V^\mu_i(d_V\phi^i,L)
\eea
The second variational formula is obtained by applying $d_V$ (which 
is  equivalent to taking two different variations and 
skew-symmetrizing): 
\bea
0=-d_V\phi^id_V\frac{\delta L}{\delta \phi^i}
+\partial_\mu \omega^\mu,
\eea
where the presymplectic current is defined by 
\bea
\ \omega^\mu=d_V V^\mu_i(d_V\phi^i,L).
\eea
This formula is contracted with the evolutionary vector field 
$R^j_f\frac{\partial}{\partial \phi^j}$, i.e., 
a gauge transformation
\bea
0=-R^i_fd_V\frac{\delta L}{\delta \phi^i}+
d_V\phi^i
\partial_{(\mu)}R^j_f
\frac{\partial^S}{\partial \phi^j_{(\mu)}}\frac{\delta L}{\delta
  \phi^i}
+\partial_\mu i_{R_f}\omega^\mu,
\eea
where $i_{R_f}\omega^\mu$ denotes the contraction of $\omega^\mu$ with
the evolutionary vector field
$R^j_f\frac{\partial}{\partial \phi^j}$.
To the second term, we apply the formula
for the commutator of an Euler-Lagrange derivative and
an evolutionary vector field
(see e.g. \cite{Barnich:2000zw} equation (6.43)). Using the fact
that $R^j_f\frac{\delta L}{\delta \phi^j}$ is a total
divergence and is thus annihilitated by the Euler-Lagrange derivative, we
get 
\bea
0=-R^i_fd_V\frac{\delta L}{\delta \phi^i}-
d_V\phi^i (-\partial)_{(\lambda)}[
\frac{\partial^S R^j_f}
{\partial \phi^i_{(\lambda)}}\frac{\delta L}{\delta
  \phi^j}]
+\partial_\mu i_{R_f}\omega^\mu.
\eea
Applying repeatedly Leibniz' rule to the second term, we obtain
\bea
0=-d_V(R^i_f\frac{\delta L}{\delta
  \phi^i})+\partial_\mu [i_{R_f}\omega^\mu+ t^{\mu,1}],
\eea
where $t^{\mu,1}$ is in vertical degree $1$ and vanishes
weakly as it is linear in the Euler-Lagrange
derivatives $\delta L/\delta\phi^i$ and their derivatives. 
Finally, using (\ref{sec1cur}), we deduce
\bea
\partial_\mu (i_{R_f}\omega^\mu+t^{\mu,1}
-d_V (S^{\mu i}_\alpha(\frac{\delta  L}{\delta\phi^i}, f^\alpha))=0,
\eea
Hence, there exists an $n-2$ form $r^{[\nu\mu],1}$ 
in vertical degree $1$ such that 
\bea
d_V(S^{\mu i}_\alpha(\frac{\delta
  L}{\delta\phi^i},
f^\alpha))=i_{R_f}\omega^\mu+t^{\mu,1}+\partial_\nu
r^{[\nu\mu],1},
\eea
Evaluating at a solution $\bar\phi(x)$ 
of the Euler-Lagrange equations of motion, this gives
\bea
S^{\mu i}_\alpha|_{\bar\phi(x)}(d_V\frac{\delta
  L}{\delta\phi^i}|_{\bar\phi(x)},
f^\alpha|_{\bar\phi(x)})(d^{n-1}x)_\mu
=i_{R_f}\omega^\mu|_{\bar\phi(x)}(d^{n-1}x)_\mu+\partial_\nu
r^{[\nu\mu],1}|_{\bar\phi(x)}(d^{n-1}x)_\mu,
\eea
Since our results imply that the the left hand side of this 
equation reduces to the exterior derivative of a conserved $n-2$ form
if and only if the $f^\alpha|_{\bar\phi(x)}$ are asymptotic
reducibility parameters, we get in this case, 
\bea
-d_H( d_V k^{[\nu\mu]}_{f}(d^{n-2}x)_{\nu\mu}
\longrightarrow i_{R_f}\omega^\mu|_{\bar\phi(x)}(d^{n-1}x)_\mu-
d_H(r^{[\nu\mu],1}|_{\bar\phi(x)}(d^{n-2}x)_{\nu\mu})
\eea
In other words, it is possible to subtract the exterior derivative of 
an $n-2$ form 
$r^{[\nu\mu],1}|_{\bar\phi(x)}(d^{n-2}x)_{\nu\mu}$ 
from the presymplectic $n-1$ form contracted with a gauge
transformations in order to get asymptotically 
the exterior derivative of a conserved $n-2$ form if and only if the gauge
parameters define asymptotic reducibility parameters. 

\subsection{Characteristic cohomology and generalized symmetries}

Algebraic and differential-geometric techniques have been used for
quite some time for the symmetry analysis of standard partial differential
equations (see e.g.
\cite{Olver:1993,Dickey:1991xa,Anderson1991,Andersonbook}
and references therein).
The application of these techniques in the
context of gauge theories, i.e., possibly degenerate Lagrangian field
theories, is more involved because of the presence of gauge
symmetries
\cite{Krasil'shchikbook,Bryant:1995,Barnich:1995db,Krasil'shchiklectures}.
Recent results show that in interacting gauge theories,
there are only few non trivial global symmetries involving the gauge fields
alone. For instance, there are none in the case of
pure four dimensional Einstein gravity \cite{Torre:1993jm,Anderson:1996eg}.
(See also \cite{Pohjanpelto:2001ua} for a recent 
classification of generalized symmetries of semi-simple 
Yang-Mills theory and compare to free electromagnetism treated in 
\cite{Lipkin,Morgan,Kibble,OConnell,Anco:2001}). 
Similarly, there are only few exact lower degree conservation laws. 
For instance, 
there are none in semi-simple Yang-Mills theory or in Einstein 
gravity (see e.g. \cite{Barnich:1995db}).

In \cite{Anderson:1996sc}, lower degree linear
characteristic cohomology for generic second-order field equations
have been 
classified directly. In particular, the equations are not assumed to
derive from a Lagrangian. 
The technical starting point for the current investigation has been 
the ``central premise'' of \cite{Anderson:1996sc} (see also 
\cite{Torre:1997cd}) 
that
cohomological techniques ``can be successfully adapted to the analysis
of asymptotic conservation laws''. 
We have made more restrictive assumptions,
namely, that the field equations are Lagrangian and that a generating
set of gauge symmetries is known, and have suitably extended the 
approach of \cite{Anderson:1996sc} by introducing additional cohomological
tools of the BRST-antifield formalism. This has allowed us 
to flesh out the general classification
theorem of \cite{Anderson:1996sc}. 

\addcontentsline{toc}{section}{Conclusion}

\section*{Conclusion}

Using cohomological tools as a guiding line,
we have investigated in detail asymptotic symmetries and
conservation laws, their relation and their algebra.  

For simplicity and clarity, 
we have restricted the present investigation to the case of
irreducible gauge theories. The extension to reducible 
gauge theories or non trivial topology 
along these lines is straightforward, 
because the associated cohomological techniques are well under control
(see e.g. \cite{Andersonbook,Barnich:1995ap}).

We stress that the cohomological methods used in this paper are 
technical tools which are not necessary but
rather convenient to derive the results. In particular
they motivate the definitions of equivalence 
classes of asymptotic
symmetries and conservation laws and 
facilitate the proof of the
various statements. This is because they take into account in a
natural way the ambiguities inherent in the definitions. 
The various definitions and results are stated in this paper 
both in terms of equivalence classes of quantities that involve only 
the original fields and also in terms of local BRST cohomology classes. 
The latter have the advantage that they are manifestly invariant 
under field redefinitions, changes in the 
description of the generating set of gauge transformations and elimination 
of auxiliary and generalized auxiliary fields, because these 
operations do not modify the local BRST cohomology. 

The most essential prerequisites for the validity of 
the results presented in this work are that the theory 
is asymptotically linear and the ``asymptotic
acyclicity properties'',
as they play a central role in the cohomological analysis.

Sufficient conditions on the asymptotic behaviour of 
fields and gauge parameters 
have been given that guarantee bijective correspondence 
between equivalence classes of 
asymptotic reducibility parameters and asymptotically 
conserved $n-2$ forms, finite charges, a well defined algebra 
and finite central 
charges. 
It would be of interest to investigate in how far 
these desirable properties still hold for more 
relaxed assumptions, or for different formulations of boundary 
conditions.

We have tested in this paper our general approach in the non trivial case 
of three dimensional anti-de Sitter gravity and have been able to
reproduce in a straightforward way the results originally obtained by 
Hamiltonian methods in \cite{Brown:1986ed}. In the future, we hope to
report on new applications of the present analysis, in particular 
for models that have not been previously treated in the
canonical framework.

\addcontentsline{toc}{section}{Acknowledgments}

\section*{Acknowledgments}

G.B. wants to thank Ian Anderson and Charles Torre for
drawing his attention to reference \cite{Anderson:1996sc} and
the relevance of cohomological methods in
the study of asymptotic symmetries and conservation laws. Useful
discussions by F.B. with Sebasti\'an Silva and 
by G.B. with Rodrigo Aros, Xavier Bekaert, Marc Henneaux, 
Christi\'an Mart\'{\i}nez and 
Ricardo Troncoso are gratefully acknowledged. 

During 
various stages of this work, G.B. has enjoyed the kind hospitality 
of the group 
``Formal Geometry and Mathematical Physics'' of the Utah State University
and of the Max-Planck-Institute for Mathematics in the Sciences in
Leipzig.  

This research has been partially supported by the ``Actions de
Recherche Concert{\'e}es" of the ``Direction de la Recherche
Scientifique - Communaut{\'e} Fran{\c c}aise de Belgique", by
IISN - Belgium (convention 4.4505.86),  by
Proyectos FONDECYT 1970151 and 7960001 (Chile) and 
by the European Commission RTN programme HPRN-CT-00131, in which G.B.
is associated to the K.U. Leuven. 

\appendix

\mysection{Appendix}

\subsection{Conventions and notation}

We assume for notational
simplicity that all fields $\phi^i$ are 
(Grassmann) even.

Consider $k$-th order derivatives 
$\frac{\6^k\phi^i(x)}{\6 x^{\mu_1}\dots\6
  x^{\mu_k}}$
of a field $\phi^i(x)$. The corresponding jet-coordinate
is denoted by $\phi^i_{\mu_1\dots\mu_k}$.
Because the derivatives are symmetric under 
permutations of the derivative indices $\mu_1,\dots,\mu_k$,
these jet-coordinates are not independent, but
one has $\phi^i_{\mu\nu}=\phi^i_{\nu\mu}$ etc.
Local functions are smooth functions depending on the
coordinates $x^\mu$ of the base space $M$, the fields
$\phi^i$, and a finite number of the jet-coordinates 
$\phi^i_{\mu_1\dots\mu_k}$. Local 
horizontal forms involve in addition
the differentials $dx^\mu$ which we treat as
anticommuting (Grassmann odd) variables, 
$dx^\mu dx^\nu=-dx^\nu dx^\mu$.

As in \cite{DeDonder1935,Andersonbook}, we define derivatives 
$\6^S/\6\phi^i_{\mu_1\dots\mu_k}$ that act on the 
basic variables through
\bea
&&\frac{\6^S \phi^j_{\nu_1\dots\nu_k}}{\6\phi^i_{\mu_1\dots\mu_k}}
=\delta^j_i\, \delta^{\mu_1}_{(\nu_1}\dots\delta^{\mu_k}_{\nu_k)}\ ,\quad
\frac{\6^S \phi^j_{\nu_1\dots\nu_m}}{\6\phi^i_{\mu_1\dots\mu_k}}
=0\quad\mbox{for $m\neq k$},
\nonumber\\[6pt]
&&\frac{\6^S x^\mu}{\6\phi^i_{\mu_1\dots\mu_k}}=0,\quad
\frac{\6^S dx^\mu}{\6\phi^i_{\mu_1\dots\mu_k}}=0,
\label{derivS}
\eea
where the round parantheses denote symmetrization
with weight one, 
\[
\delta^{\mu_1}_{(\nu_1}\delta^{\mu_2}_{\nu_2)}
=\frac 12 (\delta^{\mu_1}_{\nu_1}\delta^{\mu_2}_{\nu_2}
+\delta^{\mu_1}_{\nu_2}\delta^{\mu_2}_{\nu_1})\quad
\mbox{etc.}
\] 
For instance, the definition gives explicitly (with
$\phi$ any of the $\phi^i$):
\[
\frac{\6^S\phi_{11}}{\6\phi_{11}}=1\ ,\quad
\frac{\6^S\phi_{12}}{\6\phi_{12}}=
\frac{\6^S\phi_{21}}{\6\phi_{12}}=\frac 12\ ,\quad
\frac{\6^S\phi_{112}}{\6\phi_{112}}=\frac 13\ ,\quad
\frac{\6^S\phi_{123}}{\6\phi_{123}}=\frac 16\ .
\]
We note that the use of
these operators takes automatically
care of many combinatorical factors which arise
in other conventions, such as those used in
\cite{Olver:1993}.

The vertical
differential is defined by \eqref{vertdiff}
with Grassmann odd
generators $d_V\phi^i_{\mu_1\dots\mu_k}$, so that $d_V^2=0$. 
The total derivative is the vector field denoted by 
$\partial_\nu$ and acts on local functions according to 
\bea
\partial_\nu=\frac{\partial}{\partial x^\nu}+\sum_{k=0}
\phi^i_{\mu_1\dots\mu_k\nu}\,\frac{\partial^S}{\partial
  \phi^i_{\mu_1\dots\mu_k}}\ . \label{totder}
\eea
Here $\sum_{k=0}$ means the sum over all $k$, from
$k=0$ to infinity, with
the summand for $k=0$ given by
$\phi^i_\nu\6/\6\phi^i$, i.e., by definition $k=0$ means
``no indices  $\mu_i$''. Furthermore we are using
Einstein's summation convention over repeated indices,
i.e., for each $k$ there is a summation
over all tupels $(\mu_1,\dots,\mu_k)$.
Hence, for $k=2$, the sum over $\mu_1$ and $\mu_2$
contains both the tupel $(\mu_1,\mu_2)=(1,2)$
and the tupel $(\mu_1,\mu_2)=(2,1)$.
These conventions
extend to all other sums
of similar type. 

The horizontal differential on horizontal forms is 
defined by $d_H=dx^\nu\partial_\nu$. It is extended to the vertical
generators in such a way that $\{d_H,d_V\}=0$. 

A vector field of the form
$Q^i\partial/\partial\phi^i$, for $Q^i$ a set of local functions,
is called an evolutionary vector field with characteristic $Q^i$.
Its prolongation which acts on local functions is 
\bea
\delta_Q
=\sum_{k=0}
(\partial_{\mu_1}\dots\partial_{\mu_k}Q^i)
\, \frac{\partial^S}{\partial\phi^i_{\mu_1\dots\mu_k}}\ .
\label{evvf}
\eea 
More details on the variational bicomplex can be found 
for instance in the textbooks
\cite{Olver:1993,Saunders:1989,Dickey:1991xa,Andersonbook}.

The set of multiindices is simply the
set of all tupels $(\mu_1,\dots,\mu_k)$, including
(for $k=0$) the empty tupel. The tuple with one element is denoted by
$\mu_1$ without round parentheses, while a generic tuple is denoted by
$(\mu)$. The length, i.e., the number of individual
indices, of a multiindex $(\mu)$ is denoted by
$|\mu|$. 
We use Einstein's summation convention also for repeated
multiindices as in \cite{Andersonbook}. For instance, 
an expression of the type
$(-\partial)_{(\mu)} K^{(\mu)}$ stands for a free sum
over all tupels $(\mu_1,\dots,\mu_k)$
analogous to the one in (\ref{totder}),
\[
(-\partial)_{(\mu)} K^{(\mu)}=\sum_{k=0}(-)^k 
\partial_{\mu_1}\dots\partial_{\mu_k}K^{\mu_1\dots\mu_k}.
\] 

If
$Z=Z^{(\mu)}\partial_{(\mu)}$ is a differential operator, 
its adjoint is defined by 
$Z^+=(-\partial)_{(\nu)}[
Z^{(\nu)}\cdot]$ and its `components' are
denoted by $Z^{+(\mu)}$, i.e.,
$Z^+=Z^{+(\mu)}\partial_{(\mu)}$. 
Furthermore, we assume that
the Euler-Lagrange equations of motion and their total derivatives
$\partial_{(\mu)}{\delta L}/{\delta \phi^i}$
satisfy suitable
regularity conditions, 
such that a local function $f$ vanishes
when evaluated on {\em every} solution $\phi^i(x)$
of the Euler-Lagrange equations of motion 
($f|_{\phi(x)}=0$ whenever $({\delta L}/{\delta \phi^i})|_{\phi(x)}=0$)
if and only if
$f=G^i{\delta L}/{\delta \phi^i}$ for some differential operators
$G^i=G^{i(\mu)}\partial_{(\mu)}$.

\subsection{Higher order Lie-Euler operators}

Except for a different notation, we follow in this and the next
subsection \cite{Andersonbook}. 
The higher order Lie-Euler operators
$\delta/\delta\phi^i_{\mu_1\dots\mu_k}$ can be defined through 
the formula
\bea
\forall Q^i:\quad
\delta_Q f= \partial_{(\mu)}\Big[
Q^i\,\frac{\delta f}{\delta \phi^i_{(\mu)}}
\Big].
\eea
Explicitly,
\bea
\frac{\delta f}{\delta \phi^i_{(\mu)}}=\left(\begin{array}{c}
|\mu|+|\nu| \\ |\mu| \end{array}\right)(-\partial)_{(\nu)}\,
\frac{\6^S f}{\6\phi^i_{((\mu)(\nu))}}\ ,
\label{higherEOp}
\eea
or, equivalently,
\bea
\frac{\delta f}{\delta \phi^i_{\mu_1\dots\mu_k}}=
\sum_{l=0}\left(\begin{array}{c}
 k+l \\ k \end{array}\right)(-)^l\partial_{\nu_1}\dots\partial_{\nu_l}\,
\frac{\6^S f}{\6\phi^i_{\mu_1\dots\mu_k\nu_1\dots\nu_l}}\ ,
\eea
i.e., there is a summation over $(\nu)$
in (\ref{higherEOp}) by Einstein's
summation convention for repeated multiindices, and
the multiindex $((\mu)(\nu))$ is the tupel
$(\mu_1,\dots,\mu_k,\nu_1,\dots,\nu_l)$ when 
$(\mu)$ and $(\nu)$ are the tupels
$(\mu_1,\dots,\mu_k)$ and $(\nu_1,\dots,\nu_l)$, respectively.
Note that the sum contains only finitely many nonvanishing
terms whenever $f$ is a local function:
if $f$ depends only
on variables with at most $M$ ``derivatives'', i.e., on the
$\phi^i_{(\rho)}$ with $|\rho|\leq M$, the only possibly
nonvanishing summands are those with
$|\nu|\leq M-|\mu|$ ($l\leq M-k$).
Note also that $\delta/\delta\phi^i$ is the Euler-Lagrange
derivative.

The crucial property of these operators is that they ``absorb total
derivatives'',
\bea
&|\mu|=0:& \frac{\delta (\partial_\nu f)}{\delta \phi^i}=0,\label{eA7}\\
&|\mu|>0:&
\frac{\delta (\partial_\nu f)}{\delta \phi^i_{(\mu)}}
=\delta^{(\mu}_\nu\frac{\delta f}{\delta 
\phi^i_{(\mu^\prime))}}\ ,\quad (\mu)=(\mu(\mu^\prime)),\label{eA8}
\eea 
where, e.g., 
\[
\delta^{(\mu}_\nu\frac{\delta f}{\delta 
\phi^i_{\lambda)}}=\frac{1}{2}\big(\delta^{\mu}_\nu\frac{\delta f}{\delta 
\phi^i_{\lambda}}+\delta^{\lambda}_\nu\frac{\delta f}{\delta 
\phi^i_{\mu}}\big).
\]

\subsection{Contracting homotopy for the horizontal complex}

Define 
\bea
\rho^p_{H,\phi}(\omega^{p})=\int_0^1dt \
\frac{|\mu|+1}{n-p+|\mu|+1}\ \partial_{(\mu)}\Bigg(
\phi^i \Big[
\frac{\delta}{\delta\phi^i_{((\mu)\nu)}}\,\frac{\partial
\omega^{p}}{\partial dx^\nu}\Big][x,t\phi]\Bigg)
\label{phihomotopy}
\eea
for $\omega^{p}$ a horizontal $p$-form (note that
there is a summation over $(\mu)$ by Einstein's
summation convention).
Then:
\bea
&0\leq p< n:&
\omega^p[x,\phi]-\omega^p[x,0]=
\rho^{p+1}_{H,\phi}(d_H\omega)+d_H(\rho^p_{H,\phi}\omega);
\label{cc1}\\
&p=n:&
\omega^n[x,\phi]-\omega^n[x,0]=\int_0^1dt\ \phi^i\big[
\frac{\delta\omega^n}{\delta\phi^i}\big][x,t\phi]
+d_H(\rho^n_{H,\phi}\omega^n).
\eea

\subsection{Proof of theorem \ref{thmalgebra}}
\label{appendixproof}

\paragraph{Proof of equations (\ref{asympcom})
and (\ref{centralcharge1}).}

Using \eqref{asygauge} and \eqref{convergencec}, we have
$\delta_{\tilde f_1}\tilde k_{\tilde f_2}\asy \delta^0_{\tilde
  f_1}\tilde k_{\tilde
  f_2}
+\delta^1_{\tilde f_1}\tilde k_{\tilde
  f_2}$. According to \eqref{i1.25}, $\delta^1_{\tilde f_1}\tilde k_{\tilde
  f_2}\stackrel{\approx^{\rm free}}{\asy}\tilde
k_{[\tilde f_1,\tilde f_2]_M}+d_H(\
)$. Integrating over the boundary ${\cal C}^{n-2}$ and using the
definition \eqref{defcharge} togther with \eqref{lineom} 
then gives directly 
(\ref{asympcom}) and (\ref{centralcharge1}).

\paragraph{Proof of equations (\ref{AsymmGeneral}) and (\ref{Asymm}).}
Consider the $(n-1)$-form 
\[
s_f[\varphi,\5\phi(x)]=s^\mu_f[\varphi,\5\phi(x)](d^{n-1}x)_\mu
\]
where $s^\mu_f[\varphi,\5\phi(x)]$ is
the weakly vanishing Noether current of the free theory 
(\ref{smalls}) for arbitrary field independent
gauge parameters $f^\alpha$ (rather than for
asymptotic reducibility parameters).
We apply formula (\ref{cc1})  to this $(n-1)$-form, using
the contracting homotopy
(\ref{phihomotopy}) for the $\varphi$.
Since each term in
$s_f[\varphi,\5\phi(x)]$ is linear and homogeneous in
the $\varphi$ and their
derivatives, one has $s_f[0,\5\phi(x)]=0$ (and the integral over $t$
can be evaluated trivially). 
Furthermore, $\rho^{n-1}_{H,\varphi}s_f[\varphi,\5\phi(x)]$
is nothing but $-\4k_f[\varphi,\5\phi(x)]$.
Hence, we obtain from (\ref{cc1}):
\bea
s_f[\varphi,\5\phi(x)]=-d_H \4k_f[\varphi,\5\phi(x)]
+\rho^n_{H,\varphi}d_H s_f[\varphi,\5\phi(x)].
\label{cc2}
\eea
We now apply a transformation $\delta^0_{f'}$ 
($\delta^0_{f'}\varphi^i=R^{i0}_\alpha f^{\alpha\prime}$)
with arbitrary field independent gauge parameters $f^{\alpha\prime}$
to (\ref{cc2}). The
facts that $\delta^0_{f'}$ is a
gauge symmetry of $L^\mathrm{free}$ and that
$\delta^0_{f'}\varphi^i$ does not depend on the
$\varphi$ or their derivatives implies
\bea
\delta^0_{f'}\, \frac{\delta L^\mathrm{free}}{\delta\varphi^i}=0.
\label{cc3}
\eea
[This can be verified using
formula (6.43) in \cite{Barnich:2000zw}, for instance.]
As $s_f[\varphi,\5\phi(x)]$ depends on the $\varphi$
and their derivatives only via the $\delta L^\mathrm{free}/\delta\varphi^i$,
(\ref{cc3}) implies $\delta^0_{f'}s_f[\varphi,\5\phi(x)]=0$.
Hence, applying $\delta^0_{f'}$ to (\ref{cc2}), one obtains
\bea
d_H \4k_f[R^0_{f'},\5\phi(x)]
=\delta^0_{f'}\rho^n_{H,\varphi}d_H s_f[\varphi,\5\phi(x)]=
\delta^0_{f'}\rho^n_{H,\varphi}[\delta^0_f\varphi^iL^\mathrm{free}_id^nx],
\label{cc4}
\eea
where $L^\mathrm{free}_i\equiv {\delta
  L^\mathrm{free}}/{\delta\varphi^i}$. 
The point is now that the last expression 
on the right hand side of (\ref{cc4})
is skew symmetric under the exchange of $f$ and $f'$,
\bea
\delta^0_{f'}\rho^n_{H,\varphi}[\delta^0_f\varphi^iL^\mathrm{free}_id^nx]
=-\delta^0_{f}\rho^n_{H,\varphi}[\delta^0_{f'}\varphi^iL^\mathrm{free}_id^nx]
\label{cc5}
\eea
and thus also 
\bea
\delta^0_{f'}\rho^n_{H,\varphi}d_H s_f[\varphi,\5\phi(x)]=
-\delta^0_{f}\rho^n_{H,\varphi}d_H s_{f'}[\varphi,\5\phi(x)],
\label{cc5bis}
\eea
owing to the properties of $\rho^n_{H,\varphi}$.
Let us postpone the demonstration of
(\ref{cc5}) and first
complete the proof of the theorem, assuming that 
(\ref{cc5}) holds.
Equations (\ref{cc4}) and (\ref{cc5bis}) imply
\bea
d_H (\4k_f[R^0_{f'},\5\phi(x)]
+\4k_{f'}[R^0_{f},\5\phi(x)])=0.
\label{cc66}
\eea
As $f^\alpha$ (and $f^{\alpha\prime}$) are
arbitrary functions, one can apply the contracting homotopy 
$\rho^{n-1}_{H,f}$.  Using \eqref{cc1} and
$\4k_{0}[R^0_{f^\prime},\5\phi(x)]=0
=\4k_{f'}[0,\5\phi(x)]$ gives
\bea
\4k_f[R^0_{f'},\5\phi(x)]
+\4k_{f'}[R^0_{f},\5\phi(x)]=
d_H\rho^{n-2}_{H,f}\Big(\4k_f[R^0_{f'},\5\phi(x)]
+\4k_{f'}[R^0_{f},\5\phi(x)]\Big),
\label{cc6}
\eea
proving \eqref{AsymmGeneral}.
Integration of (\ref{cc6}) over ${\cal C}^{n-2}$ yields
\bea
K_{f',f}+K_{f,f'}=0,
\label{cc7}
\eea
for any 
$f^\alpha(x)$ and $f^{\alpha\prime}(x)$.
This implies (\ref{Asymm}) and shows that the skew
symmetry of $K$ is actually not restricted to
asymptotic
reducibility parameters but holds for to
general gauge parameters. Notice also that this
proof of (\ref{Asymm}) uses only the standard algebraic Poincar\'e
lemma rather than its asymptotic version (\ref{7.66}), i.e.,
(\ref{cc7}) holds independently of assumptions on the
boundary conditions.

\paragraph{Proof of equation (\ref{Cocycle}).}
The proof of (\ref{Cocycle}) is now very easy.
(\ref{1.18}) implies
\bea
[\delta_{\4f_1},\delta_{\4f_2}]Q_{\4f_3}
\approx \delta_{[\4f_1,\4f_2]_P}Q_{\4f_3}\ ,
\label{cc8}
\eea
We first evaluate the commutator on the
left hand side of (\ref{cc8}) using
(\ref{asympcom}), and then extract from the
result the $\varphi$-independent part. We obtain
from (\ref{asympcom}):
\[
\delta_{\4f_1}(\delta_{\4f_2}Q_{\4f_3})\sim
\delta_{\4f_1}Q_{[\4f_2,\4f_3]_M}\sim
Q_{[\4f_1,[\4f_2,\4f_3]_M]_M}+
K_{\4f_1,[\4f_2,\4f_3]_M}-N_{[\4f_1,[\4f_2,\4f_3]_M]_M} .
\]
Hence, the $\varphi$-independent part of the
left hand side of (\ref{cc8}) is
\[
K_{\4f_1,[\4f_2,\4f_3]_M}-K_{\4f_2,[\4f_1,\4f_3]_M}-
N_{[\4f_1,[\4f_2,\4f_3]_M]_M}+N_{[\4f_2,[\4f_1,\4f_3]_M]_M}.
\]
The $\varphi$-independent part of the right hand side
is $\sim (K_{[\4f_1,\4f_2]_M,\4f_3}-N_{[\4f_1,\4f_2]_M,\4f_3]_M})$,  
where we used
again (\ref{asympcom}).
Since (\ref{cc8}) gives an exact equality for the
$\varphi$-independent part (the 
equation-of-motion-terms do not contribute to this
part since they are at least linear in the $\varphi$), 
and since the terms involving the normalization vanish 
separately because of the Jacobi identity for $[\cdot,\cdot]_M$,
(which corresponds to $(\delta^{CE})^2 N=0$),  
we obtain
\bea
K_{\4f_1,[\4f_2,\4f_3]_M}-K_{\4f_2,[\4f_1,\4f_3]_M}
=K_{[\4f_1,\4f_2]_M,\4f_3}\ .
\label{cc9}
\eea
Equation (\ref{Cocycle}) follows immediately from
(\ref{cc9}) and (\ref{cc7}).

\paragraph{Direct demonstration of equation (\ref{cc5}).}
Writing out $\rho^n_{H,\varphi}$ in the left hand side of (\ref{cc5})
gives
\bea
&&
\rho^n_{H,\varphi}
(\delta^0_f\varphi^iL^\mathrm{free}_i d^nx)
= \omega_f^\nu (d^{n-1}x)_\nu,
\nonumber\\
&&
\omega_f^\nu=
\left(\begin{array}{c}|\mu|+|\rho|+1\\ |\mu|+1\end{array}\right)
\6_{(\mu)}
\Big[
\varphi^i
(-\6)_{(\rho)}
\frac{\6^S \delta^0_f\varphi^jL^\mathrm{free}_j}
{\6\varphi^i_{((\mu)(\rho)\nu)}}
\Big].
\label{nn3}
\eea
Equation (\ref{cc5}) can be directly verified by evaluation of (\ref{nn3})
for a general quadratic Lagrangian $L^\mathrm{free}$.
Up to a total divergence which
can be neglected because it gives no contribution to the
Euler-Lagrange derivatives, every Lagrangian $L^\mathrm{free}$
takes the form
\bea
L^\mathrm{free}=\varphi^i a_{ij}^{(\mu)}\varphi^j_{(\mu)}\ ,
\label{nn4}
\eea
where $a_{ij}^{(\mu)}$
are $x$-dependent coefficient functions (of the background fields and their
derivatives). The Euler-Lagrange derivatives of $L^\mathrm{free}$ are
\bea
L^\mathrm{free}_i=
a_{ij}^{(\mu)}\varphi^j_{(\mu)}
+(-)^{|\mu|+|\rho|}
\left(\begin{array}{c}|\mu|+|\rho|\\ |\rho|\end{array}\right)
\varphi^j_{(\mu)} 
\6_{(\rho)}a_{ji}^{((\mu)(\rho))}.
\label{nn6}
\eea
One now inserts (\ref{nn6}) in (\ref{nn3})
and verifies (\ref{cc5}) by direct computation.
This reduces to an exercise
in binomial coefficients.
The binomial coefficients which occur are those
in (\ref{nn3}), those in (\ref{nn6}) and those
coming from distributing the
derivatives in $\6_{(\mu)}$ and $(-\6)_{(\rho)}$
occurring in (\ref{nn3}).
The computation is  straightforward but the formulas
become involved.
Let us explicitly demonstrate it
for a Lagrangian with two derivatives
because it involves all characteristic
features (the cases without or with only one derivative are
rather trivial),
\beann
L^\mathrm{free}=\varphi^i a_{ij}^{\mu\nu}\varphi^j_{\mu\nu}\ .
\eeann
Its Euler-Lagrange derivatives are
\beann
L^\mathrm{free}_{i}=
\varphi^j_{\mu\nu}(a_{ij}^{\mu\nu}+a_{ji}^{\mu\nu})
+2\varphi^j_\mu \6_\nu a_{ji}^{\mu\nu}
+\varphi^j\6_\mu\6_\nu a_{ji}^{\mu\nu}.
\eeann
Since these contain no third or higher order
derivatives of the $\varphi$, the only nonvanishing
contributions to the sums in (\ref{nn3}) are
those with $(|\mu|,|\rho|)\in\{(0,0),(1,0),(0,1)\}$,
\beann
\omega_f^\nu&=&
\varphi^i\,\frac{\6^S \delta^0_f\varphi^jL^\mathrm{free}_j}{\6\varphi^i_\nu}
+\6_\mu\Big[\varphi^i\,
\frac{\6^S \delta^0_f\varphi^jL^\mathrm{free}_j}{\6\varphi^i_{\mu\nu}}
\Big]
-2\varphi^i\6_\rho\,
\frac{\6^S \delta^0_f\varphi^jL^\mathrm{free}_j}{\6\varphi^i_{\rho\nu}}
\\
&=&
\varphi^i\,\frac{\6^S \delta^0_f\varphi^jL^\mathrm{free}_j}{\6\varphi^i_\nu}
+\varphi^i_\mu\,
\frac{\6^S \delta^0_f\varphi^jL^\mathrm{free}_j}{\6\varphi^i_{\mu\nu}}
-\varphi^i\6_\mu\,
\frac{\6^S\delta^0_f\varphi^jL^\mathrm{free}_j}{\6\varphi^i_{\mu\nu}}\ .
\eeann
One explicitly obtains
\beann
\omega_f^\nu=
2\varphi^i\delta^0_f\varphi^j
\6_\mu a_{ij}^{\mu\nu}
+\varphi^i_\mu\delta^0_f\varphi^j
(a_{ij}^{\mu\nu}+a_{ji}^{\mu\nu})
-\varphi^i\6_\mu[(a_{ij}^{\mu\nu}+a_{ji}^{\mu\nu})\delta^0_f\varphi^j]
\\
\Longrightarrow\quad
\delta^0_{f'}\omega_f^\nu=
\delta^0_{f'}\varphi^i\delta^0_f\varphi^j
\6_\mu a_{ij}^{\mu\nu}
+\delta^0_{f'}\varphi^i_\mu\delta^0_f\varphi^j
(a_{ij}^{\mu\nu}+a_{ji}^{\mu\nu})-(f\leftrightarrow f'),
\eeann
which demonstrates (\ref{cc5}) for this case.
Analogously one can verify  (\ref{cc5}) for a 
Lagrangian with any other fixed number of derivatives which
then implies (\ref{cc5}) because every
Lagrangian $L^\mathrm{free}$ is a linear combination of 
such particular Lagrangians.

\addcontentsline{toc}{section}{References}

\providecommand{\href}[2]{#2}\begingroup\raggedright\endgroup

\end{document}